%% file: Bayesian_Hierarchical_Modeling_on_Covariance_Valued_Data.tex
\documentclass[11pt]{article}
 \usepackage{amsmath, amsthm, amssymb, bbm, setspace,bigints}
 \usepackage[margin=1 in]{geometry}
\usepackage{caption}
\usepackage{subcaption}
\usepackage[toc,page]{appendix}
\usepackage{cite}         
\usepackage[pdftex]{graphicx}
\usepackage{pdfpages}
\usepackage{epstopdf}
\usepackage{booktabs}
\usepackage{enumerate}
\usepackage{ragged2e}
\usepackage{csquotes}
\usepackage{textcomp}
\usepackage{hhline}
\usepackage{multirow}
\usepackage{xcolor,colortbl}
\usepackage{url}
\usepackage{chngcntr}
\usepackage{minitoc}
\usepackage{mathtools}
\usepackage{natbib}

\newcommand{\be} {\begin{eqnarray*}}
\newcommand{\ee} {\end{eqnarray*}}

\def\T{{ \mathrm{\scriptscriptstyle T} }}

\def\mb{\mathbb}

\def\tr{{\rm tr\,}}

\definecolor{Gray}{gray}{0.85}
\definecolor{LightCyan}{rgb}{0.88,1,1}

\newcolumntype{a}{>{\columncolor{Gray}}c}
\newcolumntype{b}{>{\columncolor{white}}c}

\theoremstyle{definition}
\title{ Bayesian Hierarchical Modeling on Covariance Valued Data}
\author{ {\bf Satwik Acharyya} \\ Department of Statistics, Texas A\&M University, \\ College Station, TX, \\email: satwik@stat.tamu.edu \\
{ \bf Zhengwu Zhang} \\ Department of Biostatistics and Computational Biology, University of Rochester Medical Center, \\ Rochester, NY, \\ email : zhengwu\_zhang@urmc.rochester.edu \\
{\bf Anirban Bhattacharya}\\ Department of Statistics, Texas A\&M University, \\ College Station, TX, \\email: anirbanb@stat.tamu.edu\\
{\bf Debdeep Pati} \\Department of Statistics, Texas A\&M University,\\ College Station, TX, \\email: debdeep@stat.tamu.edu\\
}
\date{ \hspace{1.5cm} \today}    

\begin{document}
\maketitle
\doparttoc % Tell to minitoc to generate a toc for the parts
\faketableofcontents % Run a fake tableofcontents command for the partocs

%\doublespacing

\textbf{Abstract} : Analysis of structural and functional connectivity (FC) of human brains is of pivotal importance for diagnosis of cognitive ability. The Human Connectome Project (HCP) provides an excellent source of neural data across different regions of interest (ROIs) of the living human brain. Individual specific data were available from an existing analysis \citep{dai2017discovering} in the form of time varying covariance matrices representing the brain activity as the subjects perform a specific task. As a preliminary objective of studying the heterogeneity of brain connectomics across the population, we develop a  probabilistic model for a sample of covariance matrices using a scaled Wishart distribution.  We stress here that our data units are available in the form of covariance matrices, and we use the Wishart distribution to create our likelihood function rather than its more common usage as a prior on covariance matrices. Based on empirical explorations suggesting the data matrices to have low effective rank, we further model the center of the Wishart distribution using an orthogonal factor model type decomposition. We encourage shrinkage towards a low rank structure through a novel shrinkage prior and discuss strategies to sample from the posterior distribution using a combination of Gibbs and slice sampling. The efficacy of the approach is explored in various simulation settings and exemplified on several case studies including our motivating HCP data. We extend our modeling framework to a dynamic setting to detect change points.

\section{Introduction}
Functional connectomes play a critical role in determining how the brain responds to everyday tasks and life's challenges \citep{park2013structural,glasser2016multi,jbabdi2015measuring}. In recent years, there has been an abundance of literature focusing on understanding the variation of functional connectomes in healthy and diseased people and their relationships to various covariates and phenotypes \citep{smith2015positive,finn2015functional,zhang2018relationships}. Such interests are inspired and propelled by large scale neuroimaging studies, such as the Human Connectome Project (HCP) \citep{van2013wu,glasser2016human}, the Alzheimer's Disease Neuroimaging Initiative (ADNI) \citep{weiner2010alzheimer} and the UK Biobank \citep{miller2016multimodal}.  In this article, we focus our attention to functional connectome (FC) inferred from functional magnetic resonance imaging (fMRI) data that measures the blood oxygen level dependent (BOLD) contrast signals of each brain voxel. As opposed to the anatomical axon connections (also referred as structural connectome), FC quantifies functional dependences between brain regions through correlations or covariances of BOLD signals. Conventional FC is often represented as a covariance or correlation matrix of fMRI data over a long recording time \citep{friston2011functional,hutchison2013dynamic}, where the matrix size equals the number of ROIs being considered. 
 
 While FC is assumed to be fixed or static over time in earlier studies,  there is an abundance of evidence \citep{hutchison2013dynamic,Monti2014427,Hindriks2016242} in recent studies showing that FC is a dynamic process. The dynamic FC (dFC) is represented as a time series of short-term FCs which are calculated using functional MRI data over small time intervals. Due to limitations of the fMRI BOLD (blood oxygen level dependent) contrast signals,  fMRI signals are not directly analyzed \citep{glover2011overview, turner2016uses}. The most popular way to transfer BOLD  signals into something that is reasonable to analyze is to calculate coherence between different brain regions, e.g., correlation or covariance.  We choose to use the covariance matrix, which in general carries more information than the correlation matrix. The goal of this paper is to understand and infer on the structure of dFC and detect change points in the dFC as the subjects perform a specific action. We first model the short-term FC using a scaled Wishart distribution and  then generalize the static model to a hierarchical model of a time series of covariance matrices.  Our final goal is to detect  and compare individual specific change points along the dFC based on this hierarchical model.

As argued before, a first step towards change point detection is to model a population of covariance matrices. This is entirely different from covariance matrix estimation from multivariate data, which is a well-studied problem; see \citep{leonard1992bayesian,daniels1999nonconjugate,pati2014posterior} as some representative examples of Bayesian inference for covariance matrices and \citep{Pourahmadi2011} for a more comprehensive review.  In the covariance estimation context, the observational data vectors are directly available and the goal is to characterize the dependence amongst the different variables in the data from multiple independent and identically distributed samples. On the other hand, our {\em observational units} are covariance matrices corresponding to different individuals observed over time, which we shall henceforth refer to as {\em covariance-valued data}. Hierarchical models for capturing heterogeneity in multiple related groups based  on covariance matrices are relatively fewer in number  \citep{flury1984common,flury1987hierarchy,boik2002spectral,franks2019shared,hoff2009hierarchical}. \cite{schott1999test,schott2001some} developed hypothesis testing methods based on covariance structures. In a Bayesian context, \cite{daniels2006bayesian,pourahmadi2007simultaneous,barnard2000modeling} considered parsimonious modeling of covariance matrices, which were extended to a  longitudinal setting by \cite{gaskins2013nonparametric,das2014semiparametric,gaskins2014sparsity,gaskins2016covariance}. A parallel sequence of works proposes modeling of fMRI data-matrices via Gaussian graphical modeling techniques \citep{stingo2013integrative,warnick2018bayesian}. 
In contrast to the existing studies, we built a hierarchical model on observed covariance valued datasets to detect individual specific change points. The literature on probabilistic modeling for covariance-valued data in the time series context \citep{golosnoy2012conditional,gourieroux2009wishart,yu2017generalized} is focused on maximum likelihood estimation using a non-central Wishart distribution as the likelihood.  For example, the dataset considered in \cite{yu2017generalized} comprises of low-dimensional (5 by 5) daily realized covariance (RCOV) matrices for 5 stocks observed across 2274 time points.   This single time series sequence is modeled using a generalized conditional autoregressive Wishart (GCAW) model. In presence of smaller number of parameters and a huge collection of time points, maximum likelihood estimation is a natural choice for model fitting.  On the other hand, since we are dealing with 10 by 10 covariance matrices observed over 26 time points for 500 individuals, it is important to borrow information across individuals and seek for a parsimonious modeling framework. 

%To the best of our knowledge, there is no previous work on the theoretical development of probabilistic modeling for such data and associated inferential techniques. %Bayesian inference of a covariance matrix have also gained in popularity   where the observation units are vectors and the covariance matrix is treated as an unknown parameter.   

To that end, we develop a suite of hierarchical modeling techniques for covariance-valued data to provide insight into the structural connectivity of human brains. We use a scaled version of the Wishart distribution to model the covariance-valued observations. While the Wishart distribution is commonly used as a prior distribution on inverse-covariance or precision matrices in Bayesian inference, its usage as a likelihood is novel in the Bayesian context to best of our knowledge. The presence of a modest number of observations further necessitates structured modeling of the center of the Wishart distribution, which itself is a covariance matrix. Based on empirical evidence of low effective ranks of the data matrices, we modeled the center of the Wishart model using an orthogonal factor model type decomposition and encouraged shrinkage towards a low rank structure through the development of a novel shrinkage prior. We use a combination of Gibbs and slice sampling to sample from the posterior distribution whose steps are mostly standard. 

%{\color{blue} We developed an efficient Markov chain Monte Carlo algorithm to sample from the posterior distribution based on an algorithm (Hoff, 2009b) to draw samples from a class of distributions on the Stiefel manifold.} 
%In a Bayesian inference for a covariance matrix,  a natural conjugate prior for the multivariate normal distribution is the inverse Wishart distribution (\cite{Bar2000Wishart}). In contrast, our likelihood for covariance matrices is proposed to be a suitably structured and scaled Wishart distribution. 

Our primary objective is to explore the dynamic nature of FC between different brain regions during performances of certain tasks. A dynamical FC model provides an overall architecture of how the brain functions as the individual perform certain tasks.  An important scientific goal is to identify change points  \citep{Barry93} in the time series of covariances that split the data into contiguous segments. Difference in the change points across individuals are indicative of behavioral and cognitive differences \citep{dai2017discovering}. To address this,  we extend our hierarchical model to accommodate a single or multiple change points in a fully Bayesian framework.  A novel combination of existing MCMC algorithms  renders sampling from the joint posterior distribution tractable.   The change point model is then implemented on both the HCP and the ADNI datasets to extract scientifically meaningful conclusions. For the HCP dataset, we studied the change point pattern during the motor task and discovered the primary FC change point occurs when people switch the movement from hand and foot to the tongue. For the ADNI dataset, we compared FCs in two groups of older people (supernormal subjects and normal controls) and found that  supernormal subjects have higher strength of connectivity within posterior regions or between posterior and anterior regions of their brain.

In this paper, we begin with an illustration of our motivating data set in section \ref{sec:realdata} followed by a model for covariance matrices (section \ref{sec:indep}), hierarchical covariance model (section \ref{HCOV_model}) and hierarchical change point model (section \ref{HCP_model}). Results of detailed simulation study are provided for each of the  three models. In section \ref{Real_HCP_data}, we provide the results obtained from our motivating HCP dataset under hierarchical change point model followed by model validation in section \ref{HCP_model_validation}. We studied a dynamic extension of our hierarchical model in Appendix (section \ref{Dynamic_change_point_model_section}). Section \ref{Hcp_model_sensitivity} in Appendix contains some additional results on sensitivity and robustness analysis of the hierarchical change point model. 

\section{Data description}\label{sec:realdata}
We utilize functional MRI data from two large datasets, ADNI \citep{weiner2010alzheimer} and HCP \citep{van2013wu} to illustrate the proposed method. 
ADNI was initiated by National Institute on Aging, the
National Institute of Biomedical Imaging and Bioengineering, the Food and Drug Administration, and some private pharmaceutical companies and non-profit organizations. 
ADNI assesses clinical, imaging, genetic and bio specimen biomarkers through the process of normal aging to early mild cognitive impairment, to late mild cognitive impairment, to dementia or Alzheimer's disease (AD). 
Participants were recruited across North America to participant in three phases of the study: ADNI1, ADNI GO and ADNI2.  A variety of imaging and clinical assessments were conducted for each participant. Results were then shared by ADNI through the Laboratory of Neuro Imaging's Image Data Archive (\texttt{https://ida.loni.usc.edu/}). In our study, we focus on a subset of healthy subjects that were previously identified in \citep{lin2017cingulate}.  These subjects were AD free but were clustered in two groups. The first group is called supernormals who exhibited excellent episodic memory and executive function. The other group is age-matched healthy control subjects. All their resting-state fMRI data were collected using a 3.0 Tesla Phillips MRI
with an echo-planar imaging sequence (spatial resolution
= $3 \times 3 \times 3$ mm$^3$). Structural images were obtained using an MPRAGE
sequence (spatial resolution $1 \times 1 \times 1$ mm$^3$), which
were then used for registration during preprocessing. Across individuals, the first 10 volumes were discarded to
avoid potential noise related to the equilibrium of the scanner
and participant's adaptation process. The remaining 130
volumes were preprocessed using slice time correction and
head motion correction. The images were then registered to
each individual's own structural image, normalized to the
Montreal Neurological Institute (MNI) standard space and spatially
smoothed using a Gaussian kernel (FWHM = 4 mm). We utilized the automated anatomical labeling (AAL) \citep{tzourio2002automated} to percolate the whole brain into $116$ regions of interest (ROIs).

\begin{figure}[h!]
\centering
\includegraphics[scale=0.6]{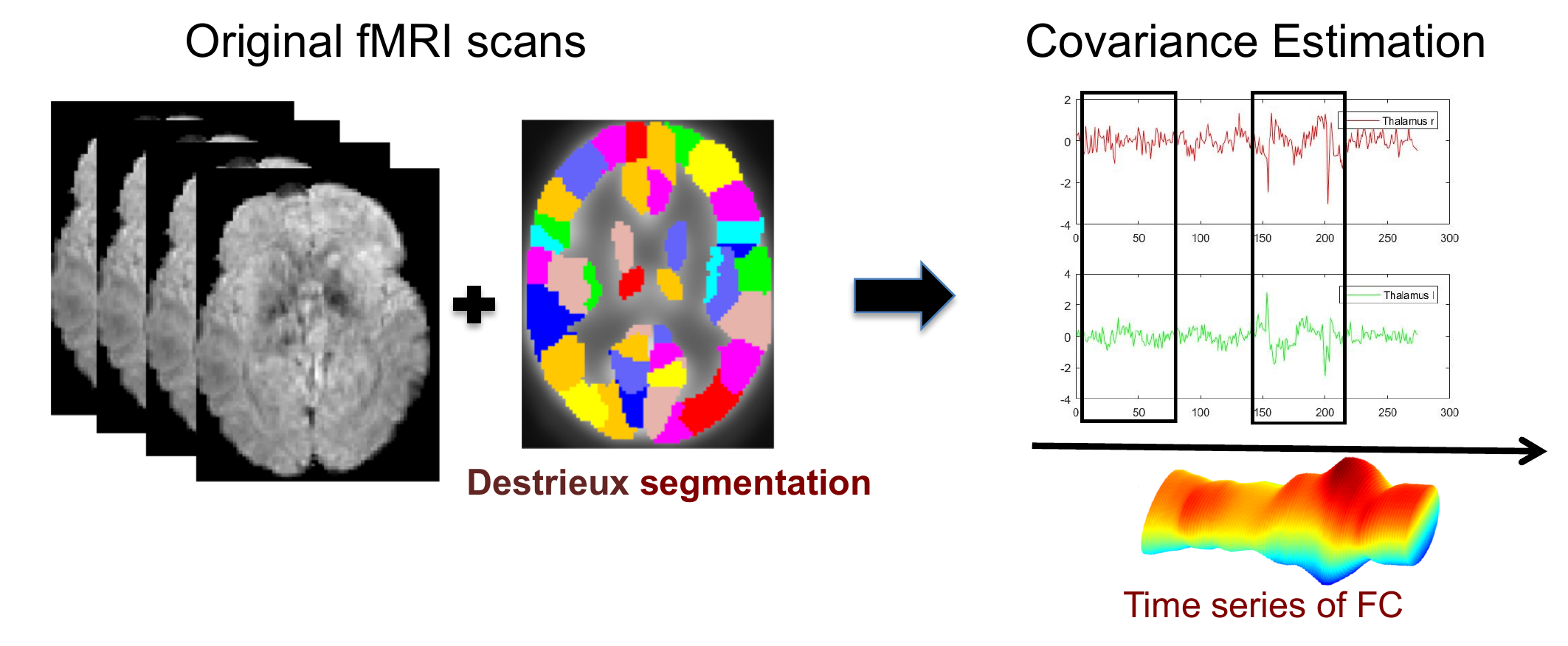}
\caption{\em{ An overview of preprocessing steps for extracting dynamic FC from fMRI data.}}
\label{fig:pipeline}
\end{figure}

The HCP project aims at characterizing human brain connectivity in $> 1,000$ healthy adults and to enable detailed comparisons between brain circuits, behavior and genetics at the level of individual subjects.  The HCP raw and preprocessed data can be easily accessed through ConnectomeDB (\texttt{http://www.humanconnectome.org}). The high-quality imaging data and the easy accessibility make it an ideal dataset for this paper.  Majority of the HCP fMRI data were acquired at 3T with a $2 \times 2 \times 2$ mm$^3$ resolution. Preprocessing steps using the HCP pipeline \citep{glasser2013minimal,glasser2016human} were performed before any data analysis, e.g., removing spatial distortions, realigning volume to compensate for subject motion, registering the fMRI to the structural MRI, reducing the bias field, normalizing the 4D image to a global mean, masking the data with the final brain mask and aligning the brain to a standard space. Figure \ref{fig:pipeline} provides an overview of the preprocessing steps. 
The Destrieux atlas \citep{Destrieux2010} was used to percolate cortical regions into 74 nodes per hemisphere. 
Similar to \cite{dai2017discovering}, for the fMRI BOLD signal in each ROI, we first calculate a mean time series, and then we utilize a sliding window method to calculate a covariance trajectory   $ \{ S_{it} \}_{\substack{i \in  [n], t \in [T] }}^{\!^{{{k \in [K]}}}}$ for subject $i$ at $t$-th window based on selected ROIs.  Therefore, $ S_{it} $ is a $p \times p$ covariance matrix representing the short time functional connectome, where p denotes number of ROIs.

\section{Hierarchical Modeling for Covariance Dataset}
Since the covariance matrices are observed for multiple individuals and time points, a natural course of action is to build a parsimonious model that borrows strength across all observational units. We first discuss an independence model with scaled Wishart distribution for covariance-valued data that serves as the basic building block for the forthcoming extensions. Both the central and non-central Wishart distributions have full support on the space of covariance matrices. However, the non-central Wishart distribution has more parameters making prior elicitation more complicated. Moreover, the presence of key parameters inside a hypergeometric function complicates posterior computation; in particular, the nice conjugate structure that we exploit throughout the article is lost. Due to these reasons, we consider a central Wishart likelihood through the rest of the paper. 

%from a computational perspective, non-central Wishart density appears to be more complicated with more parameters to deal with and the formula clearly is not amenable for prior straightforward prior elicitation, parsimonious representation and posterior computation. Henceforth, our likelihood will be based on  based on central Wishart distribution. 

Motivated by a pattern we observe in the functional connectivity data, the mean structure of the independence model is encouraged to shrink towards low rank matrices via a parsimonious shrinkage prior. We develop an MCMC algorithm to fit the independence model to data and show its efficacy in a simulation study. Next, the independence model is extended to a Bayesian hierarchical model to incorporate multiple individuals, allowing for subject specific deviations from a common mean structure. Fitting the hierarchical model requires sampling from a class of distributions on the Stiefel manifold which can be done efficiently using the algorithm in \S 3.3 of \cite{hoff2009simulation}; refer to \S  \ref{HCOV_model} herein for more details. The hierarchical model leads to our eventual goal of detecting subject specific change points in the functional connectivity data.

\subsection{Independence Model}\label{sec:indep}
We begin by describing the details of the independence model. Let $\{S_j\}_{j=1}^N$ be a collection of independent and identically distributed $p \times p$ covariance matrices. We probabilistically model the $S_j$s using a Wishart distribution, which is arguably the most recognized distributional family for covariance matrices. We shall use the standard $W_p(\nu, V)$ notation to denote the Wishart distribution on the space of  $p \times p$ positive definite matrices, with degrees of freedom $\nu > p-1$ and a $p \times p$ positive definite scale matrix $V$. The density $W_p(\nu, V)$ distribution has a density (in $X$) proportional to 
$$
|V|^{-\nu/2} \, |X|^{(\nu-p-1)/2} \, e^{-\mbox{tr}(V^{-1} X)/2}. 
$$
Specifically, we use a scaled Wishart distribution $W_{p}(\phi, \phi^{-1}\Omega)$ to model the $S_j$s, 
\begin{align}\label{eqModel}
S_j \overset{ind.}{\sim} W_{p}(\phi , \phi^{-1}\Omega) , \quad j = 1 \ldots N. 
\end{align}
The introduction of the parameter $\phi$ in the scale matrix is to decouple its presence in both the mean and covariance. For $S_1\sim W_p(\phi, \Omega)$, one has 
$$
\mathbb{E}[S_{1}] = \phi \Omega, \quad \text{Var}(S_{1,_{ij}}) = \phi (\omega_{ij}^{2} + \omega_{ii} \omega_{jj}), \quad \text{Cov}(S_{1,_{ij}},S_{1,_{kl}}) = \phi(\omega_{ik} \omega_{jl}+\omega_{il} \omega_{jk}),
$$  
whereas for $S_1 \sim W_p(\phi, \phi^{-1} \Omega)$, 
$$
\mathbb{E}[S_{1}] = \Omega, \quad \text{Var}(S_{1,_{ij}}) = \phi^{-1} (\omega_{ij}^{2} + \omega_{ii} \omega_{jj}), \quad \text{Cov}(S_{1,_{ij}},S_{1,_{kl}}) = \phi^{-1}(\omega_{ik} \omega_{jl}+\omega_{il} \omega_{jk}). 
$$
Thus, in the parameterization we work with, $\Omega$ is the population mean. We henceforth fix $\phi$ at $(p+1)$ and validate our assumption in Appendix \ref{variable_phi}. Now, we focus our attention on modeling the mean $\Omega$. %We provide a validation of setting $\phi$ to be $p+1$ in section \eqref{variable_phi} of Appendix.

An unstructured $p \times p$ covariance matrix has $p(p+1)/2$ free elements, (e.g. in the HCP dataset, $p = 10$ leading to a total number of $55$ parameters and in the ADNI dataset, $p = 7$ in the present case leading to $28$ parameters).  Given that we only have a modest number of time points, it is important to make meaningful structural assumptions on $\Omega$ to reduce the effective number of parameters to be estimated. We conducted an exploratory analysis to find patterns in the data matrices that could direct us towards a parsimonious model. The data matrices and their inverses did not contain any obvious sparsity pattern. Next, we investigated the effective ranks of the data matrices. For a $p \times p$ positive definite matrix $A$ with eigenvalues $s_1(A) \ge s_2(A) \ge s_p(A) \ge 0$, its effective or intrinsic rank \citep{Vershy11},
$$
r_e(A) :\,= \frac{ \sum_{k=1}^p s_k(A) }{ s_1(A) }
$$
is the ratio of its trace and largest eigenvalue. The effective rank satisfies $1 \le r_e(A) \le \mbox{rank}(A)$, so that it always provides a lower bound to the actual rank. Further, the effective rank is a smooth function of its argument. For example, consider the class of matrices 
$$
M_\lambda = u u^\T + \lambda I_p
$$
for $\lambda > 0$ and $u$ a $p$-dimensional vector of unit length. The matrices $M_\lambda$ increasingly get close to being rank deficient as $\lambda \downarrow 0$, however, this is not captured by the rank as $\mbox{rank}(M_\lambda) = p$ for any $\lambda > 0$. On the other hand, $r_e(M_\lambda) = 1 + (p-1)\lambda/(1 + \lambda)$, which smoothly decays to $1$ as $\lambda \downarrow 0$. These features 
render the effective rank a suitable measure to capture the intrinsic dimensionality of a matrix and indicate potential near rank-deficiencies.

\begin{figure}[h!]
\centering
 \includegraphics[scale=0.25]{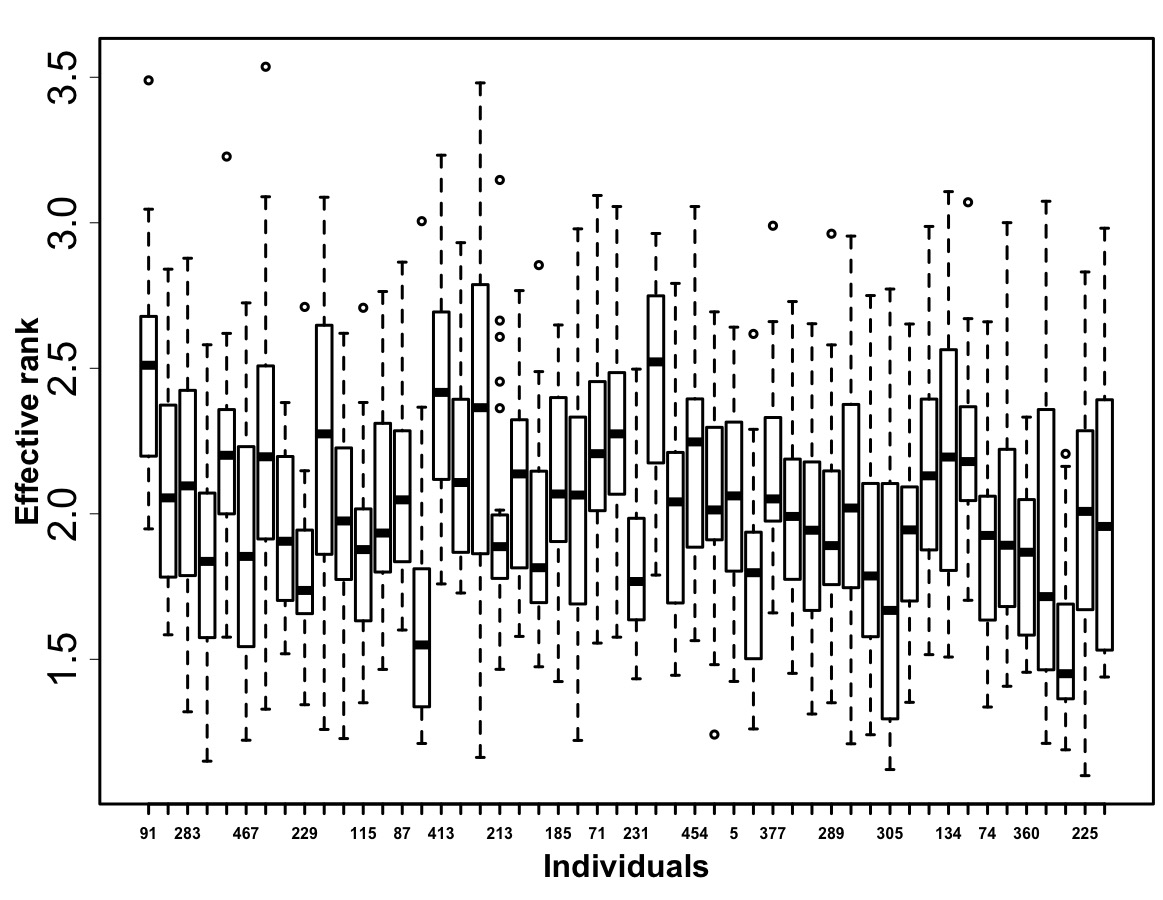}
 \caption{{\em Boxplot of effective ranks of the data matrices for the motor task of HCP across the $26$ time points for $50$ randomly chosen individuals. Each boxplot corresponds to a separate individual.}}
 \label{fig:effrank}
\end{figure}

Figure \ref{fig:effrank} shows boxplots of the effective ranks of the data matrices across the 26 time points for 50 randomly selected individuals from motor task of HCP dataset. It is evident that the $10 \times 10$ data matrices have low effective rank, with the bulk of the empirical effective rank distribution between $1.5$ and $3$. This observation motivated us to consider an orthogonal factor model type decomposition \citep{hoff2009hierarchical} for $\Omega$ to exploit the near low-rank structure as
\begin{align}\label{eqOmega}
\Omega = VDV^{\T} + \sigma^{2} I_{p},
\end{align}
where $V \in \mb R^{p \times r^\ast}$ for some $r^\ast \le p$ is a semi-orthogonal matrix satisfying $V^{\T}V = I_{r^{\ast}}$, and $D= \text{diag}(d_{1},\dots, d_{r^{\ast}})$ is a diagonal matrix with non-negative diagonal entries. Such a decomposition readily satisfies the positive definiteness constraint on $\Omega$. Different variance components are typically employed in factor analysis to account for variables may with different scales. However, since all the entries represent functional connectivities, we make the simplifying assumption of using the same $\sigma^{2}$ for all the components. 

We operate in a Bayesian framework to perform inference based on the posterior distribution of the model parameters. Before proceeding to  describe our prior specifications, it is important to discuss the role of $r^\ast$ in what follows. In a fully Bayesian framework, one may treat $r^\ast$ as a parameter which designates the effective rank of $\Omega$ and assign it a prior distribution; the discrete uniform distribution on $\{1, \ldots, p\}$ being a default choice. Under this prior, the posterior distribution of $r^\ast$ is proportional to the marginal likelihood of the data given $r^\ast$, which is intractable in the present context. While it is possible to sample $r^\ast$ inside a larger trans-dimensional MCMC algorithm such as the reversible jump MCMC (RJMCMC), its implementation remains computationally challenging, especially when considering extensions to the hierarchical modeling setup later on. Moreover, the effective rank does not have a clear biological interpretation in our real application and is purely a modeling device to induce parsimony. Based on these considerations, we undertake a shrinkage approach rather than explicit selection of the rank. Specifically, we set $r^\ast$ to a conservative upper bound, with $p$ being a default choice, and encourage a subset of the diagonal entries of $D$ to shrink towards zero. If $\mathcal{A} \subset \{1, \ldots, p\}$ denotes the active subset, that is, the subset of diagonal entries of $D$ that are left unshrunk, then $V D V^\T \approx V_{\mathcal{A}} D_{\mathcal{A}} V_{\mathcal{A}}^\T$, where $V_{\mathcal{A}}$ denotes the $p \times |\mathcal{A}|$ sub-matrix of $V$ corresponding to the columns in $\mathcal{A}$, and $D_{\mathcal{A}}$ denotes the corresponding $|\mathcal{A}| \times |\mathcal{A}|$ diagonal sub-matrix of $D$. This leads to an approximately low rank decomposition under the posterior, which is sufficient for our purpose. In the factor modeling context, \cite{ABDD12} considered a shrinkage prior on the factor loadings matrix rather than placing a prior on the number of factors, e.g., as in \cite{Lopes04}. We have a very different shrinkage mechanism as our shrinkage operates on the diagonal matrix $D$. 

Fixing $r^\ast$, the unknown parameters in our model are $(V,D,\sigma^{2})$ with parameter space $\mathcal{V}_{p,r^{\ast}} \otimes \mathcal{D}_{r^{\ast}} \otimes \mathbb{R}_{+}$, where $\mathcal{V}_{p,r^{\ast}}$ denotes the Stiefel manifold of $p \times r^\ast$ semi-orthogonal matrices, and $\mathcal{D}_{r^{\ast}}$ the collection of $r^{\ast}$ dimensional diagonal matrices with non-negative entries. 
The likelihood function for the parameters is given by 
\begin{align}\label{eqlikelihood}
\displaystyle L(V,D,\sigma^{2}) =  |\Omega|^{- \frac{N\phi}{2}} \prod_{j=1}^{N} \exp\bigg\{-\frac{\phi}{2} \tr(\Omega^{-1} S_j)\bigg\}.
\end{align} 

We now discuss prior choices on the parameters. For computational convenience, we reparameterize to $(V, \tilde{D}, \sigma^2)$ where $\tilde{D} = D/\sigma^2$, so that $\Omega = \sigma^2 (V \tilde{D} V^\T + I_p)$. We place a uniform prior on $V$ supported on the Stiefel manifold $\mathcal{V}_{p, r^\ast}$, and an inverse-gamma $\text{IG}(\alpha_{\sigma} , \beta_{\sigma})$ prior on $\sigma^2$. To set up our sparsity favoring shrinkage prior on the diagonal entries $\tilde{d}_h$s of $\tilde{D}$, first decompose 
\begin{align}\label{eqD}
\tilde{d}_h = \tau \lambda_h, \quad h = 1, \ldots, r^\ast. 
\end{align}
In \eqref{eqD}, $\tau$ plays the role of a global shrinkage parameter while the $\lambda_h$s allow for coordinate specific deviations, much in the spirit of the global-local shrinkage priors popularly used in regression (\cite{carvalho2010horseshoe}). We place independent half-Cauchy priors on the $\lambda_h$s, $\lambda_h \overset{ind.} \sim \mbox{Ca}_+(0, 1)$, with density proportional to $1/(1+t^2) \, I_{0, \infty}(t)$. The half-Cauchy prior is a popular choice as a prior distribution of shrinkage parameters due to its positive density at zero and heavy tails (\cite{Polson2012}; \cite{carvalho2010horseshoe}). We complete the prior specification by placing a half-Cauchy prior truncated to $(0, 1)$ on $\tau$. Truncating the prior on the global parameter leads to better identifiability and is recommended by \cite{VanderPas2014} in the context of the horseshoe prior. The multiplicative prior on the $\tilde{d}_h$s can also be interpreted as an additive one-way ANOVA type decomposition in the logarithmic scale, 
$$
\log \tilde{d}_h = \mu + \beta_h, \ \mu = \log(\tau), \ \beta_h = \log(\lambda_h), \ h = 1, \ldots, r^\ast, 
$$
with grand mean $\mu$ and main effects $\beta_h$s. The posterior computation is also conveniently carried out in the logarithmic scale, which we describe next. 

\

We develop a fully automated and easy to implement Markov chain Monte Carlo algorithm to sample from the joint posterior distribution of $(V, \tilde{D}, \sigma^2)$ given the data. Specifically, we use a combination of Gibbs sampling with slice sampling and Metropolis-within-Gibbs to iteratively sample from the full-conditional distribution of each parameter block given the rest. The sampler iterates through the following steps; the derivations are deferred to the Appendix (section \ref{derivation_IND}). We use the notation $[\theta \mid -]$ to denote the full conditional distribution of a parameter. 
\begin{itemize}
\item Sample $V$ from its matrix Bingham$(S^{N}, \phi E^{-1}/2\sigma^{2})$ full-conditional distribution. The matrix Bingham$(A, B)$ distribution has a density with respect to the uniform distribution on the Stiefel manifold given by 
$$
p_{B}(X \mid A,B) \propto \text{etr}(BX^{\T}AX), 
$$
where $A$ and $B$ are symmetric and diagonal matrices, respectively. In our case, $S^{N}(= \sum_{j=1}^{N} S_{j})$ is a symmetric matrix by definition and $E = (\tilde{D}^{-1} + I_{r^{\ast}})$
The matrix Bingham distribution is conveniently sampled using the \texttt{R} package \texttt{rstiefel} \citep{rstiefel_Hoff}. 

%The conditionally conjugate prior on $V$ provides the full-conditional as $(V \mid -) \sim $ Bingham$(S, \phi E^{-1}/2\sigma^{2})$.
\item Update the $\{ \beta_{h} \}$s from their independent full conditional distributions using slice sampling. Set $M = V^{T}S^{N}V$ and consider the transformation $w_{h} = (1+\beta_{h} \mu)^{-1}$ for $h = 1, \ldots, r^\ast$. Then sample
\\ $[ u_{h} \mid w_{h}, - ] \sim$ Uniform$[ 0, \{ \mu^{2} + (( 1-w_{h})/w_{h} )^{2} \}^{-1} ]$, 
\\ $[w_{h} \mid u_{h}, - ] \sim$ Gamma(shape $=N\phi/2 - 1$, rate $= \phi M_{hh} /(2 \sigma^{2})$) truncated to the region $[\{1+ \sqrt{ (1/u_{h})-\mu^{2}} \}^{-1}, \infty]$, 
and set $\beta_j = (1 - w_h)/(w_h \mu)$. 

\item  To sample $\mu$, propose $\mu^{*} \sim N( \mu , s^{2})$ and compute the Metropolis ratio 
\begin{align*}
\alpha(\mu,\mu^{*}) = \frac{ \Pi(\mu^{*} \mid - ) }{ \Pi(\mu \mid - ) }
\end{align*} 
where $\Pi(\mu \mid - )$ denotes the full-conditional of $\mu$. Accept $\mu^{*}$ with probability $\min\{  \alpha(\mu,\mu^{*}) , 1 \}$.

\item Sample $\sigma^{2}$ from its inverse-gamma full conditional distribution as   
\begin{align*}
[ \sigma^{2} \mid - ] \sim \text{InvGamma}\bigg(\alpha_{\sigma}-1+ \frac{Np\phi}{2} , \beta_{\sigma} + \frac{\phi \tr(QS^{N})}{2} \bigg)
\end{align*}
where $Q = ( V\tilde{D}V^{\T} + I_{p} )^{-1} $.
\end{itemize}
We observed good mixing and convergence of the above MCMC sampler based on standard MCMC diagnostics. Although not our primary motivation, one can estimate the effective rank based on a simple post-processing step of the MCMC samples for the $\{d_h\}$s. As in \cite{bhattacharya2015dirichlet,li2017variable} at each MCMC iteration, we cluster the $\{d_h\}$s into two groups using 2-means clustering and save the size of the group having the larger mean. The mode of these numbers across the MCMC iteration is then used as an estimate of the effective or intrinsic rank. We find that this approach performs well in our simulation and real examples. A more nuanced approach for post-processing was proposed by \cite{li2017variable}, which can also be used in the present context.

\subsubsection{Simulation Study for Independence Model}\label{subsec:indep_sim}
We conduct a detailed simulation study to illustrate the performance of the independence model in terms of recovering the true parameters. 
We fixed $p = 50$ and varied $N \in \{100,250,500,750,1000\}$. The true intrinsic rank of the data generating mechanism was fixed at $3$ to mimic the observation in Figure \ref{fig:effrank}. We set $\phi = p+1$, the true $\sigma_{0}^{2} = 0.25$, the true $\tilde{D}_{0} = \{1.25,2,1.55 \}$ and considered $V_{0}$ from a uniform distribution on Stiefel manifold $\mathcal{V}_{p, r^\ast}$. 100 independent datasets have been generated from the model \eqref{eqModel}. We denote the true covariance matrix $\sigma_0^2 (V_0 \tilde{D}_0 V_0^\T + I_p)$ by $\Omega_0$.

\begin{figure}[h!]
\begin{subfigure}{0.5\textwidth}
\centering
\includegraphics[scale=0.28]{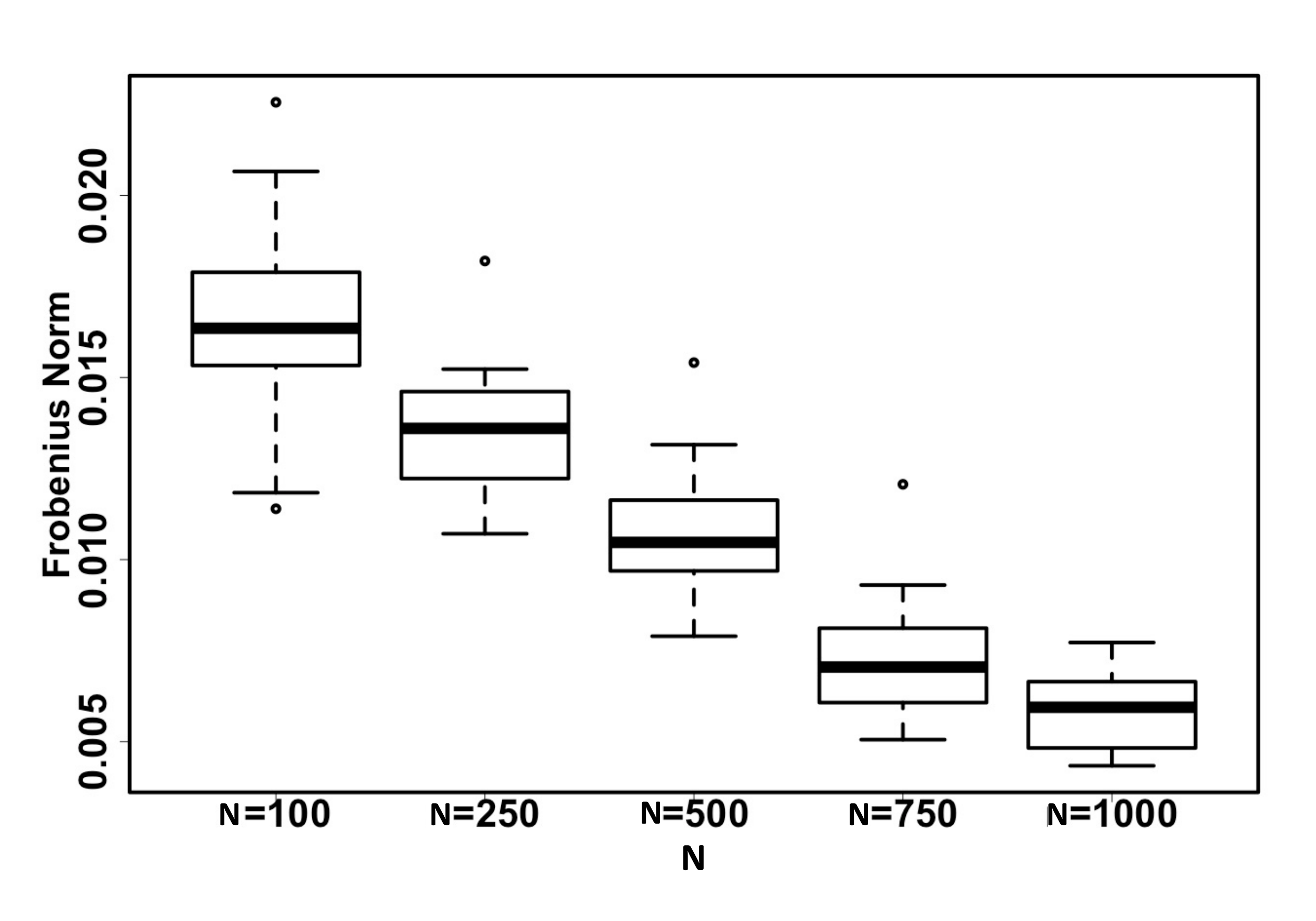}
\caption{}
\label{fig:constOmega}
\end{subfigure}\hfill
\begin{subfigure}{0.5\textwidth}
\centering
\includegraphics[scale=0.26]{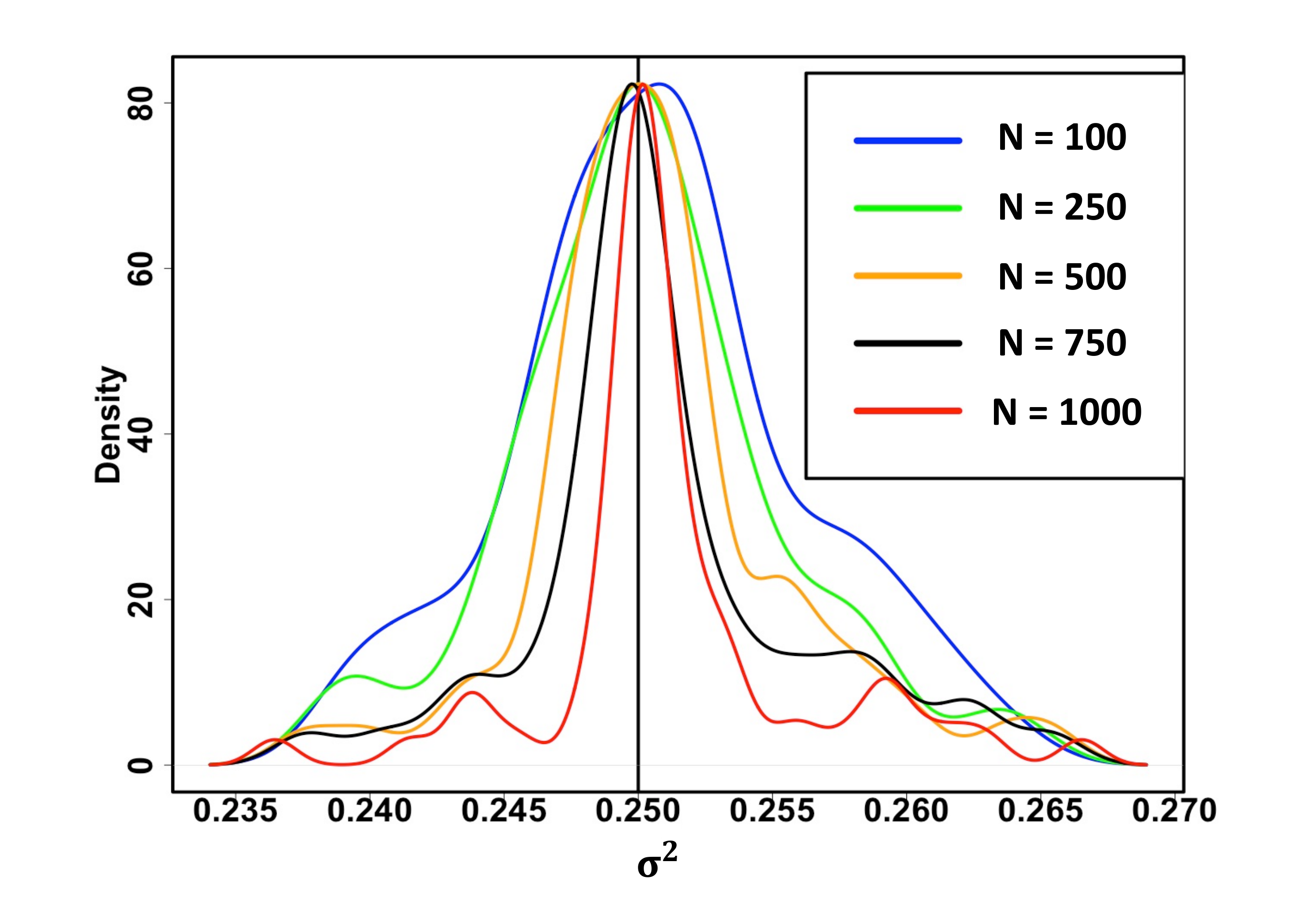}
\caption{}
\label{fig:sigma025}
\end{subfigure}
\caption{{\em Results for the independence model with $p=50$, $\sigma_{0}^{2} = 0.25$, $\tilde{D}_{0} = \{1.25,2,1.55 \}$, and  $N \in \{100,250,500,750,1000\}$. (\ref{fig:constOmega}) Boxplots of the scaled Frobenius norm difference $\|\widehat{\Omega}_B - \Omega_0\|/p$ across the 100 replicates for the different values of $N$, where $\widehat{\Omega}_B$ is the posterior mean and $\Omega_0$ is the population mean. (\ref{fig:sigma025}) Posterior density of $\sigma^{2}$ for different values of $N$ increasingly concentrate around the true value $0.25$. }}
 \label{fig:independence_model}
\end{figure}

For model fitting, we set $r^\ast = 10$; see Section \ref{Hcp_model_sensitivity} of the Appendix for a sensitivity analysis. The inverse-gamma hyperparameters $\alpha_\sigma$ and $\beta_\sigma$ were elicited in an empirical Bayes approach. Specifically, we used a method-of-moments type estimator for these hyperparameters. We ran our MCMC algorithm for 10,000 many iterations, discarding the first 5000 many iterates as burn-in. Letting $\widehat{\Omega}_B$ denote an estimate of the posterior mean based on the retained MCMC samples, we provide boxplots of the scaled Frobenius norm difference $\|\widehat{\Omega}_B - \Omega_0\|/p$ across the 100 replicates for the different values of $N$ in Figure \ref{fig:constOmega}. Here, and elsewhere, $\|A\| = \sqrt{\tr(A^\T A)}$ denotes the Frobenius norm of a matrix. As expected, both the center and spread of the boxplots tend to decrease with increasing $N$, implying the consistency of the posterior mean in recovering the population mean. Figure \ref{fig:sigma025} shows density plots of the posterior samples of $\sigma^{2}$ which increasingly concentrate around the true value, $\sigma_{0}^{2} = 0.25$, with increasing $N$.

Next, we compare the performance of the posterior mean $\widehat{\Omega}_B$ with the sample mean $\widehat{\Omega}_s = N^{-1} \sum_{j=1}^N S_i$, which is an unbiased estimator of $\Omega_0$. We consider three different norms between covariance matrices \citep{Dryden09} listed in Table \ref{table:norm_list}. 
\begin{table}[h!] 
\begin{center}
\begin{tabular}{|c|c|c|}
\hline
\textbf{Name} &\textbf{Notation} &\textbf{Form} \\
\hline
Euclidean             &$d_{E}(S_{1},S_{2})$       &$ \| S_{1} - S_{2} \|$   \\ \hline
Riemannian          &$d_{R}(S_{1},S_{2})$      &$ \| \log( S_{1}^{-1/2} S_{2} S_{1}^{-1/2} )  \|$  \\ \hline
Cholesky              &$d_{C}(S_{1},S_{2})$      &$ \| \text{chol}(S_{1}) - \text{chol}(S_{2})  \| $ \\  \hline
\end{tabular}
\end{center}
\vspace{-0.5cm}
\caption{{\em Notation and definition of distances between two covariance matrices.} }
\label{table:norm_list}
\end{table}

\begin{figure}[h!] % "[t!]" placement specifier just for this example
\begin{subfigure}{0.3\textwidth}
\centering
\includegraphics[scale = 0.125]{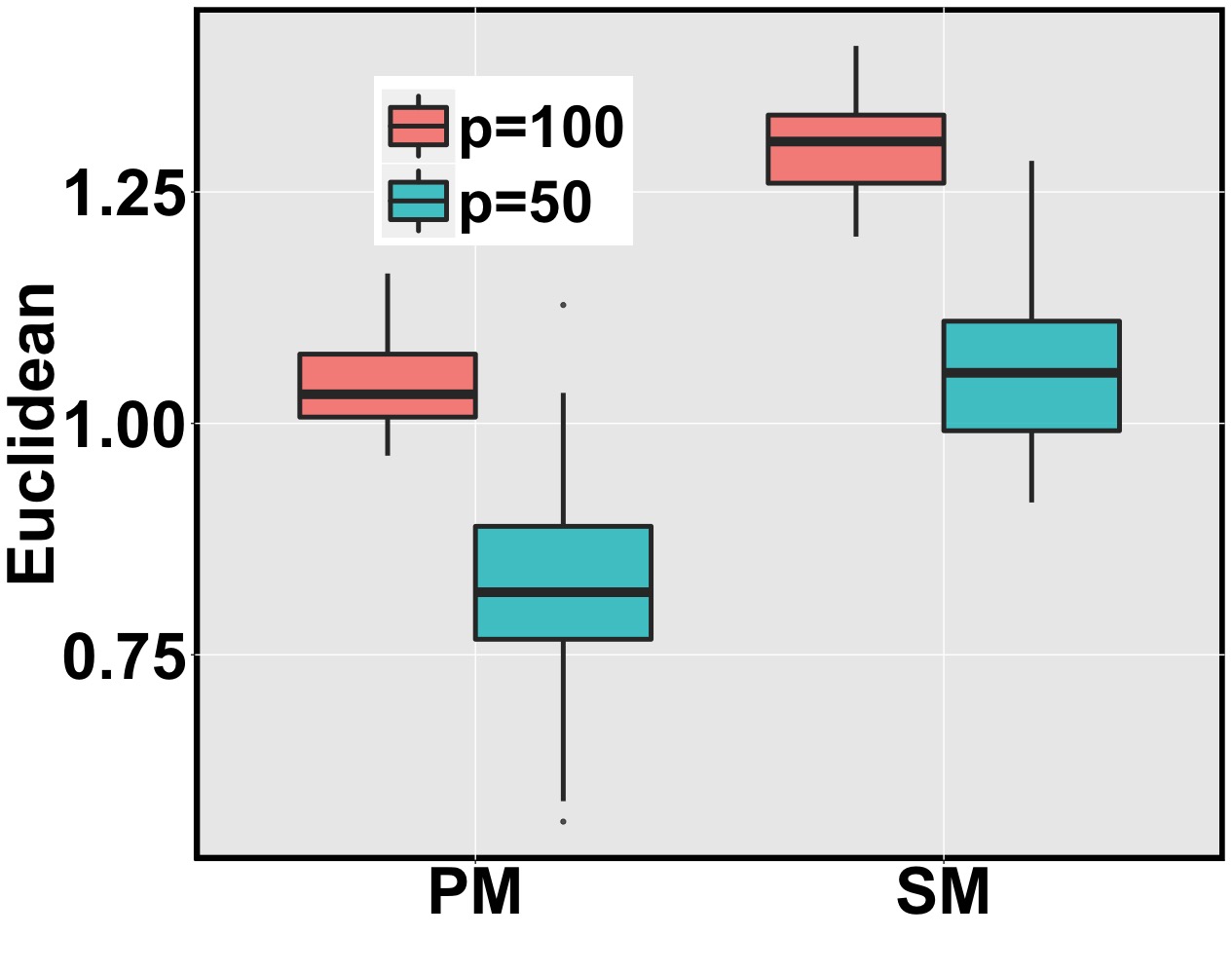}
\end{subfigure}\hfill
\begin{subfigure}{0.3\textwidth}
\centering
\includegraphics[scale = 0.125]{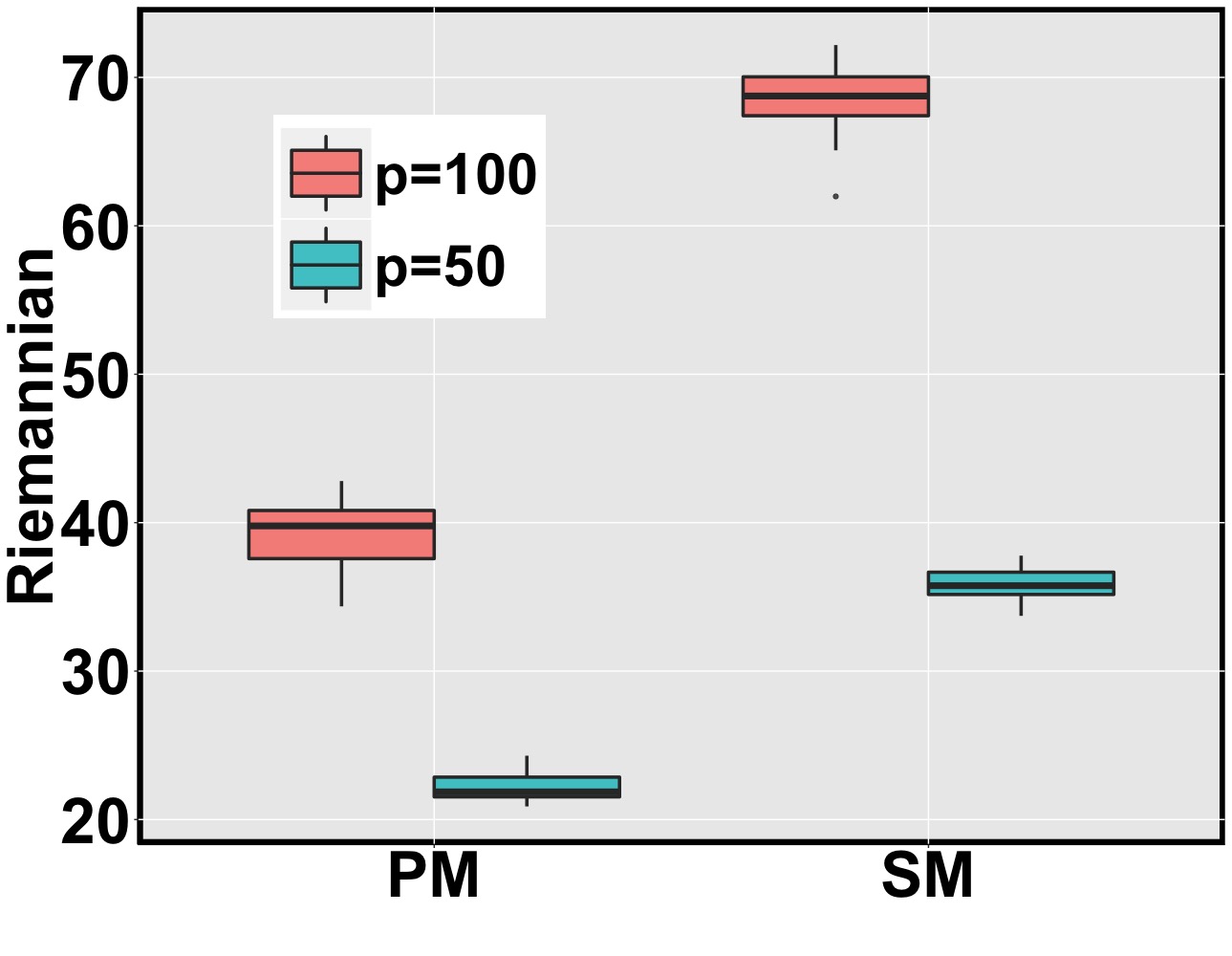}
\end{subfigure}\hfill
\begin{subfigure}{0.3\textwidth}
\centering
\includegraphics[scale = 0.125]{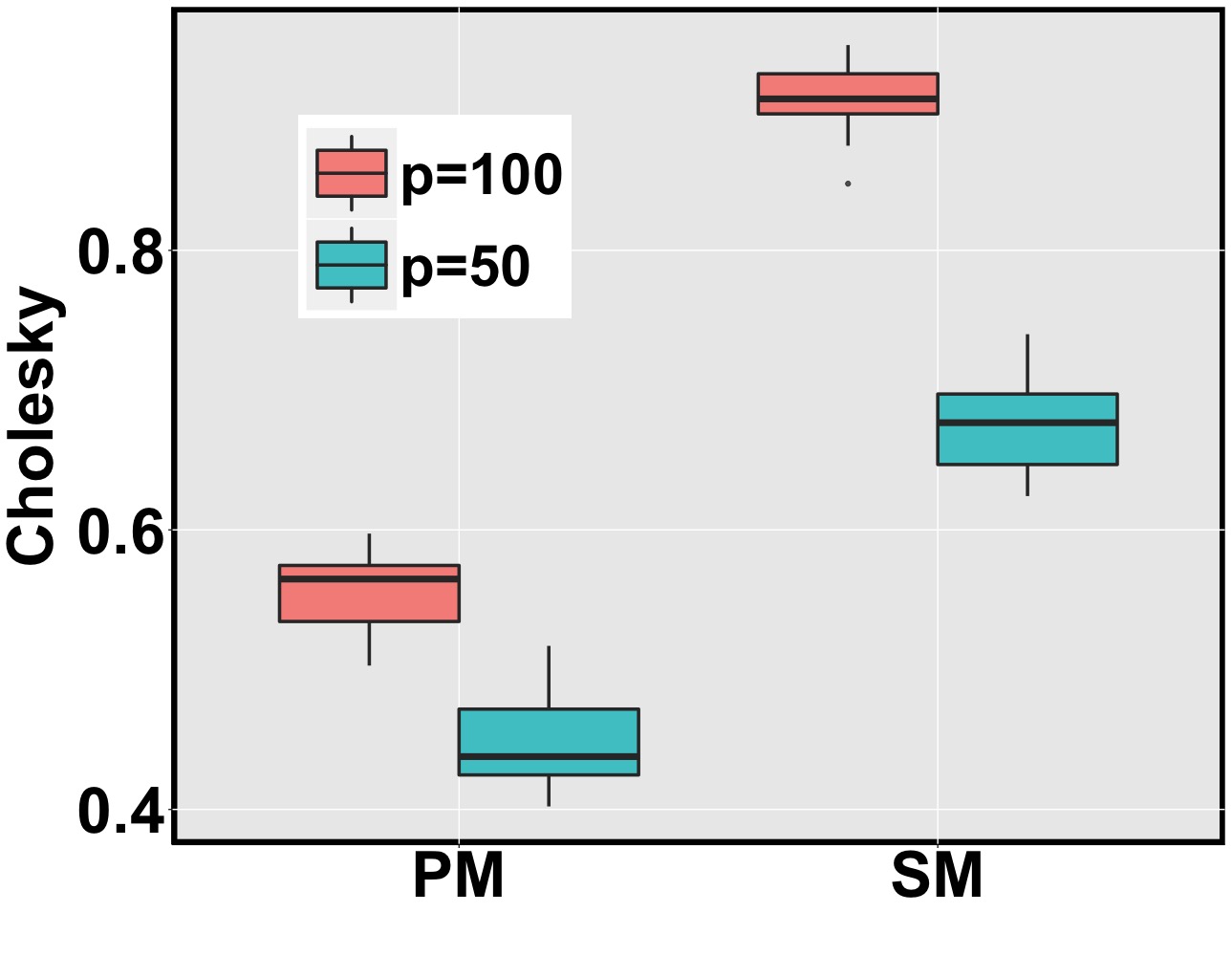}
\end{subfigure}\hfill
\caption{{\em Boxplots of $100 \times$ $d(\widehat{\Omega}_B, \Omega_0)$ and $d(\widehat{\Omega}_s, \Omega_0)$ over 100 replicates for $N=100$, where $d$ generically refers to one of the three distances in Table \ref{table:norm_list}. PM \& SM  correspond the distances for the posterior mean and sample mean respectively.} }  \label{PMvsSM}
\end{figure}

For each distance $d$, we compare $d(\widehat{\Omega}_B, \Omega_0)$ and $d(\widehat{\Omega}_s, \Omega_0)$ for $p \in \{50, 100\}$; summary measures are tabulated in Table \ref{table:mean_comparison_p_50} and Table \ref{table:mean_comparison_p_100} for $p=50$ and $p=100$ respectively. We scaled the distances except the Riemannian norm by $p$ for tabulation; the Riemannian norm is scale-invariant. Boxplots of the distances (in their original scale) across the 100 replicates are provided in Figure \ref{PMvsSM}. It is evident that the posterior mean overall provides a substantial improvement over the sample mean, especially in higher dimensional situations. 
%This shows the advantages of exploiting the underlying rank deficient structure. 

\begin{table}[h!]
\begin{center}
\footnotesize{
\begin{tabular}{|c|cc|cc|cc|}
\hline
\multirow{2}{*}{$p=50$} &\multicolumn{2}{c|}{Euclidean}  &\multicolumn{2}{c|}{Riemannian} &\multicolumn{2}{c|}{Cholesky} \\ 
\hhline{~------}
 &PM &SM  &PM &SM  &PM &SM \\\hline
$N=100$   &$1.65_{0.20}$  &$2.10_{0.22}$  &$44.4_{1.8}$  &$71.9_{2.0}$  &$0.89_{0.4}$  &$1.34_{0.7}$   \\ \hline
$N=250$   &$1.35_{0.17}$  &$1.48_{0.20}$  &$42.2_{1.5}$  &$62.8_{1.7}$  &$0.83_{0.4}$  &$1.24_{0.6}$   \\ \hline
$N=500$   &$1.01_{0.15}$  &$1.20_{0.17}$  &$36.1_{1.4}$  &$53.6_{1.7}$  &$0.76_{0.3}$  &$0.99_{0.6}$   \\ \hline
$N=750$   &$0.71_{0.09}$  &$1.08_{0.11}$  &$33.3_{1.0}$  &$48.5_{1.3}$  &$0.71_{0.1}$  &$0.90_{0.3}$   \\ \hline
$N=1000$   &$0.51_{0.06}$  &$0.93_{0.10}$  &$32.2_{0.9}$  &$44.6_{1.1}$  &$0.67_{0.1}$  &$0.82_{0.2}$  \\ \hline
\end{tabular}
}
\end{center}
\caption{{\em The average of $100 \times d(\widehat{\Omega}_B, \Omega_0)$ and $100 \times d(\widehat{\Omega}_s, \Omega_0)$ over 100 replicates for $p=50$, where $d$ generically refers to one of the three distances in Table \ref{table:norm_list}. Subscripts denote $100 \times$standard deviation across the 100 replicates. PM \& SM  correspond the distances for the posterior mean and sample mean respectively. All the displayed values are scaled by $p=50$.}}
\label{table:mean_comparison_p_50}
\end{table}

\begin{table}[h!]
\begin{center}
\footnotesize{
\begin{tabular}{|c|cc|cc|cc|}
\hline
\multirow{2}{*}{$p=100$} &\multicolumn{2}{c|}{Euclidean}  &\multicolumn{2}{c|}{Riemannian} &\multicolumn{2}{c|}{Cholesky} \\ 
\hhline{~------}
 &PM &SM &PM &SM  &PM &SM  \\\hline
$N=100$   &$1.04_{0.46}$  &$1.31_{0.41}$   &$39.3_{2.1}$  &$98.7_{2.2}$  &$0.58_{0.4}$  &$0.91_{0.5}$  \\ \hline
$N=250$   &$0.92_{0.40}$  & $1.18_{0.39}$  &$35.5_{1.8}$  &$86.1_{2.0}$  &$0.51_{0.3}$  &$0.79_{0.4}$   \\ \hline
$N=500$   &$0.68_{0.28}$  &$0.74_{0.26}$   &$31.4_{1.6}$  &$76.1_{1.8}$  &$0.46_{0.3}$  &$0.68_{0.3}$   \\ \hline
$N=750$   &$0.39_{0.20}$  &$0.48_{0.23}$   &$30.3_{1.4}$  &$69.0_{1.5}$  &$0.39_{0.2}$  &$0.61_{0.2}$   \\ \hline
$N=1000$   &$0.36_{0.17}$  &$0.45_{0.18}$ &$28.8_{1.4}$  &$63.4_{1.4}$  &$0.35_{0.1}$  &$0.54_{0.2}$ \\ \hline
\end{tabular}
}
\end{center}
\caption{{\em Same setting as in Table \ref{table:mean_comparison_p_50} with $p=100$.} }
\label{table:mean_comparison_p_100}
\end{table}

\begin{figure}[h!]
\begin{subfigure}{0.5\textwidth}
\centering
\includegraphics[scale=0.22]{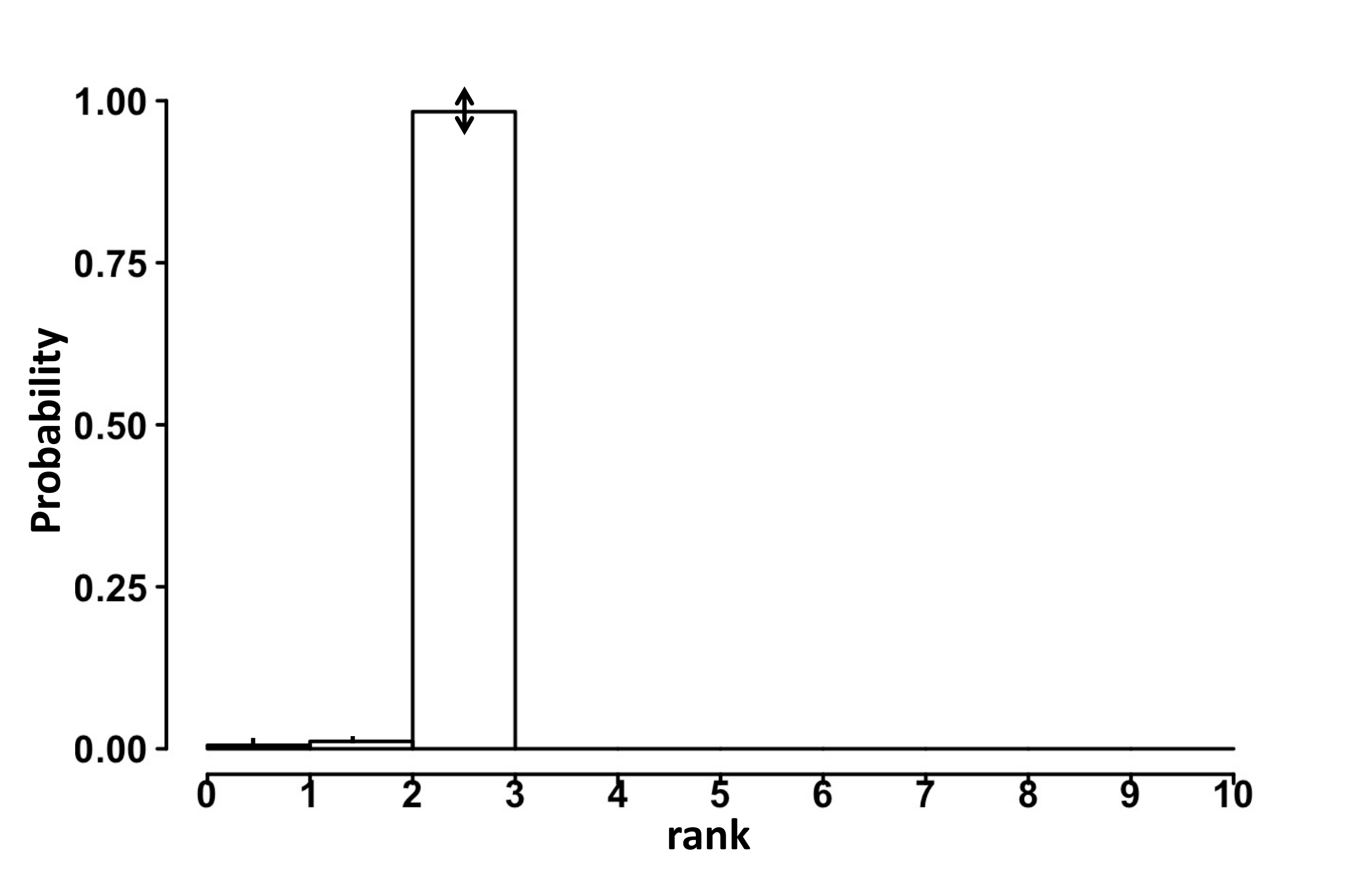}
\caption{}
\label{fig:rank_est_setting_1}
\end{subfigure}\hfill
\begin{subfigure}{0.5\textwidth}
\centering
\includegraphics[scale=0.22]{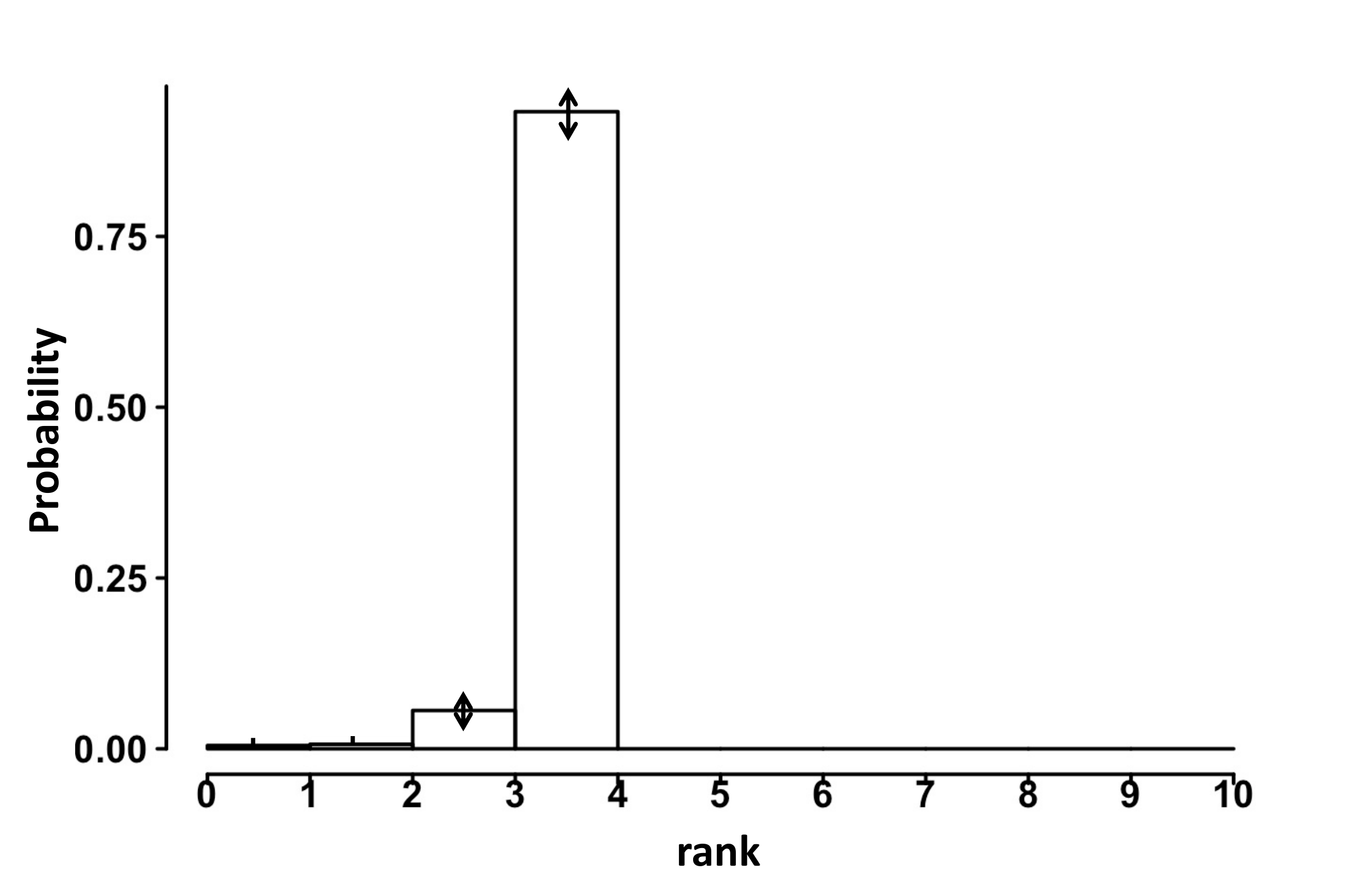}
\caption{}
\label{fig:rank_est_setting_2}
\end{subfigure}
\caption{{\em Rank estimate for Independence model with $N = 100$, $p=50$ and $\sigma_{0}^{2} = 0.25$. Lines on the bars shows the standard errors of probability for each point across replicates. The probability is defined as average posterior probability across replicates for each point. (\ref{fig:rank_est_setting_1}) Left panel plot provides the rank estimate for setting $(i)$ $\tilde{D} = \{1.25,2,1.55 \}$ which indicates high probability mass at $3$. (\ref{fig:rank_est_setting_2}) Right panel plot provides the rank estimate for setting $(ii)$ $\tilde{D} = \{0.75,1.25,2,1.55 \}$. There is a moderately high probability mass at $4$ and a significant amount of mass at $3$ because of existence of a weak signal. }}
 \label{fig:rank_est_independence_model}
\end{figure}

Finally, we illustrate the performance of the post-processing step outlined in the previous subsection to estimate the effective rank. We only consider $N = 100$ and 2 different settings of $\tilde{D}$, (i) $\{1.25,2,1.55 \}$ and (ii) $ \{0.75,1.25,2,1.55 \}$. Setting (ii) has a weaker signal strength compared to (i). Following the discussed methodology and setup, we provide the rank estimates under different scenarios in Figure \ref{fig:rank_est_independence_model} which shows high probability mass at $3$ and $4$ in Figure \ref{fig:rank_est_setting_1} and \ref{fig:rank_est_setting_2} respectively. In different simulation settings, our proposed method is able to recover the true ranks. We highlight that the simulation is performed under mild model misspecification - the true covariance matrix in our simulation study is chosen to be exactly low-rank whereas our model only assumes a near low-rank structure. Despite this, our model is able to detect the true rank.

%Independence model is able to recover the true ranks as well. 

\subsection{Hierarchical Covariance Model}\label{HCOV_model}
In this subsection, we extend the independence model to a hierarchical modeling framework encompassing all the individuals. Our hierarchical modeling framework lets the different individuals to share common parameters while allowing for subject specific deviations, striking a balance between pooling of information across different individuals while retaining flexibility. Letting $S_{it}$ denote the observed covariance matrix for individual $i$ at time $t$, we let 
\begin{equation} \label{HCovModel_decomp}
\begin{rcases}
 & S_{it} \overset{ind.}\sim W_{p}(\phi, \phi^{-1} \Omega_{i}) \\
 & \Omega_{i} = VD_{i}V^{\T} + \sigma_{i}^{2}I_{p}
\end{rcases} 
\quad
\begin{array}{r@{\;}l}
t = 1, \ldots , T, \\
i = 1, \ldots , n. 
\end{array}
\end{equation}
The first line of \eqref{HCovModel_decomp} posits the same scaled Wishart model as in the previous subsection with individual specific mean $\Omega_i$. As discussed earlier, we only have data on $T = 26$ time points for each individual. On the other hand, there are a relatively larger number of individuals in the study. For this reason, rather than separately fitting the independence model for each individual, we consider a structured decomposition of $\Omega_i$ that lets $D_i$ and $\sigma_i^2$ vary across individuals, while keeping $V$ fixed. This is akin to an expansion of the $\Omega_i$s in terms of a {\em fixed dictionary} $V$, with subject specific loadings. This fixed dictionary expansion vastly reduces the number of model parameters and allows one to borrow information across individuals to estimate the common dictionary $V$. We continue using $\sigma_i^2 I$ for the residual part in the covariance decomposition for model parsimony. We later conduct model validation to show that model \eqref{HCovModel_decomp} provides an adequate fit to the data compared to separately fitting the independence model. 

We continue to use the uniform prior on the Stiefel manifold for $V$. After reparameterizing to $\tilde{D}_i$, we place independent copies of the shrinkage prior introduced earlier on the $\tilde{D}_i$s, and independent inverse-gamma priors on the $\sigma_i^2$s.  

% However, since all the entries represent functional connectivities, we make the simplifying assumption of using the same $\sigma^{2}$ for all the components.
%This idea can be view as factor model type decomposition with different loadings and noise for different individuals with identical semi-orthogonal matrix. Now we use same priors on the parameters $(V, D, \sigma^{2})$ as we have already discussed earlier in (\ref{eqV}),(\ref{eqD}),(\ref{eqDprior}) and (\ref{eqSigmaPrior}).

%\subsubsection{Posterior computation}\label{HCOV_model_post_comp}
%\textcolor{red}{I have made some changes here, more required.}
We extend the MCMC algorithm for the independence model to the hierarchical setting. 
The updates for $\tilde{D}_i$ and $\sigma_i^2$ proceed independently across $i$ exactly along same lines as before. However, since $V$ is common to all individuals, its full conditional no longer remains a matrix Bingham distribution. We show in the Appendix (section \ref{derivation_HCM}) that the full-conditional distribution of $V$ is given by 
\begin{align} \label{eqV_HCovM}
\begin{split}
& [V \mid -] \propto \exp\bigg[ \frac{\phi}{2} \sum_{i=1}^{n} \tr \bigg \{ V \frac{E_{i}^{-1} }{\sigma_{i}^{2}} V^{\T} \sum_{t=1}^{T} S_{it} \bigg \}\bigg] \propto \prod_{j=1}^{r^{\ast}} \exp( v_{j}^{\T} H_{j} v_{j} ), \\
& \text{where} \hspace{0.2cm} H_{j} = \sum_{i=1}^{n} (\frac{ \phi S_{i}^{\ast}}{ 2 e_{ij} \sigma_{i}^{2} } ), \hspace{0.2cm} \displaystyle  S_{i}^{\ast} = \sum_{t=1}^{T} S_{it}, \hspace{0.2cm} \text{and} \hspace{0.2cm} E_{i} = ( \tilde{D}_{i}^{-1} + I_{r^{\ast}} ).
\end{split}
\end{align}
We sample from the above density by adopting the Gibbs sampling scheme of \cite{hoff2009simulation} to sample from a class of matrix Bingham--von Mises--Fisher (BMF) distributions \citep{khatri1977mises}. The BMF distribution has a density on the Stiefel manifold given by $p_{\text{BMF}}(V \mid A,B,C) \propto \text{etr}(C^{\T}V + BV^{\T}AV)$. Hoff considers the case when $B$ is diagonal, noting that the general symmetric case can be handled by a transformation, when the density assumes the form
$$
p_{\text{BMF}}(V \mid A,B,C) \propto \prod_{j=1}^{K} \exp(c_j^{\T} v_j + b_{j,j} v_j^{\T} A v_j ). 
$$
Hoff used Gibbs sampling to sample from $p_{\text{BMF}}$ by alternately sampling from the full-conditional distributions 
of each column $v_j$ given the rest; see \S\,3.3 of \cite{hoff2009simulation} for details. 

The distribution of $[V \mid -]$ in equation \eqref{eqV_HCovM} is almost identical to the above BMF distribution; we have $C = 0$ in our case and matrices $H_j$ in place of $b_{j,j} A$. This minor difference is immaterial from a Gibbs sampling standpoint; the steps can be found in the Appendix: section \ref{derivation_HCM}. 

%We implement the algorithm \citep{hoff2009simulation} by setting $C=0$ and an identity matrix for $B$ which leads to a Bingham distribution \citep{bingham1974antipodally}. We have varying $H_{j}$ $(j = 1, \dots ,r^{\ast})$ in \eqref{eqV_HCovM} which does not impact the algorithm  \citep{hoff2009simulation} since we use Gibbs sampler. We modify the 3rd step of the existing algorithm \citep{hoff2009simulation} by switching to a vector variate Bingham distribution from BMF distribution and iterate through the steps (Appendix: section \ref{derivation_HCM}) to obtain samples from the density \eqref{eqV_HCovM}. 

\subsection{Simulation study for Hierarchical Covariance Model}
We conduct a replicated simulation study to illustrate the operating characteristics of the hierarchical model. We set $n = 100, T = 26$ and $p = 50$ for our simulations. The true $V_0$ is generated uniformly on the Stiefel manifold. Also, for each $i$, the diagonal entries of the true $D_{0i}$ are generated uniformly between $0$ and $5$, while the $\sigma_i^2$s are generated uniformly between $0.25$ and $0.50$. We generate 100 independent simulation replicates as above. 

We fit the hierarchical model using the MCMC outlined in the previous subsection. We set $r^\ast = 10$ and use a modification of the empirical Bayes procedure to elicit the hyperparameters $\alpha_\sigma$ and $\beta_\sigma$.
As metrics of parameter recovery, we considered
\begin{align}\label{eq:dists}
d_\Omega = \frac{1}{n} \sum_{i=1}^n \underbrace{ \, \|\widehat{\Omega}_i - \Omega_{0i} \|}_{d_\Omega^{(i)}}, \quad d_\sigma = \frac{1}{n} \sum_{i=1}^n \underbrace{|\widehat{\sigma}^2_i - \sigma_{0i}^2 |}_{d_\sigma^{(i)}}.%, \quad d_V = \| \widehat{\mathcal{P}}_V - \mathcal{P}_{V_0} \|, 
\end{align}
where $\widehat{\Omega}_i$ and $\widehat{\sigma}_i^2$ are the posterior means of $\Omega_i$ and $\sigma_i^2$ for $i = 1, \ldots, n$. $d_\Omega^{(i)}$ and $d_\sigma^{(i)}$ is an individual specific measure of the distance of the posterior mean from the truth, while $d_\Omega$ and $d_\sigma$ are average measures over all the individuals.

\begin{figure}[h!] % "[t!]" placement specifier just for this example
\centering
\includegraphics[scale = 0.13]{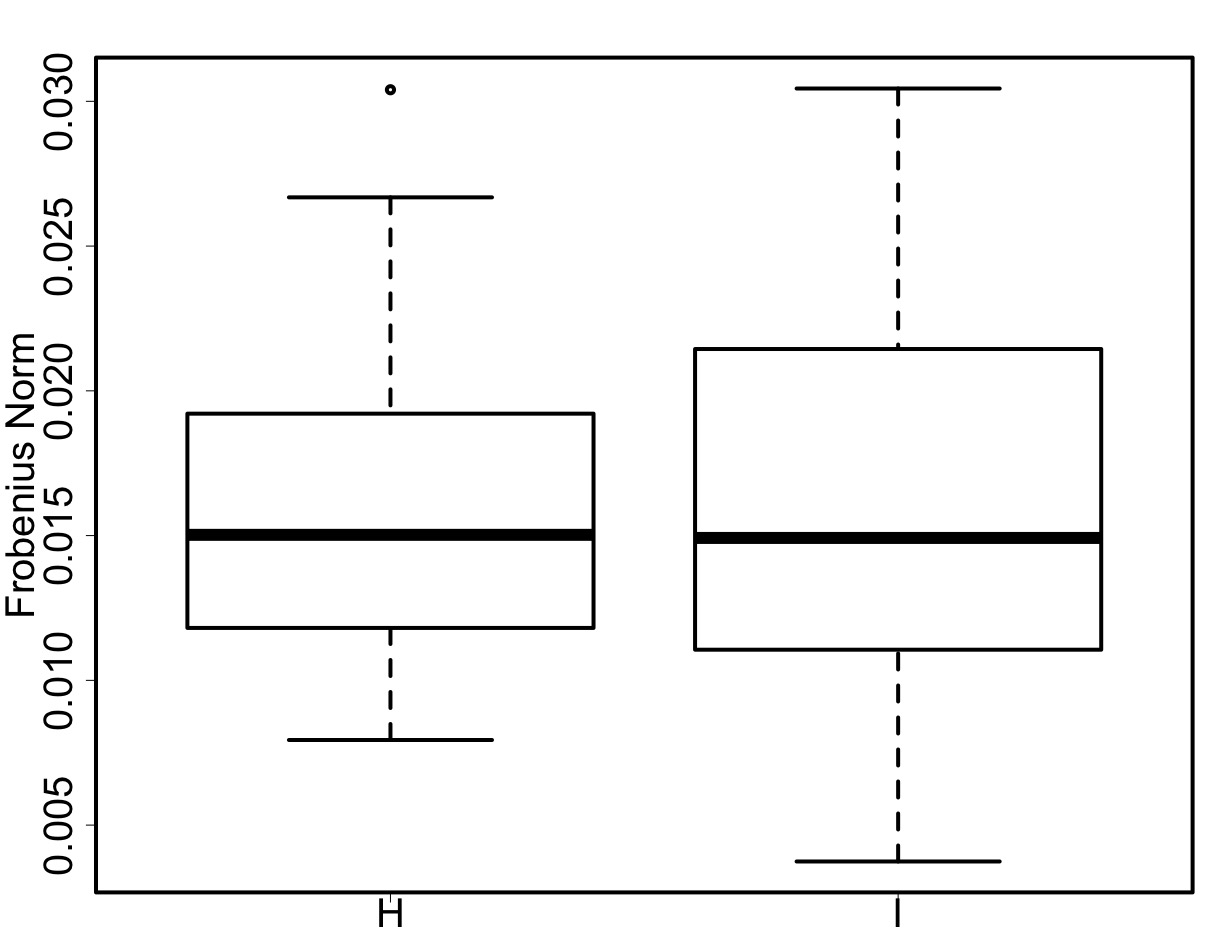}
\caption{{\em 
Boxplots of $\{d_\Omega^{(i)}\}_{i=1}^n$ defined in \eqref{eq:dists} averaged over the simulation replicates for the hierarchical (H) and independence (I) model. The boxplot for the hierarchical model shows a smaller spread due to borrowing information across individuals. 
}} 
\label{HSomega}
\end{figure}

\begin{figure}[h!] 
\begin{subfigure}{1\textwidth}
\begin{subfigure}{0.3\textwidth}
\centering
\includegraphics[scale = 0.139]{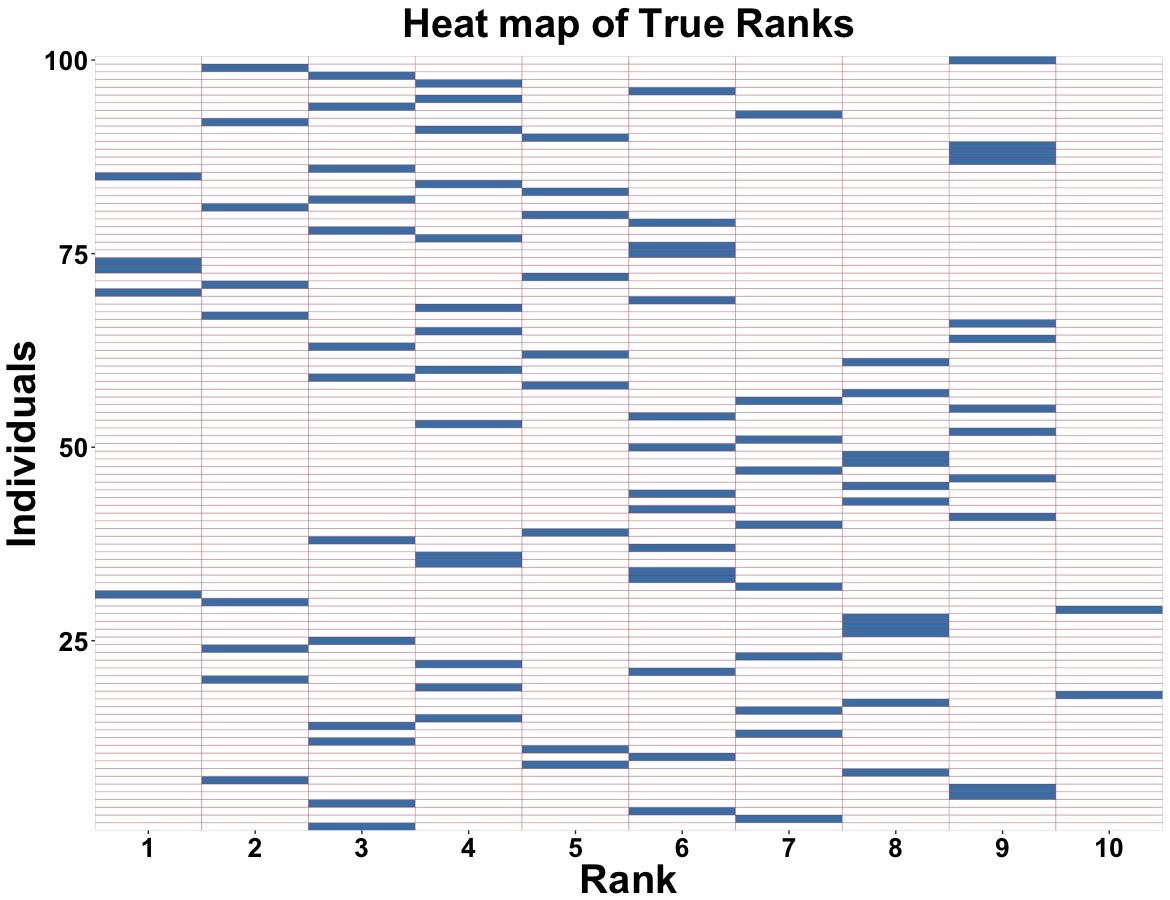}
\end{subfigure}\hfill
\begin{subfigure}{0.3\textwidth}
\centering
\includegraphics[scale = 0.139]{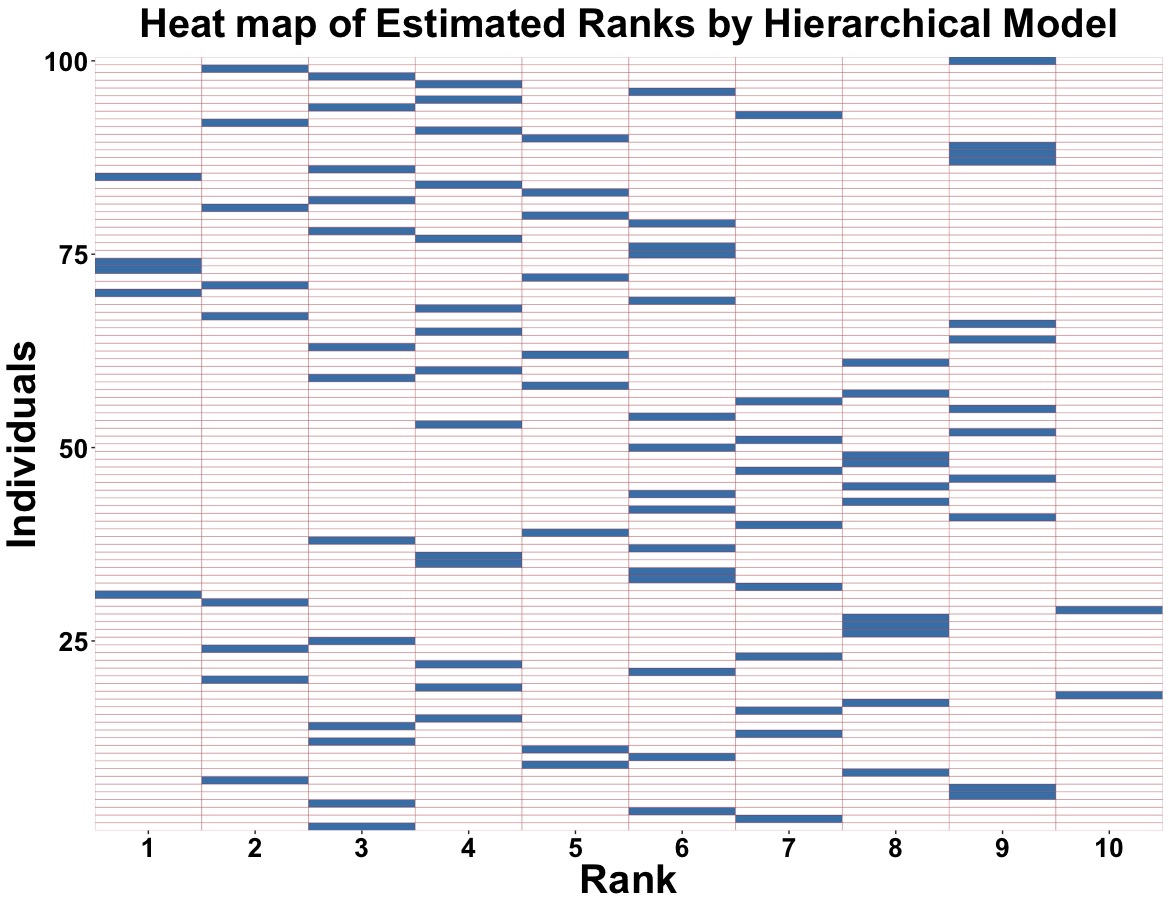}
\end{subfigure}\hfill
\begin{subfigure}{0.3\textwidth}
\centering
\includegraphics[scale = 0.139]{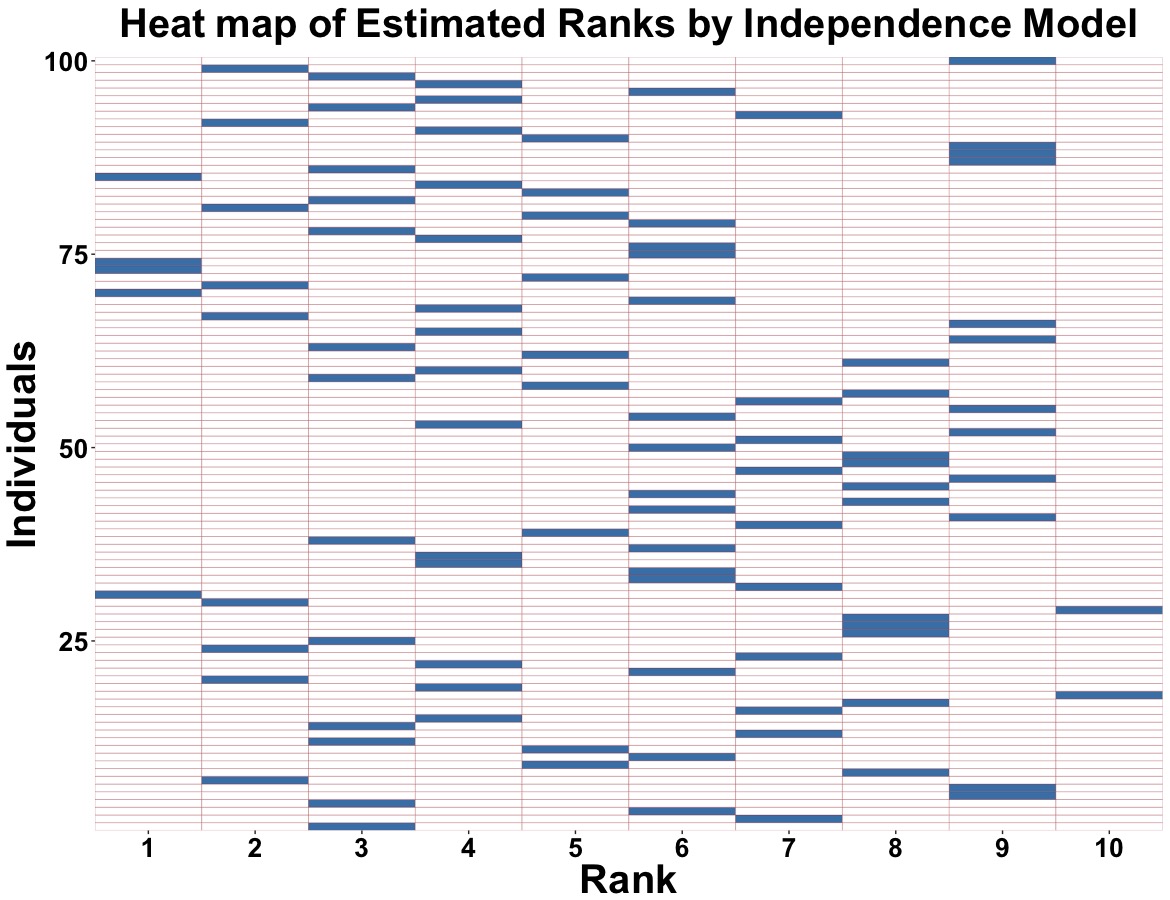}
\end{subfigure}
\end{subfigure}
\caption{ {\em Rank estimates for Hierarchical and Independence Model with $(n,p) = (100,50)$. Left panel: The true ranks across individuals shown in a heat map of the binary matrix $R = (r_{ih})$, with $r_{ih} = 1$ if the data matrix for individual $i$ has rank $h$, and $0$ otherwise. The middle and right panels correspond to the estimated ranks by the hierarchical and independence models  
respectively. } }
\label{HS_rank}
\end{figure}

As a point of comparison, we also fit the independence model in the previous subsection separately for each individual. 
Figure \ref{HSomega} shows boxplots of $\{d_\Omega^{(i)}\}_{i=1}^n$ averaged over the simulation replicates for the hierarchical and independence model. The tighter spread of the boxplot for the hierarchical model indicates the gains from borrowing information across subjects. The hierarchical model also successfully recovered the true ranks as shown in Figure \ref{HS_rank}. 

%All the displayed results shown are summarized over $100$ replicated data sets. 

\begin{figure}[h!] % "[t!]" placement specifier just for this example
\begin{subfigure}{0.45\textwidth}
\centering
\includegraphics[scale = 0.15]{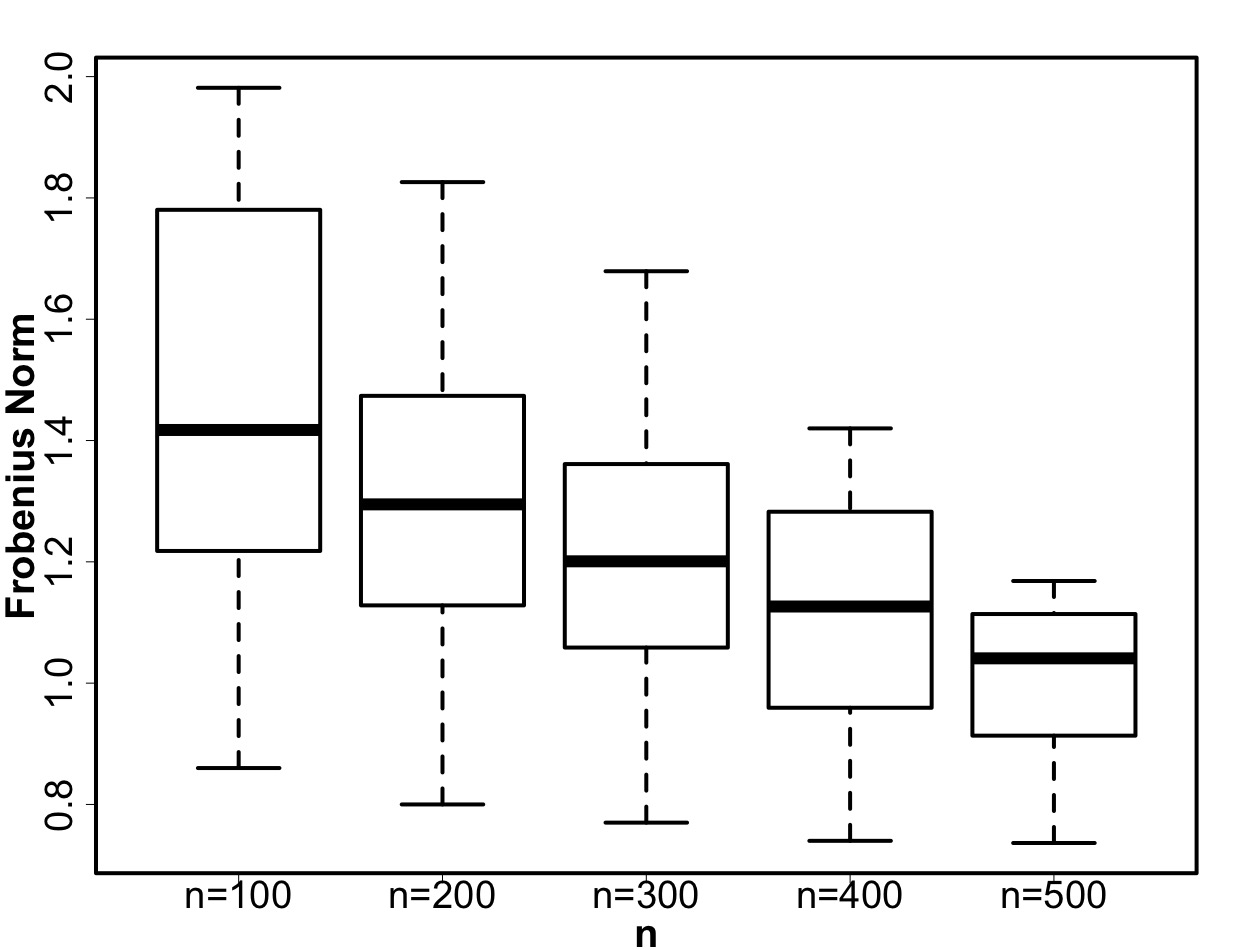}
\caption{} \label{H_sim_Omega_consistency}
\end{subfigure}\hfill
\begin{subfigure}{0.45\textwidth}
\centering
\includegraphics[scale = 0.15]{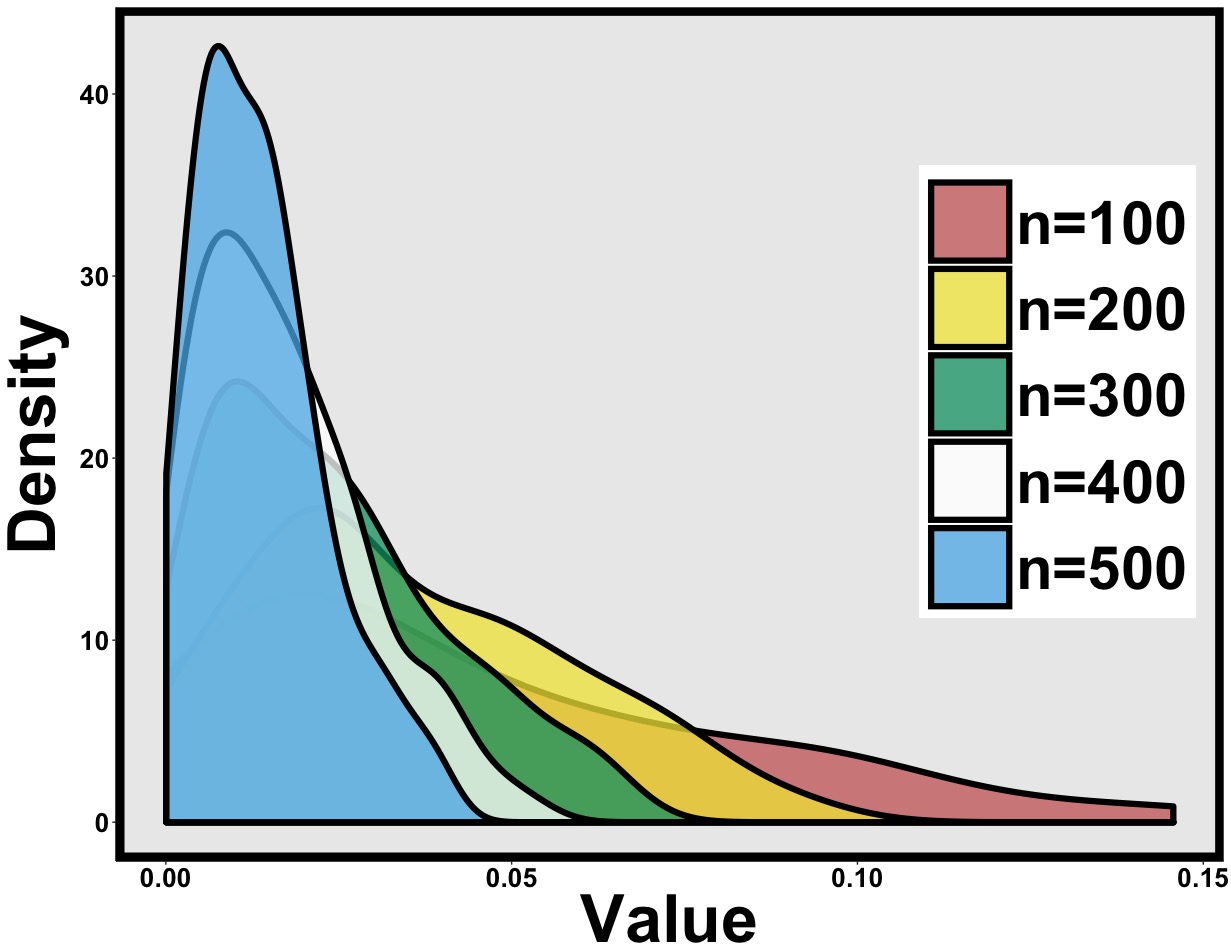}
\caption{} \label{H_sim_sigma_consistency}
\end{subfigure}
\medskip
\begin{subfigure}{0.45\textwidth}
\centering
\includegraphics[scale = 0.15]{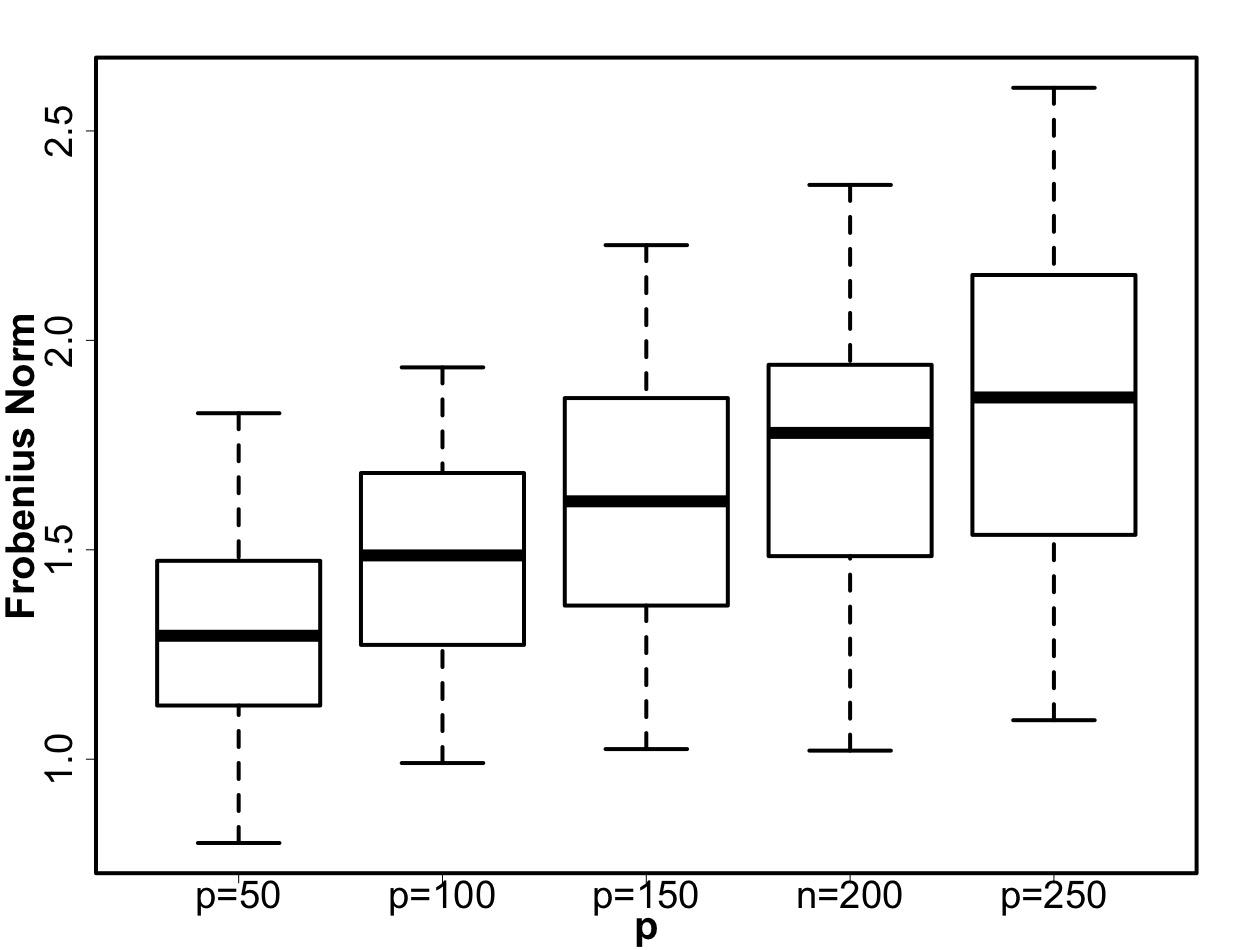}
\caption{} \label{H_sim_Omega_vs_p}
\end{subfigure}\hfill
\begin{subfigure}{0.45\textwidth}
\centering
\includegraphics[scale = 0.15]{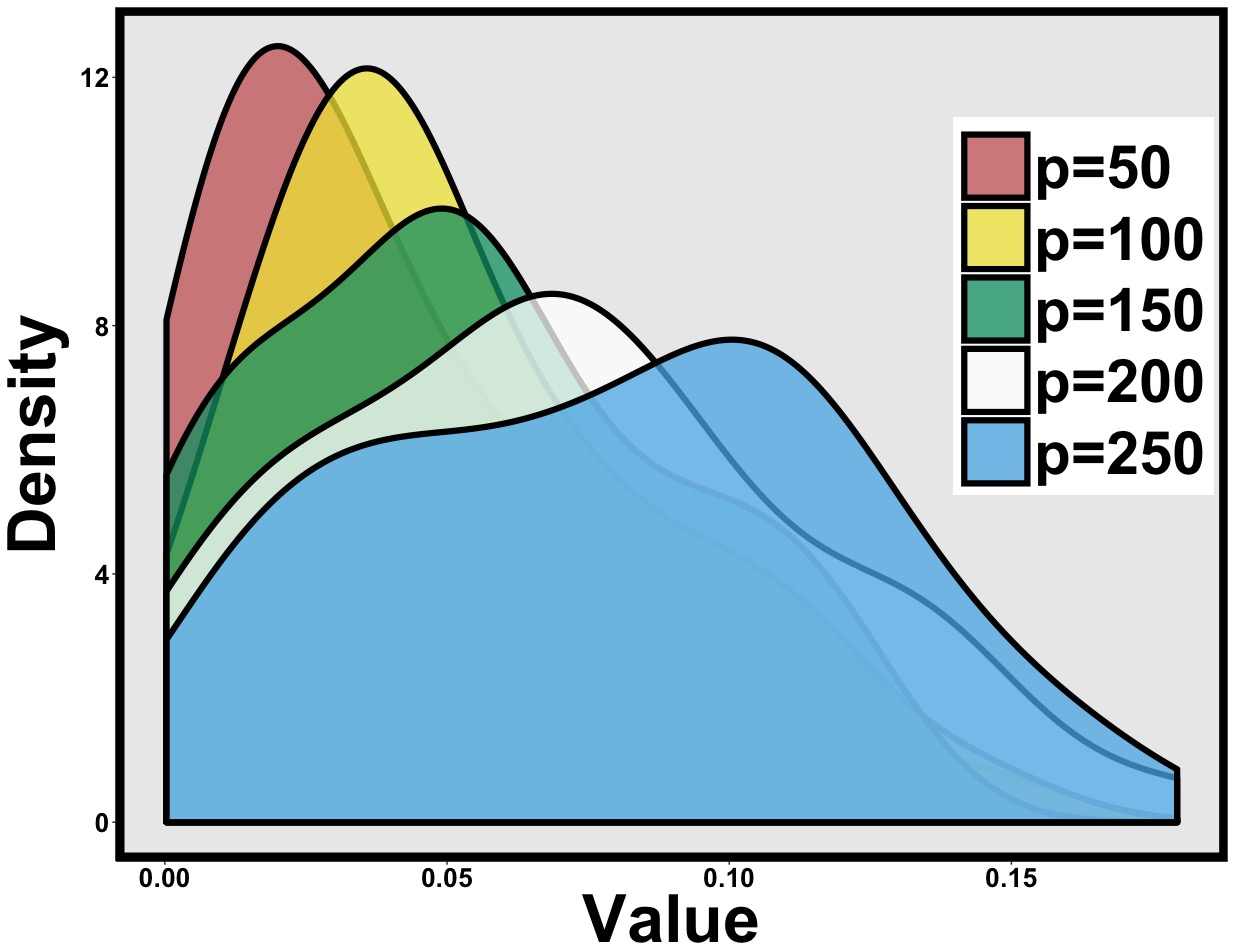}
\caption{} \label{H_sim_sigma_vs_p}
\end{subfigure}
\caption{{\em Left panel: Boxplot of $d_\Omega$ across simulation replicates for varying $n$ (top panel) and varying $p$ (bottom panel). Right panel: Density plots of $d_{\sigma}$ across simulated replicates for varying n (top panel) and varying p (bottom panel). }}
\label{H_sim}
\end{figure}

We conducted a second set of simulations by varying $n \in \{100, 200, 300, 400, 500\}$ and $p \in \{50, 100, 150, 200, 250\}$. A summary is presented in Figure \ref{H_sim}. In the top left panel, we provide the boxplot of $d_\Omega$ across the 100 simulation replicates for the different values of $n$ keeping $p$ fixed, while the bottom left panel provides the same for varying $p$ and fixed $n$. As expected, the estimation performance improves for larger $n$ and smaller $p$. We observe similar pattern in the density plots of $d_{\sigma}$ w.r.t. increasing $n$, fixed $p$ (Figure \ref{H_sim_sigma_consistency}) and fixed $n$, increasing $p$ (Figure \ref{H_sim_sigma_vs_p}).

\subsection{Real Data Analysis for ADNI dataset}
In this case study, we utilize 18 subjects' resting-state fMRI data from ADNI. Half of them are from supernormal (SN) subjects who possess excellent \citep{lin2017cingulate,lin2017identification}, and the other half are healthy control (HC) subjects. Each group contains 9 individuals with its resting state fMRI data at baseline preprocessed. From previous literature, we identified seven interesting ROIs, left occipital cortex, left occipital cortex,  left precuneus, left superior temporal cortex, right middle frontal gyrus, right parahippocampus, right thalamus (indexed as ROI $1,2,...,7$),  that are potentially linked to cognition, emotional regulation and memory. After preprocessing, we obtained a mean BOLD signal within each ROI and then applied a sliding window method to obtain a $7\times7\times24$ covariance matrix time series.  We applied our proposed hierarchical model on this dataset for different group of individuals and obtained Bayes estimates of individual specific covariance matrix $\Omega_i$ for $i = 1,..,9$. (Note that in Section 5.4 we have validated that there is no change point in these covariance trajectories.) 

\begin{figure}[h!] 
\centering
\includegraphics[scale=0.25]{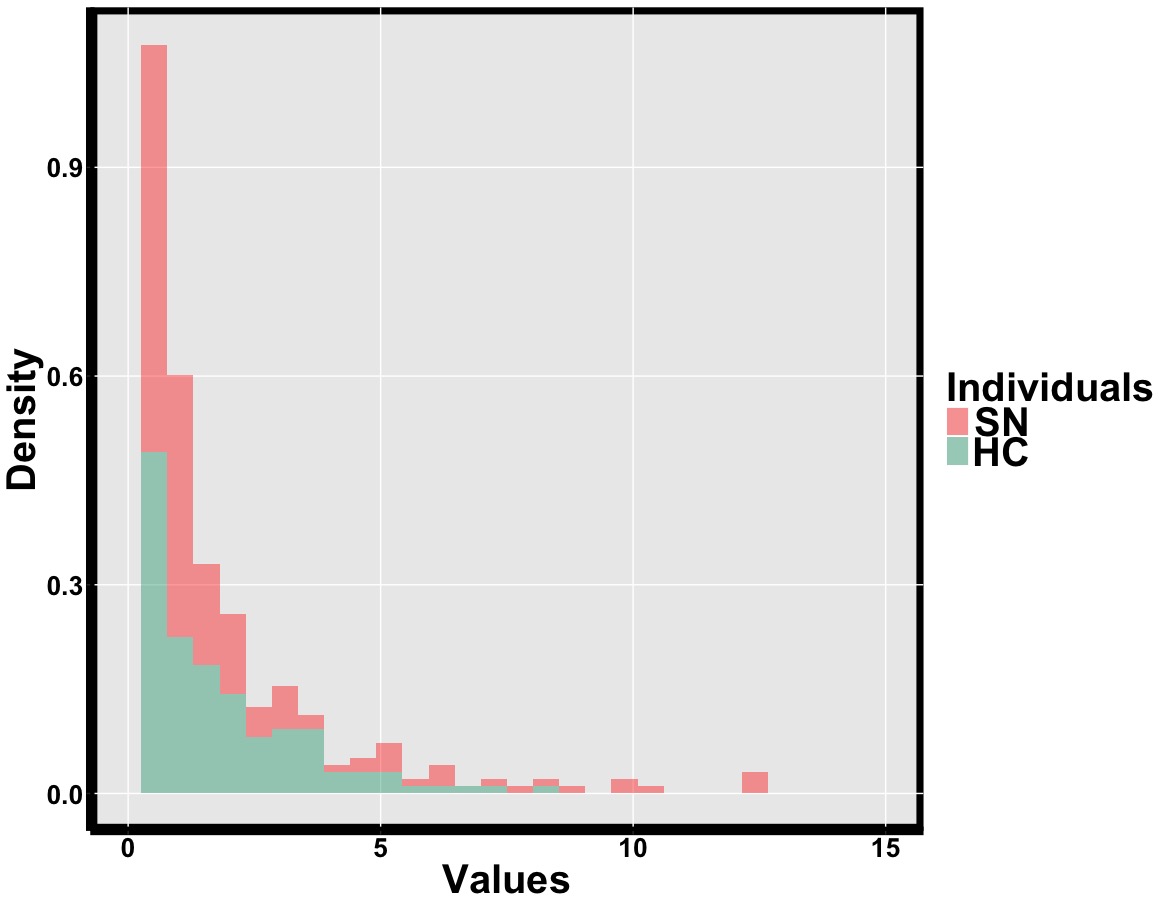}
\caption{{\em Histogram of off-diagonal elements which denote FC between ROIs among supernormals(SN) and health controls(HC). }}
\label{off_diag_SNHC}
\end{figure}

\begin{figure}[h!] 
\begin{subfigure}{1\textwidth}
\begin{subfigure}{0.45\textwidth}
\centering
\includegraphics[scale = 0.2]{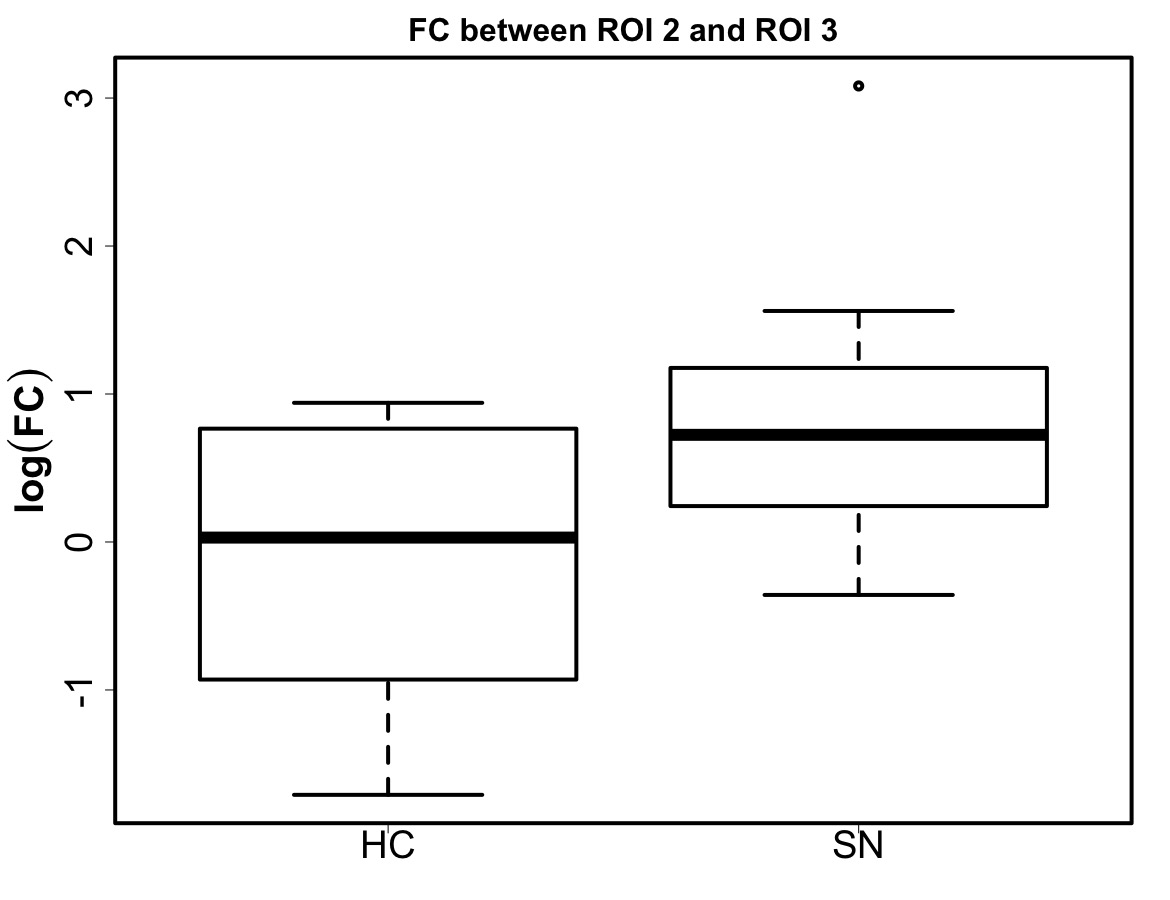}
\caption{} \label{ROI(3,6)_SN_HC}
\end{subfigure}\hfill
\begin{subfigure}{0.45\textwidth}
\centering
\includegraphics[scale = 0.2]{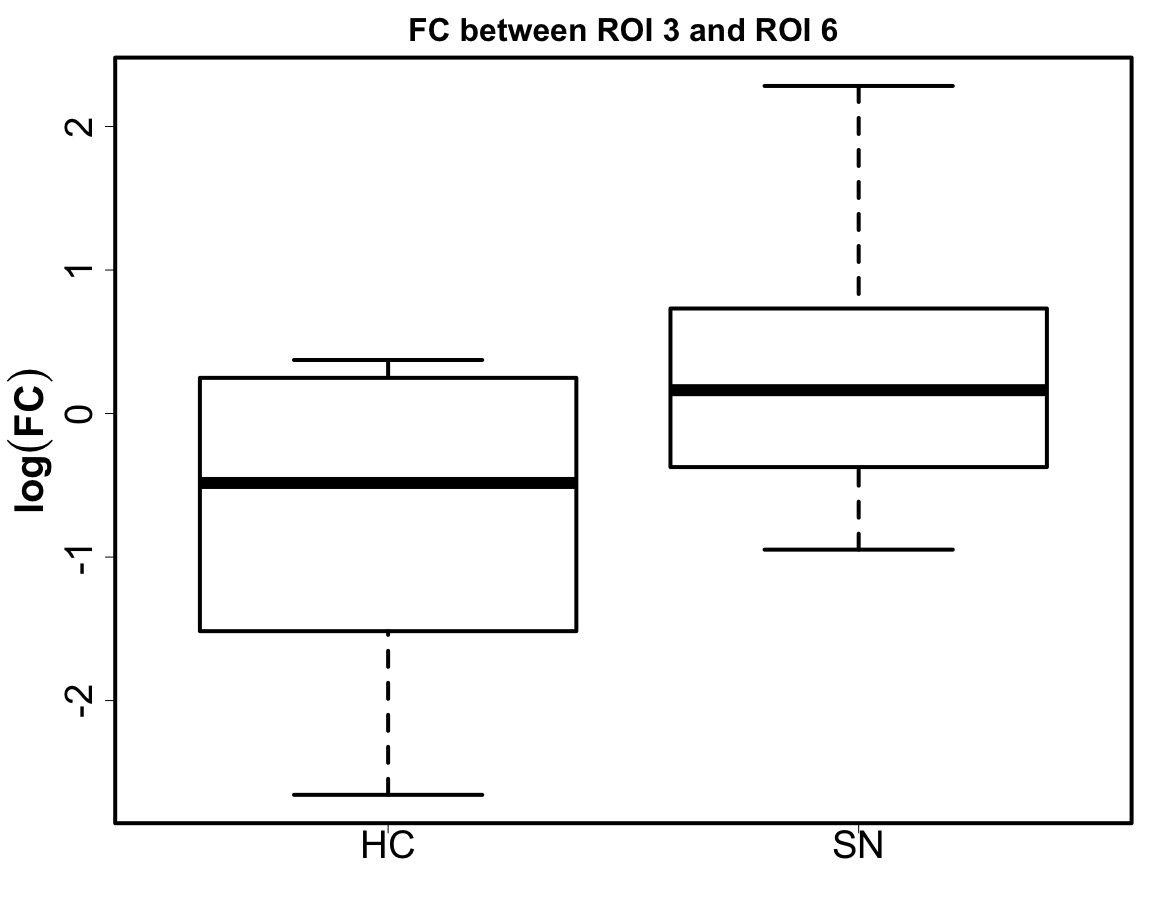}
\caption{} \label{ROI(3,6)_SN_HC}
\end{subfigure}\hfill
\end{subfigure}
\caption{ {\em Left panel represents the boxplot of $\{ \Omega_{i}[2,3], i =1 , \dots , 9 \}$ which are basically FC between ROI 2 and ROI 3 among healthy controls(HC) and supernormals(SN). Right panel represents the same between ROI 3 and ROI 6. } }
\label{SN_HC_ROI}
\end{figure}

\begin{figure}[h!] 
\begin{subfigure}{1\textwidth}
\begin{subfigure}{0.45\textwidth}
\centering
\includegraphics[scale = 0.16]{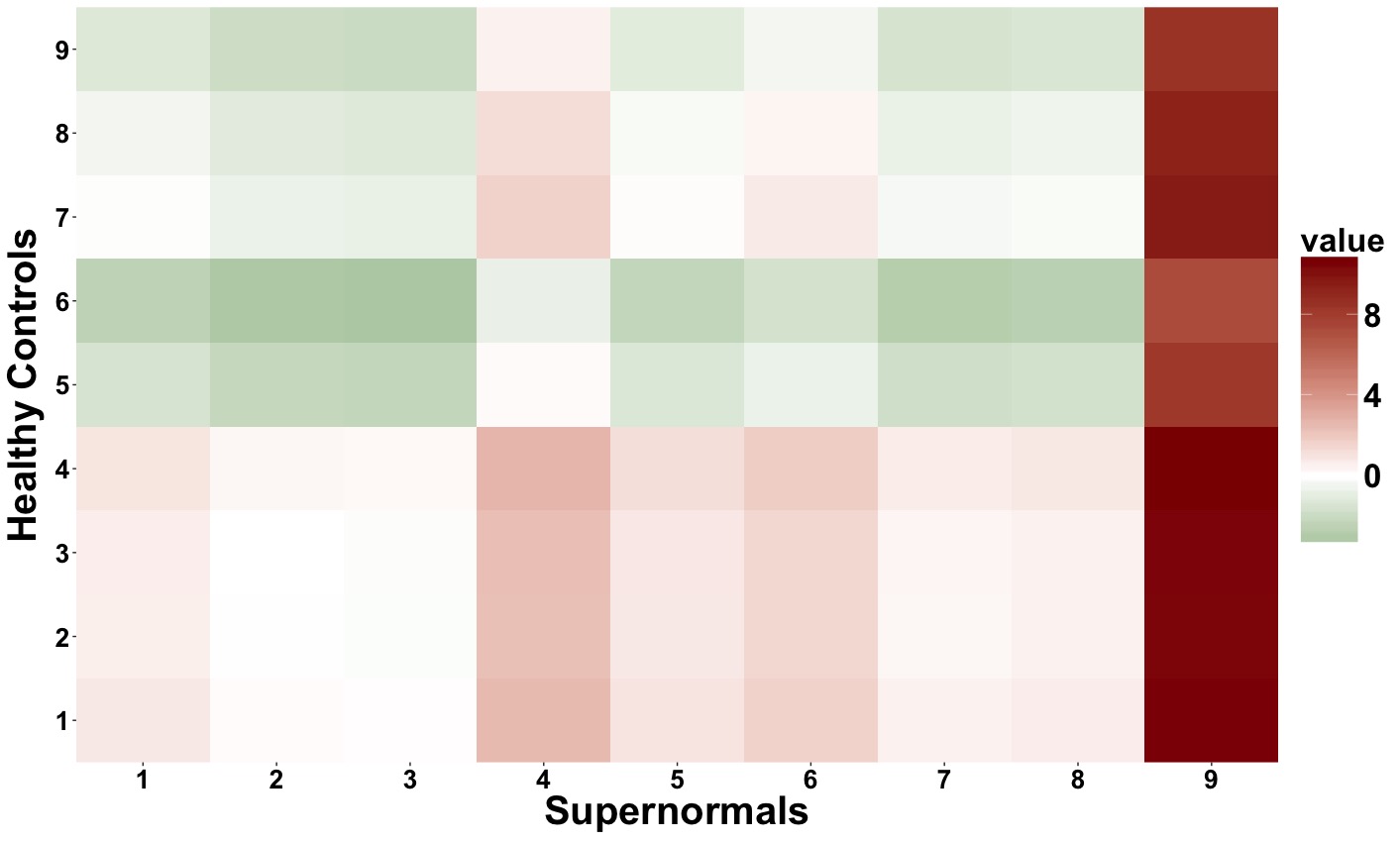}
\caption{Individual specific $\Omega$. } \label{Omega_ADNI3}
\end{subfigure}\hfill
\begin{subfigure}{0.45\textwidth}
\centering
\includegraphics[scale = 0.16]{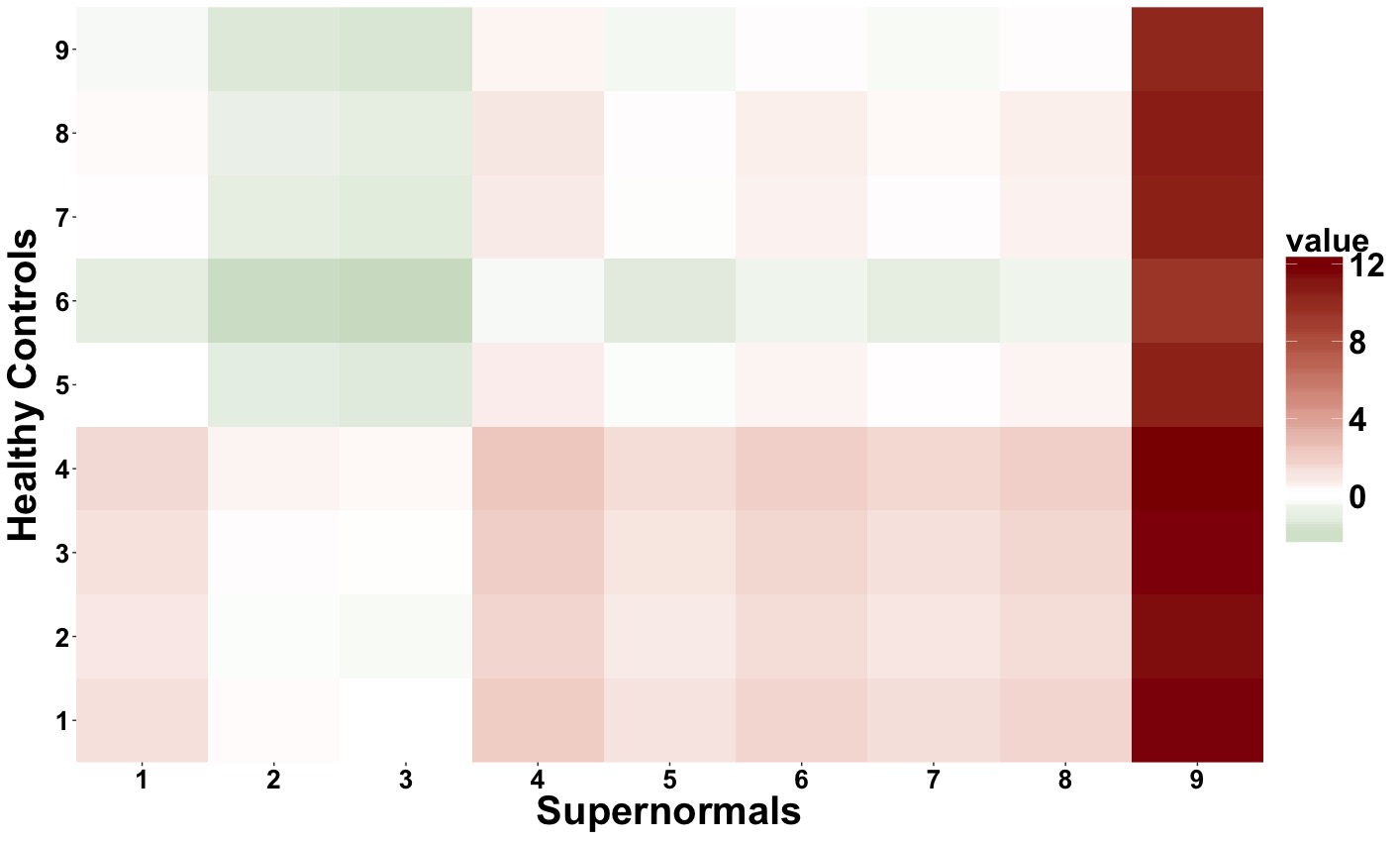}
\caption{Individual specific sample mean.} \label{sample_mean_ADNI3}
\end{subfigure}\hfill
\end{subfigure}
\caption{ \em{The magnitudes of functional connectivity between different ROIs are compared in case of hierarchical posterior estimate of $\Omega$ (left panel) and sample estimate (right panel)  for ADNI dataset consisting of supernormals and healthy controls. The $(i,j)^{th}$ element of the associated matrix for both heatmaps is measured as mean difference of absolute off-diagonal values of $i^{th}$ supernormal and $j^{th}$ healthy control.} }
\label{ADNI_diff_plot}
\end{figure}

 To compare the functional connectivity between ROIs across SN and HC,  we look at all the off-diagonal elements of $\Omega_{i} (i=1,\dots,9)$ for  both SN and HC using an overlaid histogram in Figure \ref{off_diag_SNHC}. The overlaid histograms clearly show that on an average the FC for the SN individuals is higher. 
% {\color{red}This histogram summarizes over the functional connectivity between different regions of interests for SN and HC.}
In addition, it is also important to know which one out of the $21$ pairs of regions accounts for the maximum separation in $\Omega_{i}$ between SN and HC.  A simple multiple comparison test reveals that the FC difference between SN and HC for the ROI-pairs  (2, 3)  and (3, 6) are statistically significant (p-values $0.038$ and $0.013$), where (2,3) represents a posterior regions' connection and (3,6) represents an anterior-posterior connection. This finding is in line with the literature \citep{lin2017cingulate}: SN group has higher strength of connectivity within posterior regions or between posterior and anterior regions. Box plots for ROI-pairs in Figure \ref{SN_HC_ROI} clearly show that the FC for SN is higher than HC for both the ROI-pairs. 
Next we compare overall functional connectivity between supernormals and healthy controls through  corresponding magnitudes of FC between different ROIs. Figure \ref{ADNI_diff_plot} represents the heat map of matrices associated with hierarchical posterior estimate of $\Omega$ and sample mean. Each element in the associated matrix represents the mean difference of absolute values of off-diagonal elements of $\Omega$ in Figure \ref{Omega_ADNI3} and the sample mean in Figure \ref{sample_mean_ADNI3} . These off-diagonal elements represent individual specific FC between different ROIs. The difference is slightly more evident for the posterior estimate in Figure \ref{Omega_ADNI3} than the sample mean in Figure  \ref{sample_mean_ADNI3}.

\section{Hierarchical Change point Model}\label{HCP_model}
Although a majority of previous works on modeling functional connectivity assumes stationarity \citep{friston2011functional,hutchison2013dynamic}, recent developments in Dynamic Connectivity Regression \cite{cribben2013detecting} suggest the necessity of incorporating non-stationary modeling of the time series of covariance matrices.  It is reasonable to  assume that different parts of brains will react distinctly under the effect of external stimuli, so assuming a common mean for the Wishart distribution in \eqref{HCovModel_decomp} is not warranted unless the subjects are in a resting state. Moreover, in the presence of multiple subjects, it becomes necessary to borrow information across multiple subjects while retaining some commonality features. Preliminary time series models based on sliding window technique \citep{lindquist2014evaluating} and asymptotic tests are based on a single subject and do not naturally extend to the case when multiple subjects are concerned.
 
In the following, we extend our hierarchical model in \eqref{HCovModel_decomp} to include the most simple departure from stationarity, which is accommodating a single change point in the mean of the Wishart distribution.  

Focussing on one action, the hierarchical change point model across individuals is 
\begin{equation} \label{HCpModel}
\begin{rcases}
 & S_{it}^{(1)} \sim W(\phi_{1}, \phi_{1}^{-1} \Omega_{1_{i}}), t = 1,\dots,c_{i} \\
 & S_{it}^{(1)} \sim W(\phi_{2}, \phi_{2}^{-1} \Omega_{2_{i}}), t = c_{i}+1,\dots,T
\end{rcases}
\quad
\begin{array}{r@{\;}l}
i = 1, \dots , n.
\end{array}
\end{equation}
where $c_i$ represents the change point specific to subject $i$. We  used scaled Wishart distributions with different individual specific means $\Omega_{1_{i}}$ and $\Omega_{2_{i}}$ before and after the change points respectively.  A similar orthogonal factor model type decomposition (discussed in \eqref{eqOmega}) is proposed on $\Omega_{1_{i}}$ and $\Omega_{2_{i}}$. 
\begin{equation} \label{HCpM_Omega_decom}
 \Omega_{1_{i}} = V_{1}D_{1_{i}}V_{1}^{\T} + \sigma_{1_{i}}^{2}I_{p},  \quad 
 \Omega_{2_{i}} = V_{2}D_{2_{i}}V_{2}^{\T} + \sigma_{2_{i}}^{2}I_{p}, \quad i = 1, \dots , n. 
\end{equation}
Observe that the orthogonal matrices $V_1$ and $V_2$ are fixed across individuals and thus viewed as a common dictionary on which  individual specific loadings $D_{1_{i}}$ and $D_{2_{i}}$ act on to create subject specific deviations.   We place independent uniform prior distributions on the Stiefel manifold for $V_{1}$ and $V_{2}$ along with independent global-local prior on $\tilde{D}_{1}$ and $\tilde{D}_{2}$ exactly as in \S \ref{sec:indep}. Independent inverse-gamma priors are chosen on the $\sigma_{1_{i}}^{2}$ and $\sigma_{2_{i}}^{2}$. We assumed that {\em apriori} any time point is equally probable to be a {\em change-point}, i.e., 
\begin{align} \label{HCpM_prior}
 c_{i} \sim \text{Discrete-Uniform(\{1, \ldots, T\})}.
\end{align}
An efficient Gibbs sampler is developed mimicking \S \ref{sec:indep} with an additional step to update the change-points $c_i, i=1, \ldots, n$.  A detailed calculation of the steps is provided in the Appendix (section \ref{derivation_HCM}). 

\subsection{Simulation Study for Hierarchical Change point Model}
To demonstrate the the hierarchical change point model \eqref{HCpModel} on simulated datasets, 
we consider $n=100$, $p=50$ and $T=26$ with $n_{2} = 40$ individuals having change-points 
$c_{0i} \in \{2, \ldots, T-1\}$ and the remaining individuals with size $n_{1} = 60$ having no change points. For simplicity and to develop a simulation scenario analogous to the HCP dataset,
we assume all the individuals are observed at the same time points and the boundary points cannot be considered as a candidate for a change-point.  For clarity of exposition, any parameter with subscript ``1''  correspond to the pre-change-point regime (deemed as Group 1) and the ones with subscript ``2'' corresponds to the post-change-point (Group 2).  

True individual specific ranks are generated from discrete uniform distribution spanning over $\{1, \dots, r^{\ast}=10 \}$. The true values of the diagonal matrices $\{ D_{01_{i}} \}_{i=1}^{n}$ and $\{ D_{02_{i}} \}_{i=1}^{n_{2}}$ are generated from $\text{unifrom}(0,5)$ to include a  wide range signal strengths. $\{ \sigma^{2}_{01_{i}} \}_{i=1}^{n}$ and $\{ \sigma^{2}_{02_{i}} \}_{i=1}^{n_{2}}$ are generated from $\text{uniform}(0.25,0.50)$. Using these values, $\Omega_{1}$ and $\Omega_{2}$ are constructed using the equation (\ref{HCpM_Omega_decom}) and we set $(\phi_{1}, \phi_{2}) = (p+1,p+1)$.  100 replicated datasets are then generated  from (\ref{HCpModel}). 
 
\begin{figure}[h!] % "[t!]" placement specifier just for this example
\begin{subfigure}{0.4\textwidth}
\centering
\includegraphics[scale = 0.2]{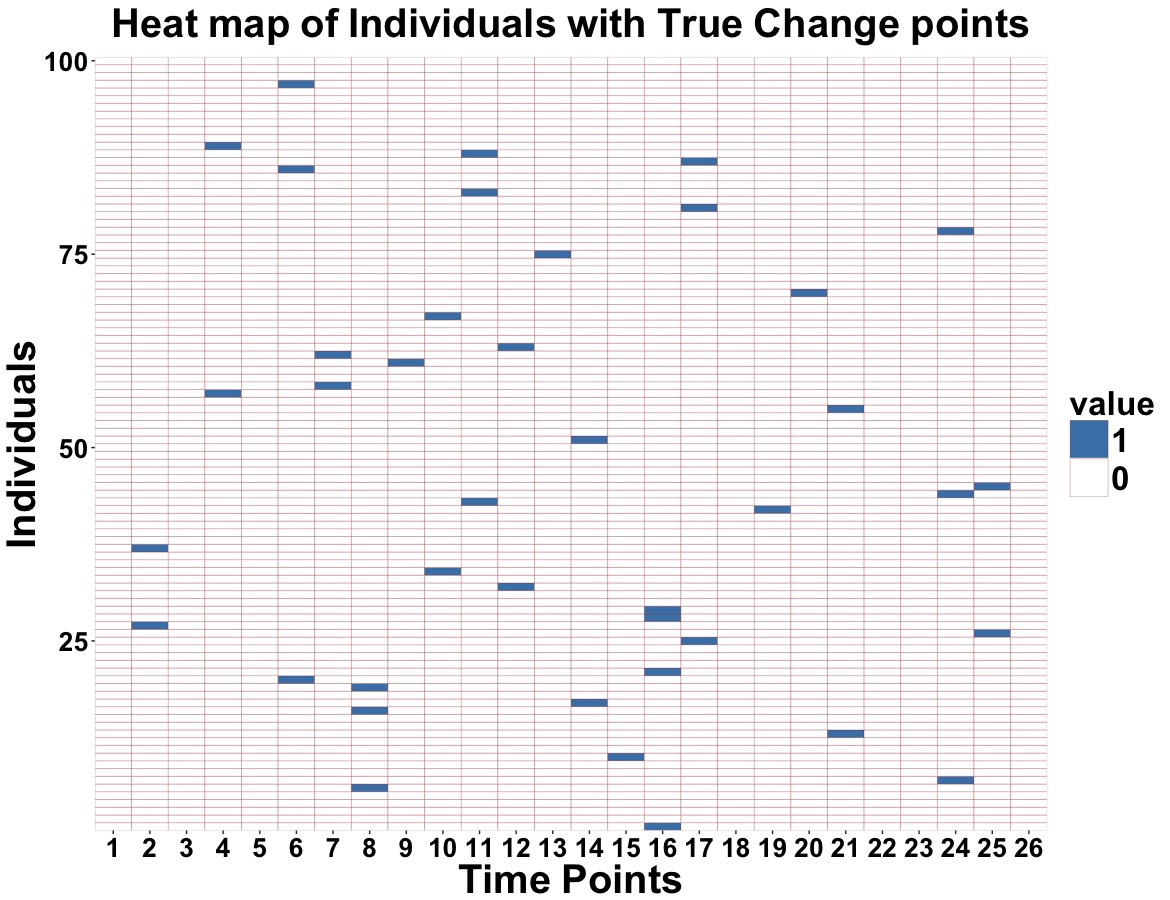}
\caption{} \label{HCpM_sim_cp_true}
\end{subfigure}\hfill
\begin{subfigure}{0.4\textwidth}
\centering
\includegraphics[scale = 0.2]{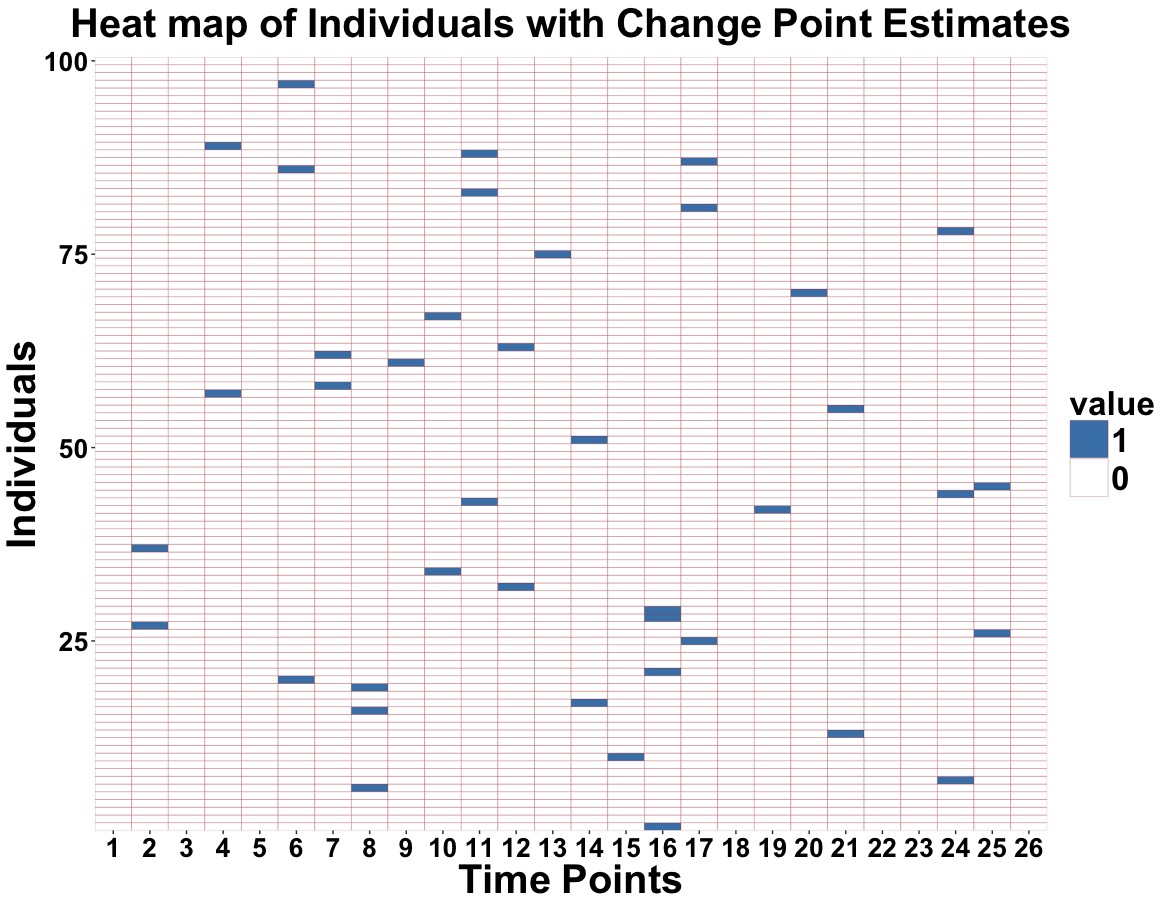}
\caption{} \label{HCpM_sim_cp_est}
\end{subfigure}

\caption{{ \em Change point estimates of Hierarchical Change Point Model with $(n,p,T) = (100,50,26)$. (\ref{HCpM_sim_cp_true}) Left panel represents the true change points with a heat map of the binary matrix $C = (c_{il})$, where $c_{il} = 1$ if the ith individual has change point at l, and 0 otherwise. (\ref{HCpM_sim_cp_est}) Right panel shows the heat map of individuals with estimated change points. }}
\label{HCpM_sim_change_point}
\end{figure}
\begin{figure}[h!] % "[t!]" placement specifier just for this example
\begin{subfigure}{0.45\textwidth}
\centering
\includegraphics[scale = 0.19]{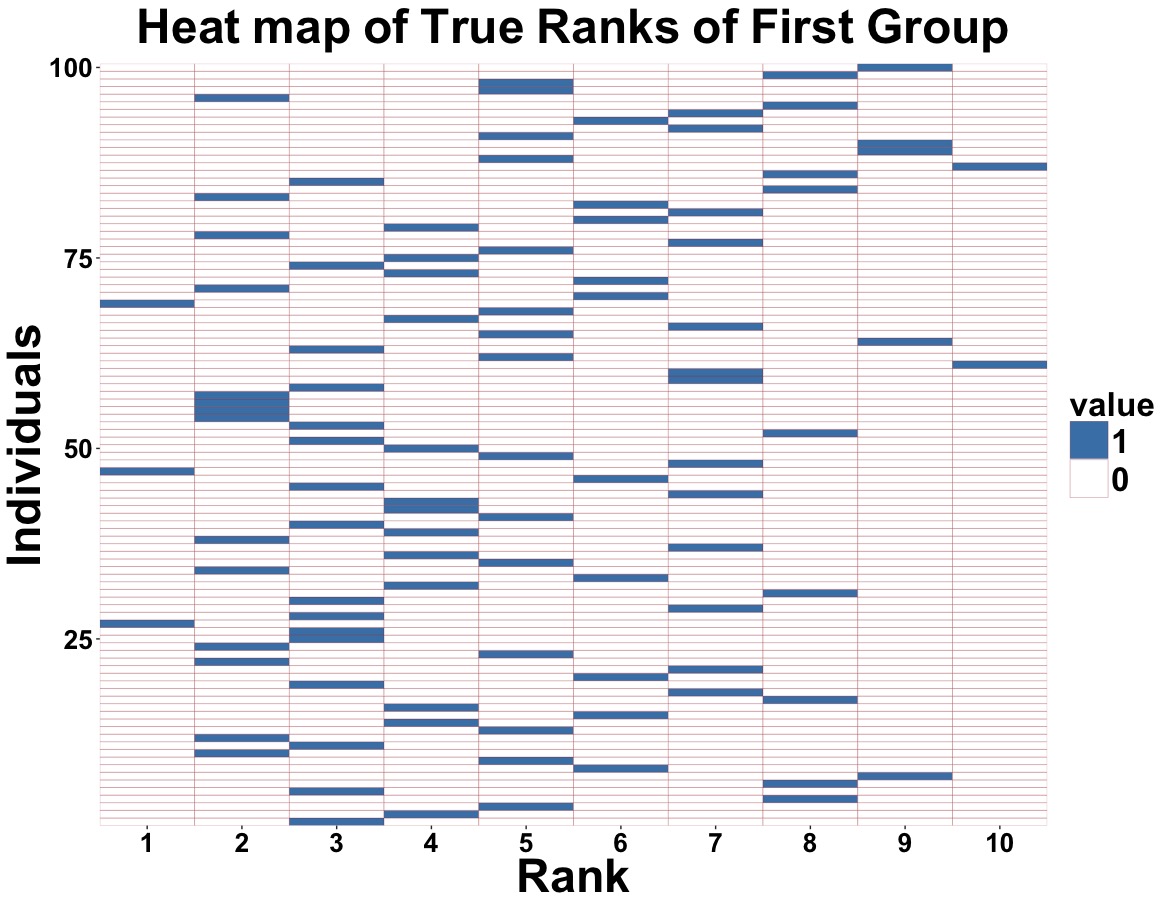}
%\caption{} \label{HCpM_sim_true_rank1}
\end{subfigure}\hfill
\begin{subfigure}{0.45\textwidth}
\centering
\includegraphics[scale = 0.19]{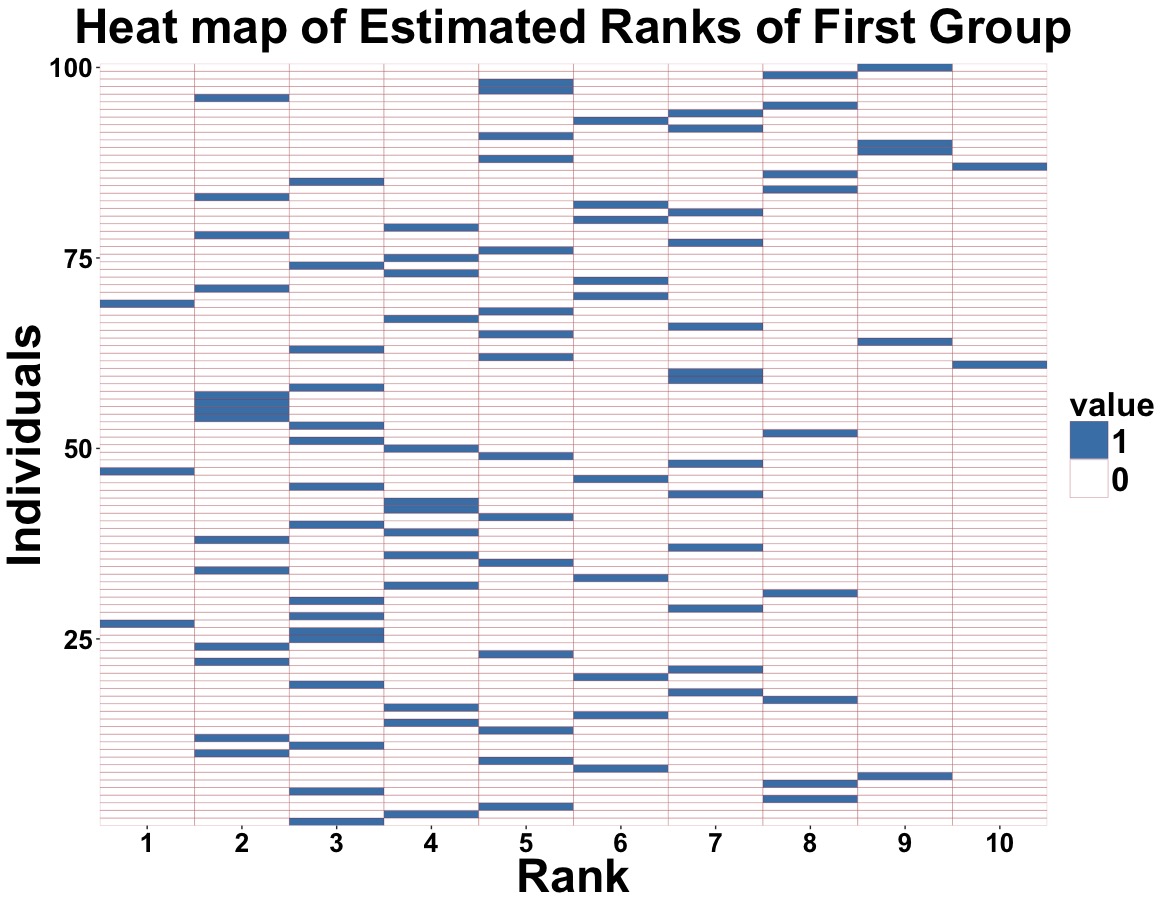}
%\caption{} \label{HCpM_sim_est_rank1}
\end{subfigure}

\medskip
\begin{subfigure}{0.45\textwidth}
\centering
\includegraphics[scale = 0.19]{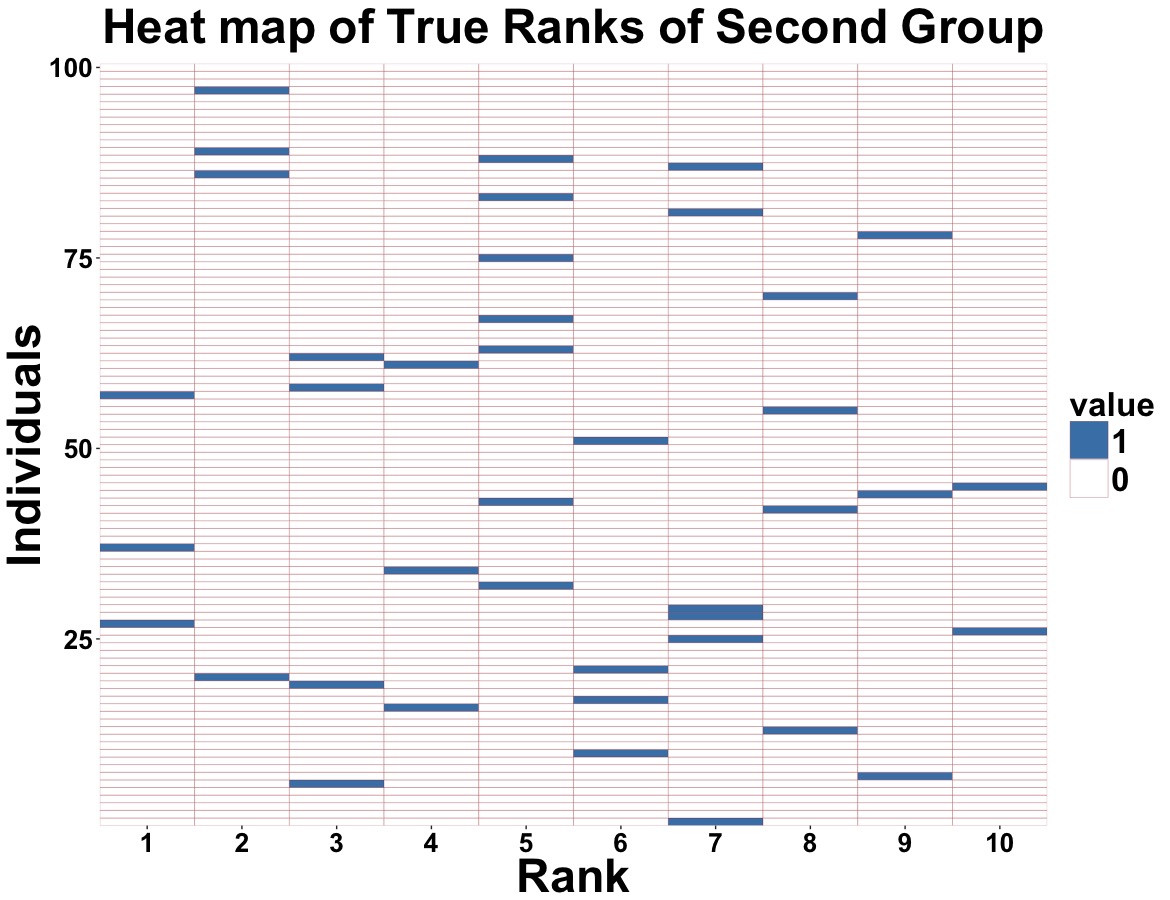}
%\caption{} \label{HCpM_sim_true_rank2}
\end{subfigure}\hfill
\begin{subfigure}{0.45\textwidth}
\centering
\includegraphics[scale = 0.19]{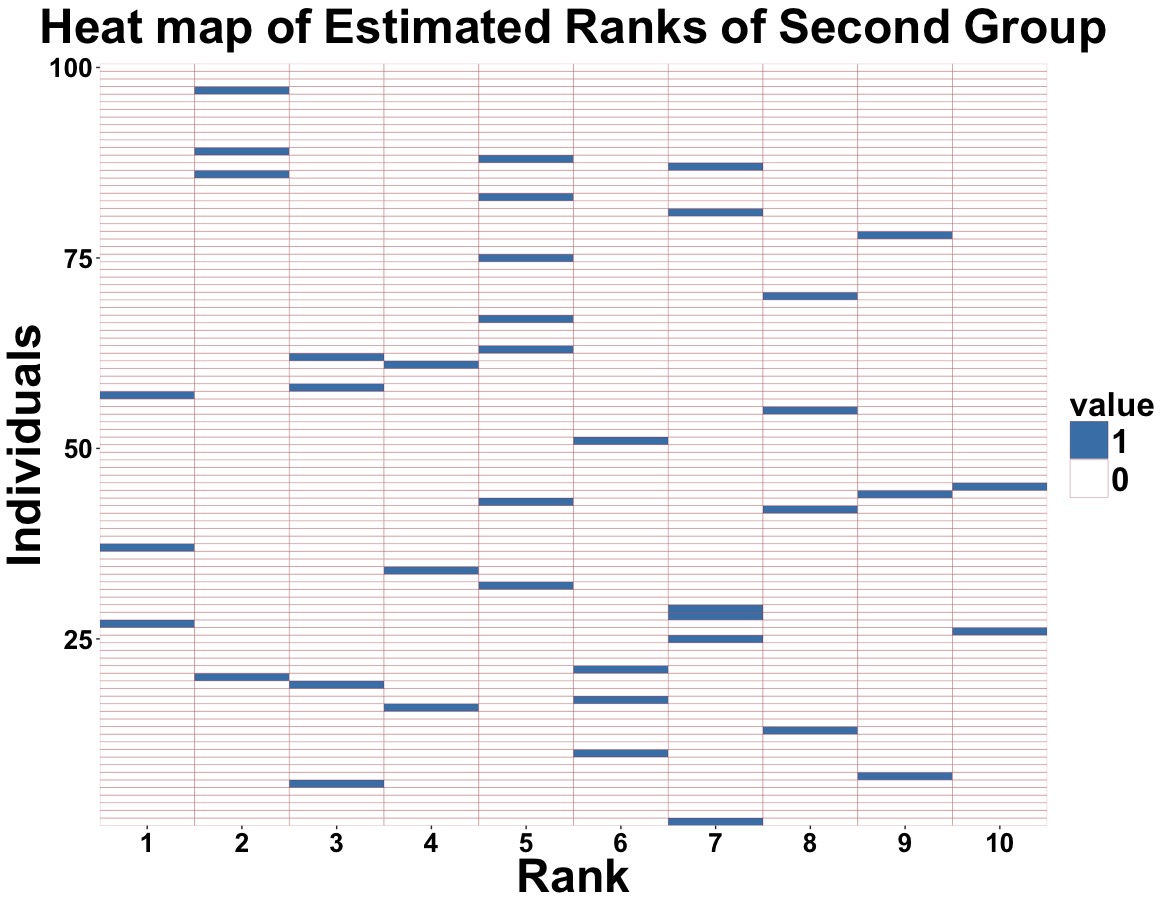}
%\caption{} \label{HCpM_sim_est_rank2}
\end{subfigure}

\caption{ {\em Rank estimates of Hierarchical Change Point Model. Upper panel: The heat of true and estimated ranks corresponding to before change points scenario. Lower panel provides the after change point picture for $n_{2} = 40$ individuals with change points. The construction of binary matrices are described in Figure \ref{HS_rank}. }} 
\label{HCpM_sim_rank}
\end{figure}
\begin{figure}[h!] % "[t!]" placement specifier just for this example

\begin{subfigure}{0.45\textwidth}
\centering
\includegraphics[scale = 0.16]{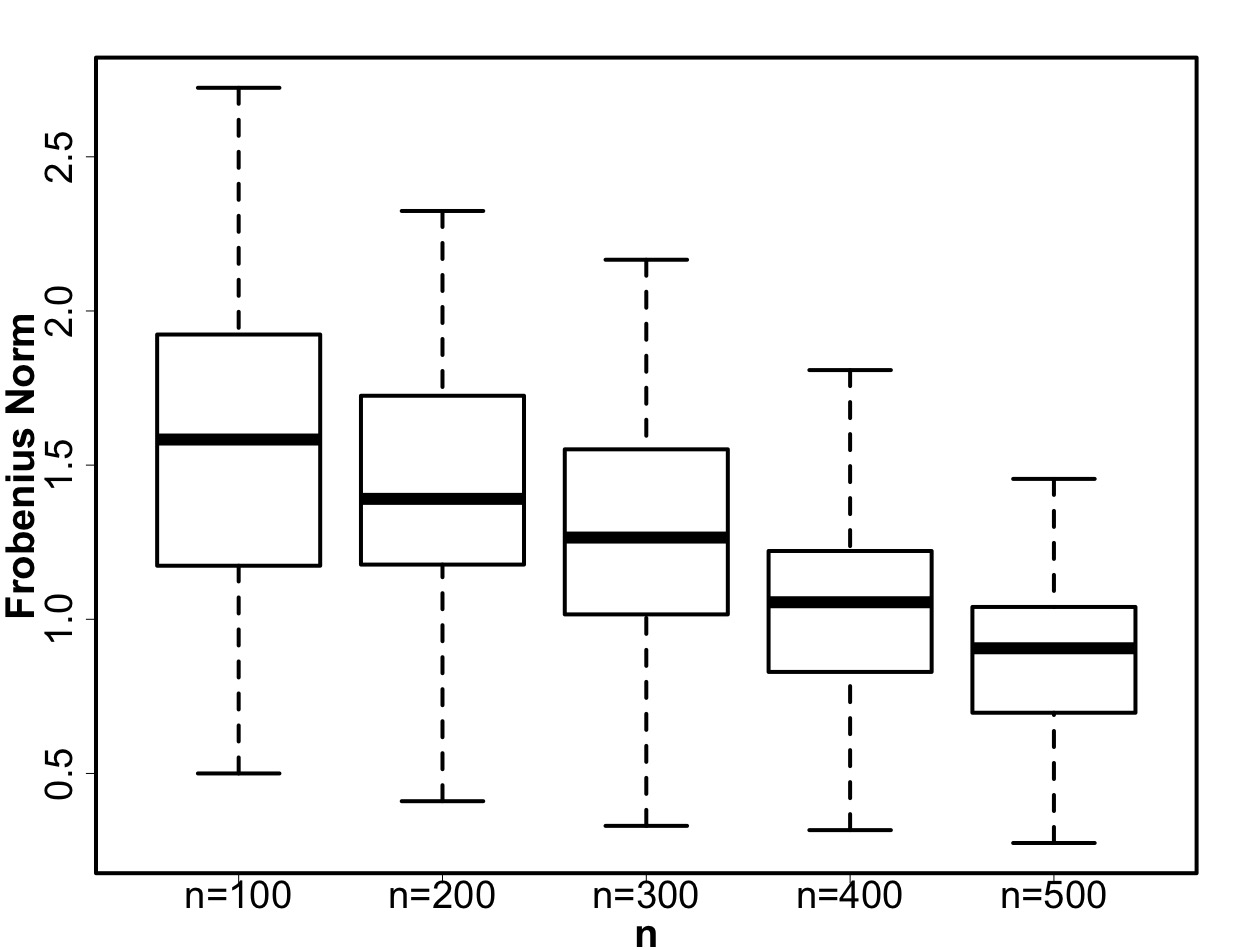}
\caption{} \label{HCpM_D_Omega1}
\end{subfigure}\hfill
\begin{subfigure}{0.45\textwidth}
\centering
\includegraphics[scale = 0.16]{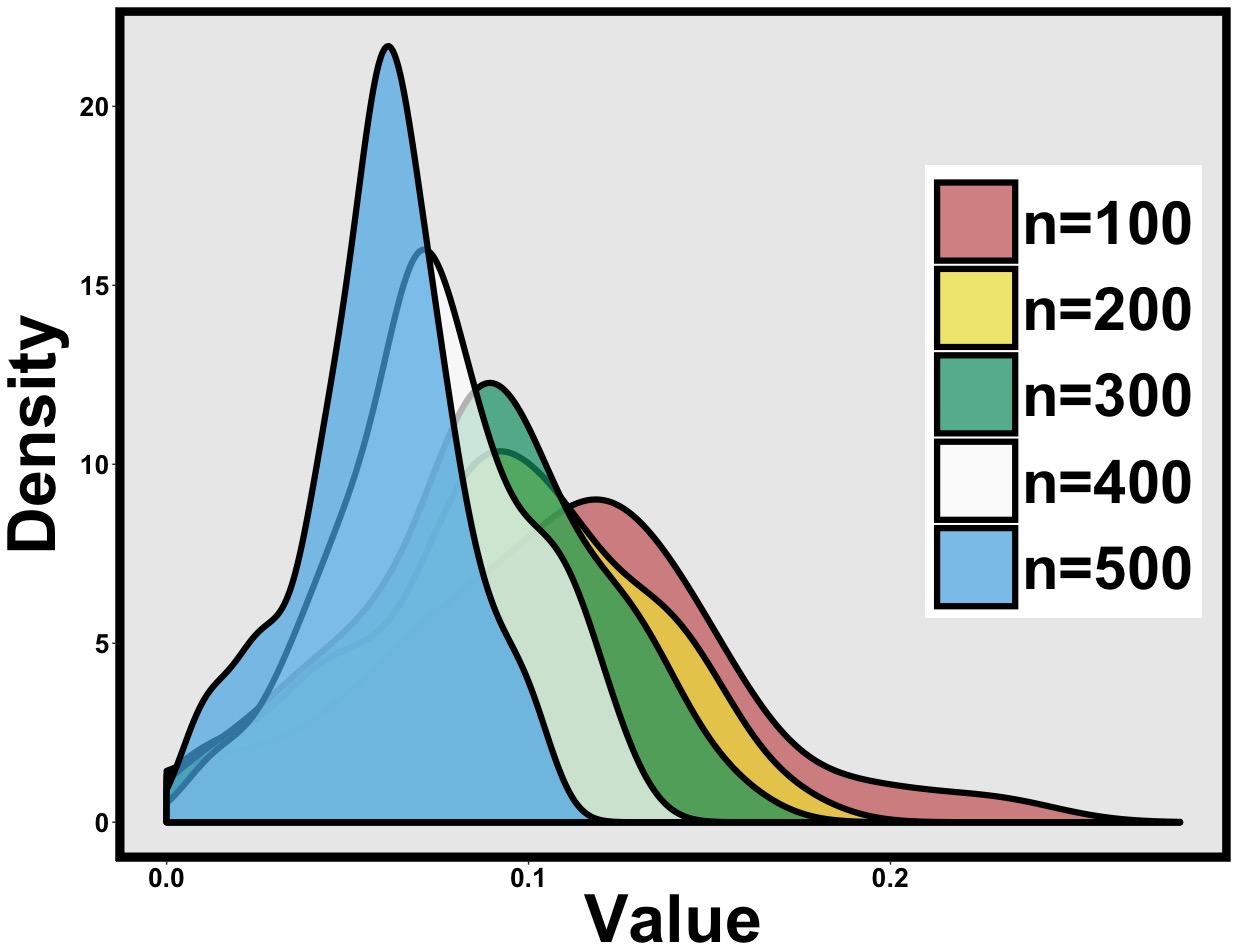}
\caption{} \label{HCpM_sim_sigma1}
\end{subfigure}

\medskip
\begin{subfigure}{0.45\textwidth}
\centering
\includegraphics[scale = 0.16]{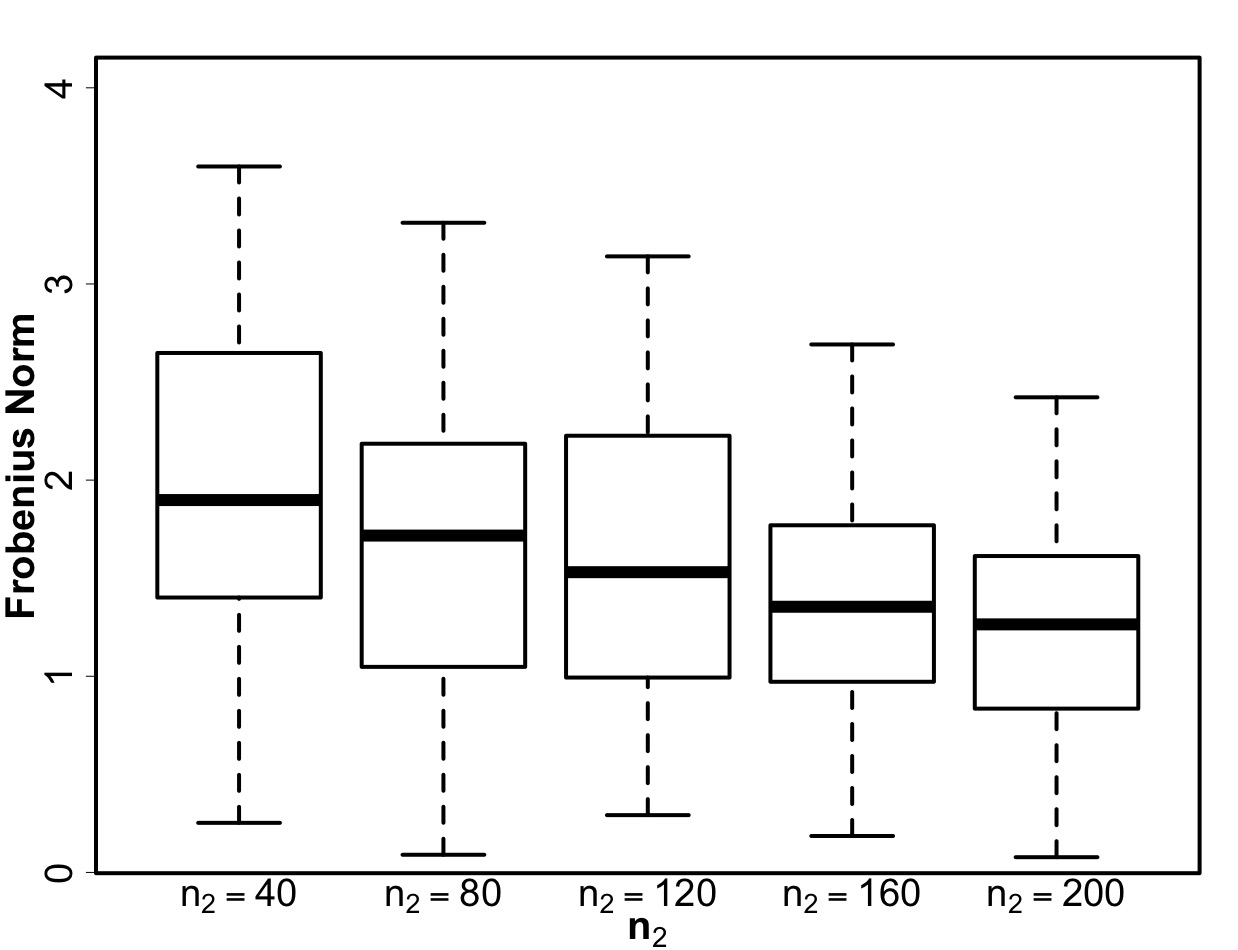}
\caption{} \label{HCpM_D_Omega2}
\end{subfigure}\hfill
\begin{subfigure}{0.45\textwidth}
\centering
\includegraphics[scale = 0.16]{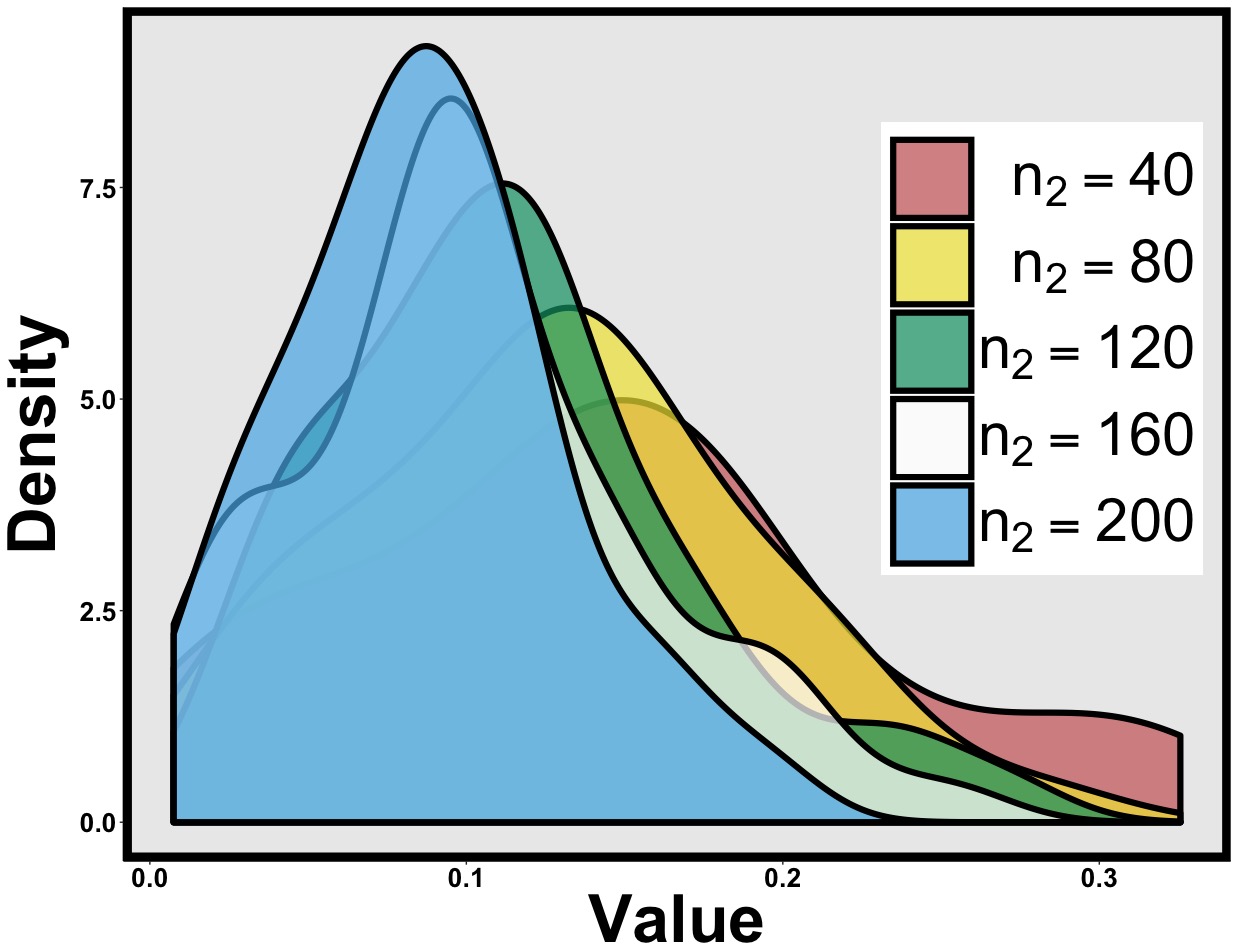}
\caption{} \label{HCpM_sim_sigma2}
\end{subfigure}

\caption{ { \em Left panel : Boxplot of $d_{\Omega_{1}}$ (upper panel) and $d_{\Omega_{2}}$ (lower panel) across simulated replicates for increasing n and $n_{2}$ respectively. Right panel : Density plot of $d_{\sigma_{1}}$ (upper panel) and $d_{\sigma_{2}}$ (lower panel) w.r.t. varying n and $n_{2}$ respectively.     } } 
\label{HCpM_sim_Omega_sigma}
\end{figure}

The MCMC is run for 5,000 iterations leaving a burn-in sample of 5,000. Subject-specific change point estimates $\hat{c}_i$ are obtained from the posterior mode of $c_i$.  
Since the focus of this section is correct detection of change-points, we only display the estimated change-points corresponding to the $n_{2} = 40$ individuals in  Figure \ref{HCpM_sim_change_point}. Our proposed model is successful to recover individual specific change points. The ranks corresponding to the covariance matrices across individuals are also estimated correctly in all the cases as presented in Figure \ref{HCpM_sim_rank}.

%Figure (\ref{HCpM_sim_change_point}) represents estimate change points for specific individuals and we can observe here our proposed method is able detect all the change points. In the exactly same setting we have provided the heat map of true and estimated rank1 and rank2 in figure \ref{HCpM_sim_rank}. To estimate rank2 we have considered only those $40$ individuals with change points. Ranks for group1 and group2 for all the corresponding individuals are recovered properly by our proposed method.

To demonstrate consistency of the estimate of $\Omega_{ji}, j=1,2; i=1, \ldots, n$ with increasing
sample size, we consider another simulation setting 
where $p$ is fixed at $50$ and $n$ takes values in the range $\{100,200,300,400,500\}$ with  $n_{2} \in \{40,80,120,160,200\}$. Figure \ref{HCpM_sim_Omega_sigma} presents the summary of the variability of the parameters  $(\Omega_{1i}, \sigma_{1i}^{2},\Omega_{2i},\sigma_{2i}^{2})$ appropriately summarized for the $n$ individuals using the metrics $d_\Omega$ and $d_\sigma$ over 100 simulated replicates. It is evident that on an average $d_{\Omega_{1}}$ (Figure \ref{HCpM_D_Omega1}) and $d_{\Omega_{2}}$ (Figure \ref{HCpM_D_Omega2}) decreases with a smaller spread with increasing $n$ and $n_{2}$ respectively. Similarly the density plots of $d_{\sigma_{1}}$ (Figure \ref{HCpM_sim_sigma1}) and $d_{\sigma_{2}}$ (Figure \ref{HCpM_sim_sigma2}) become more concentrated as $n$ increases.   

%Next we have presented the variation in parameters $(\Omega_{1}, \sigma_{1}^{2},\Omega_{2},\sigma_{2}^{2})$ for increasing $n$. Here we have considered $n=(100,200,300,400,500)$ and $n_{2} = (40,80,120,160,200)$ where $n_{2}$ denotes the individuals with change points only. In figure \ref{HCpM_sim_Omega_sigma}, we have presented variation of the parameters across subjects. In case of $\Omega_{2}$ and $\sigma_{2}^{2}$ we have only considered the individuals with change points. There is a decreasing trend in both boxplots of $D(\Omega_{1_{i}})$ and $D(\Omega_{2_{i}})$ in figure \ref{HCpM_D_Omega1} and \ref{HCpM_D_Omega2} respectively. Densities of $D(\sigma^{2}_{1_{i}})$ and $D(\sigma^{2}_{2_{i}})$ are more concentrated with increasing value of $n$ and $n_{2}$ respectively. The whole scenario is presented in figure \ref{HCpM_sim_sigma1} and \ref{HCpM_sim_sigma2}.

%\newpage
\section{Real Data Analysis for HCP Dataset}\label{Real_HCP_data}
In this section, we consider the HCP dataset \citep{van2013wu} as discussed in \S \ref{sec:realdata}. Time series of covariance matrices describing the connectivity were acquired from each subject while they were performing different tasks involving different neural systems, under resting state or external stimuli.  A quick exploratory analysis of the dataset shows the wide variation in the range of values of the covariance matrices. For the change point model \eqref{HCpModel} to be applicable, we scale each covariance matrix by the lowest singular value of that matrix as a simple variance stabilizing transformation.   Based on empirical validation from Figure \ref{fig:effrank} on small effective ranks of the covariance matrices, we applied the hierarchical change point model \eqref{HCpModel}. Task-specific summary of findings is provided below. 

\subsection{Case study for Motor Task}
\begin{figure}[h!]
\begin{subfigure}{0.45\textwidth}
\centering
\includegraphics[scale = 0.2]{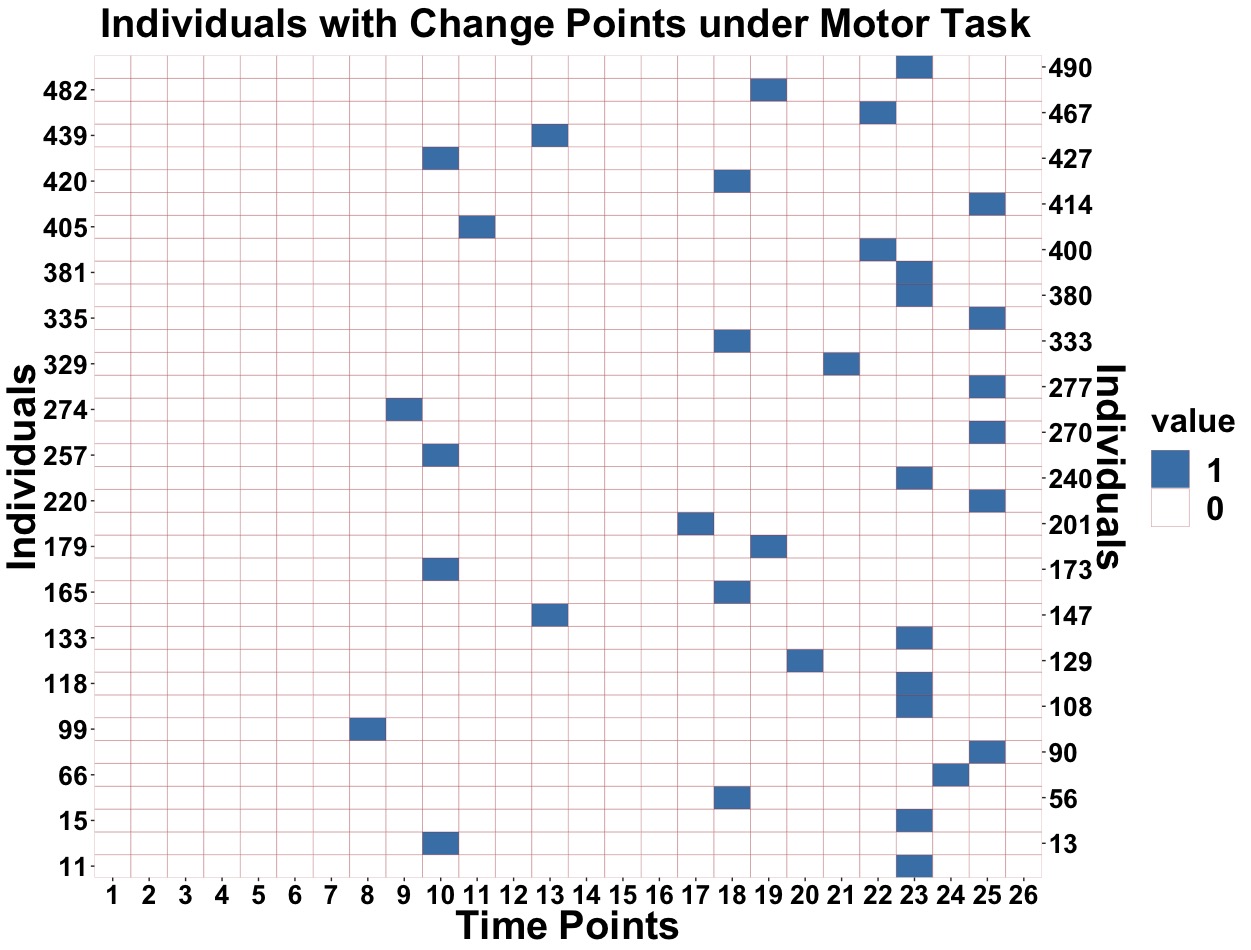}
\caption{} \label{Motor_Change_Point}
\end{subfigure}\hfill
\begin{subfigure}{0.45\textwidth}
\centering
\includegraphics[scale = 0.2]{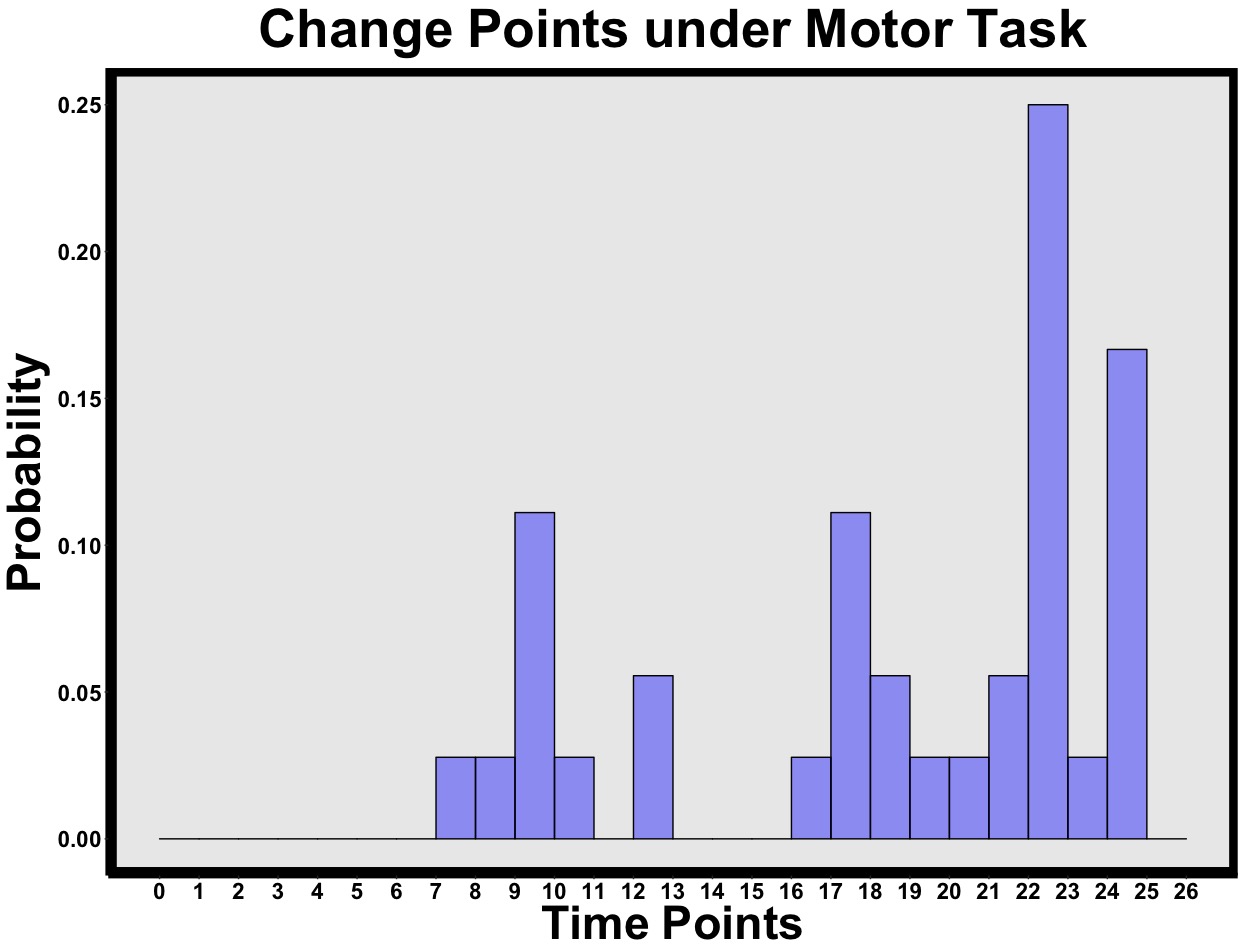}
\caption{} \label{hist_chp_Motor}
\end{subfigure}
\caption{ \em{ In Figure (\ref{Motor_Change_Point}), we have provided heat map of a binary matrix consisting of $1$ to $(i,j)^{\text{th}}$ position which corresponds to $i^{\text{th}}$ labeled individual and $j^{\text{th}}$ $(j=1,\dots,26)$ time point which is a change point for the corresponding individual and $0$ otherwise. Heat map was made with individuals consisting of change points under Motor task. Individuals are labeled on both sides of the y-axis. Figure (\ref{hist_chp_Motor}) is histogram of change points under Motor task which indicates most of the individuals have change points at $23$. }}
\label{Motor_change_point}
\end{figure}

%A detailed description of gambling task corresponding to HCP can be found in \cite{Delgado2000}. We applied hierarchical change point model on scaled data which detects $70$ individuals with change points under Gambling task. In figure \ref{Gambling_change_point} we have labeled $70$ individuals with their corresponding change points. Figure \ref{Gambling_Change_Point} consists of 70 individuals with their most eminent change point which has been detected through the hierarchical model under gambling task. Figure \ref{hist_chp_gambling} shows the over all pattern of the most eminent change points across all the individuals through a histogram plot of the change points. The histogram in figure \ref{hist_chp_gambling} shows that more than 10\% individuals have change points at $21$.  
The HCP motor task experiment was set up by Buckner and colleagues \citep{Buckner2011}. Participants are presented with visual cues that ask them to either tap their left or right fingers, or squeeze their left or right toes, or move their tongue to map motor areas. In the experiment, there are 13 blocks, with 4 hand movements, 4 foot movements, and 2 tongue movements. In addition, there are 3 15-second fixation blocks between different tasks.  We identified ten cortical ROIs related to the motor control around the motor strip area, including left and right postcentral gyrus, precentral gyrus, and central gyrus,  and generated a  $10 \times 10 $ covariance matrix time series with $26$ time points. The proposed hierarchical change point detection model is then applied with fitted ranks in $\{5, 6, 7, 8\}$.  Change points are observed for $36$ individuals; refer to the Table \ref{table:hcp_model_different_fitted_rank}. 
Figure \ref{Motor_Change_Point} shows the 36 labeled individuals from the first column of Table \ref{table:hcp_model_different_fitted_rank} with their corresponding most dominant change points. The histogram in Figure \ref{hist_chp_Motor} displaying the pattern of the change points across the individuals shows that most of the individuals have change points at time point $23$. In the experiment design, this corresponds to the time point of switching the movement from hand and foot to the tongue. We applied our methodology on the gambling task as well. A discussion on the findings is deferred to the Appendix (section \ref{HCP_gambling}). 

One obvious limitation of \eqref{HCpModel} is that it can only account for the most dominant change point. It is possible that there exists more than one change point for a specific individual under a certain task.  In the following, we extended the methodology to enable detection of multiple change points. 

%We observe change points for 36 individuals. 
%Detailed results for different rank values are provided on table \ref{table:hcp_model_different_fitted_rank}. 
%\begin{figure}[h!]
%\begin{subfigure}{0.45\textwidth}
%\centering
%\includegraphics[scale = 0.22]{Gambling_chp_R}
%\caption{} \label{Gambling_Change_Point}
%\end{subfigure}\hfill
%\begin{subfigure}{0.45\textwidth}
%\centering
%\includegraphics[scale = 0.21]{hist_chp_gambling}
%\caption{} \label{hist_chp_gambling}
%\end{subfigure}
%\caption{ \em{ Right panel shows heat map of binary matrix consisting of $1$ to $(i,j)^{\text{th}}$ position which corresponds to $i^{\text{th}}$ individual and $j^{\text{th}}$ $(j=1,\dots,23)$ time point which is a change point for the corresponding individual and $0$ otherwise. Individuals with one change point under gambling task are labeled on y-axis. In left panel we have histogram of change points under gambling task which shows most of the individuals have change point at $21$. }}
%\label{Gambling_change_point}
%\end{figure}

\subsection{Multiple Change point Analysis}\label{multiple_chp_HCP}
\begin{figure}[h!]
\centering
\includegraphics[scale = 0.2]{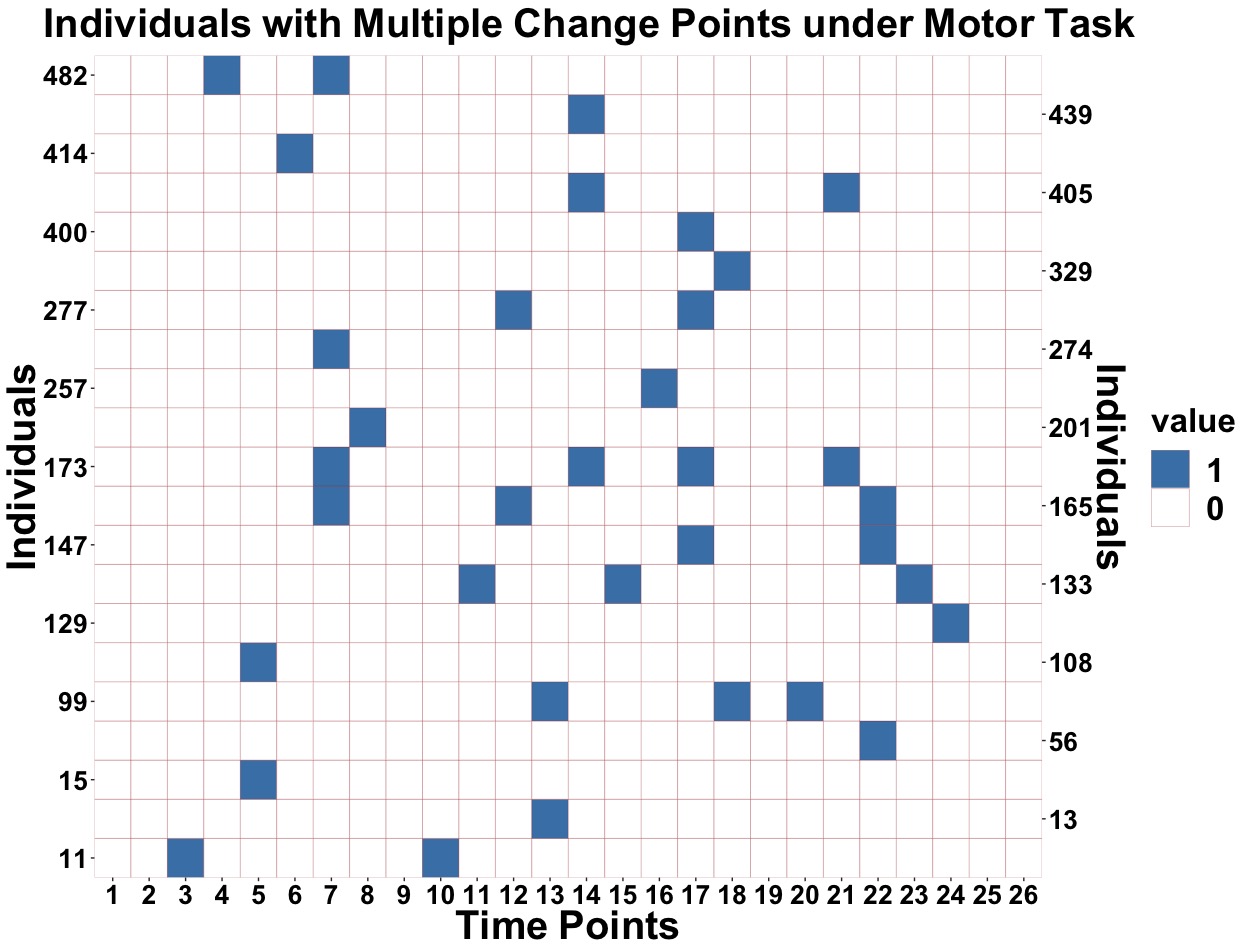}
\caption{ \em{ Multiple change points for different individuals under Motor task. On y-axis of the plot we have labeled the individuals with multiple change points and on x-axis we have time points. Blue lines denote the individual specific change points.} }
\label{Motor_Multiple_change_point}
\end{figure}

Our hierarchical change point model detects the most dominant change points along the time frame. We adapted a standard sliding window approach to detect multiple change points for different individuals. Denote by $c_{i}$ $(1<c_{i}<T)$ the first most dominant change point in the interval $\{1,\dots,T\}$ for individual $i$ which is detected through the hierarchical change point model. We slide our time window before and after the most dominant change point $c_{i}$.  We note here that applying  our change-point model over a time window containing $c_{i}$ recovers the $c_{i}$ as the most dominant change point. Hence we consider the windows $\{1, \dots , c_{i}-1\}$ \& $\{ c_{i}+1, \dots , T \}$  for further detection of the next dominant change points.  Suppose, there is a change point $c^{\ast}_{i}$ in the interval $\{1, \dots , c_{i}-1\}$. Then we again split the time window into $\{1, \dots , c^{\ast}_{i}-1 \}$ \& $\{c^{\ast}_{i} + 1, \dots , c_{i}-1 \}$ and apply the change point detection method to the two intervals separately. Same procedure is followed on the time window $\{ c_{i}+1, \dots , T \}$.  Figure \ref{Motor_Multiple_change_point} shows the individuals specific multiple change points under 
%\begin{figure}[h!]
%\begin{subfigure}{0.45\textwidth}
%\centering
%\includegraphics[scale = 0.2]{Multiple_ch_Gambling}
%\caption{\em{Gambling Task}} \label{Gambling_Multiple_change_point}
%\end{subfigure}\hfill
%\begin{subfigure}{0.45\textwidth}
%\centering
%\includegraphics[scale = 0.2]{Multiple_ch_Motor}
%\caption{\em{Motor Task}} \label{Motor_Multiple_change_point}
%\end{subfigure}
%\caption{ \em{ Multiple change points for different individuals under Gambling and Motor task respectively. On y-axis of both the plots we have labeled the individuals with multiple change points and on x-axis we have time points. Blue lines denote the individual specific change points.} }
%\label{Change_point_Multiple}
%\end{figure}
 the motor task where we considered individuals with at least two change points. W detected $21$ individuals with multiple change points under motor task which is shown in Figure \ref{Motor_Multiple_change_point}. There is no individual under motor task with more than four change points. 

\subsection{ADNI Dataset}  
 We  applied the hierarchical change point model on the previously described ADNI dataset, where the subjects are believed to be at a resting state.  As anticipated, our model did not detect any change points in the dataset for both supernormals and health control groups. To test for model adequacy, we used the Watanabe-Akaike information criterion (WAIC) \citep{gelman2014understanding}. WAIC is a fully Bayesian approach to measure model accuracy computed with log pointwise posterior predictive density and then adding a correction for the effective number of parameters to adjust for overfitting. Table \ref{WAIC_table_SNHC} provides the WAIC values for ADNI dataset with respect to the independence model \eqref{eqModel}, hierarchical covariance model \eqref{HCovModel_decomp} and the change point model \eqref{HCpModel} respectively. The WAIC values are lowest in the case of hierarchical covariance model which reaffirms the lacks of  change points in resting state fMRI dataset. Higher WAIC values for the hierarchical change-point model suggests overfitting from a more complex model where the data do not have a changepoint. 
 
 \begin{table}[h!]
\begin{center}
\begin{tabular}{|c|c|c|} \hline
Model &SN &HC\\ \hline
Independence Model &$80.32$ &$76.71$   \\ \hline
Hierarchical Covariance Model &$49.07$  &$44.63$ \\  \hline
Hierarchical Change-point Model &$68.57$  &$61.96$  \\  \hline
\end{tabular}
\end{center}
\caption{WAIC values for ADNI dataset with respect to three models which are defined in \eqref{eqModel}, \eqref{HCovModel_decomp} and \eqref{HCpModel} respectively. Reported WAIC values are in scale of $10^{2}$. }
\label{WAIC_table_SNHC}
 \end{table}

\section{Model Validation}\label{HCP_model_validation}
In the following, we first consider an adhoc graphical summary measure of the posterior to justify the extension to hierarchical covariance model from the independence model for the HCP dataset. The exploratory analyses in Figures $17$ and $18$ were conducted to attest to two aspects of our hierarchical covariance model, namely i) using subject specific $D_{i}$ and variance components, and ii) using a common semi-orthogonal matrix $V$ across different subjects. A formal model comparison is done later using the WAIC \citep{gelman2014understanding} to provide more support to our visual illustration in Figures $17$ and $18$. 

Figure $17$ shows the variation of the posterior estimate of the maximum eigenvalue and $\sigma^2$ across the individuals, where data for each individual is fitted using the independence model on the $26$ time points.  The density plot is obtained by smoothing posterior estimates across the individuals. As clearly seen, there is a substantial variability in both the variance component and the maximum eigenvalue which prompted us to consider individual specific $\sigma_{i}^{2}$ and eigenvalues $D_{i}$ for the hierarchical covariance model.

\begin{figure}[h!]
\centering
\includegraphics[scale=0.2]{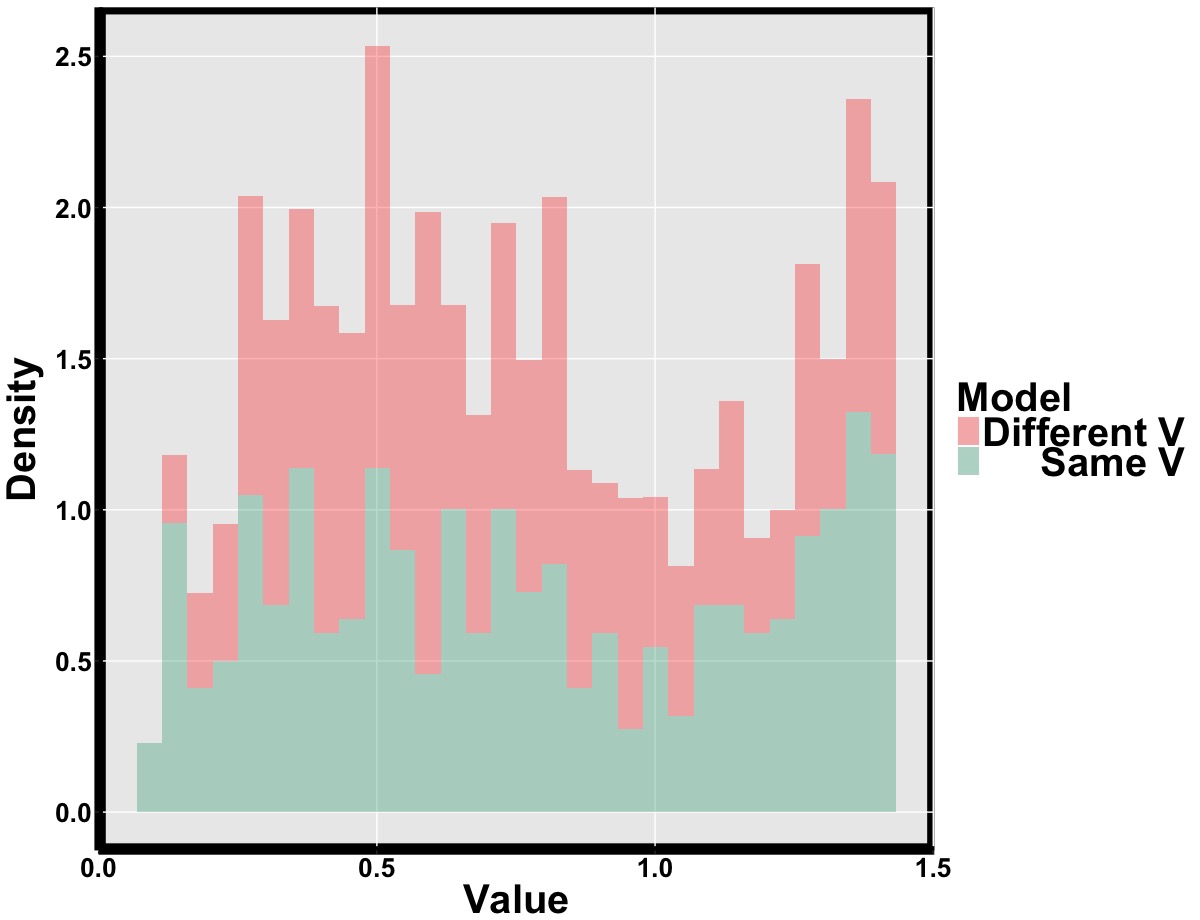}
\caption{ \em{ ``Different V" indicates the variation in $V_{i}$s which are obtained from fitting independence model separately for each individual and measured the deviation of $V_{i}$s from $\bar{V}$ which is karcher mean of individual specific $V_{i}$s. Later we have simulated a datum with $\bar{V}$, estimated $\tilde{D}$ and estimated $\sigma^{2}$ and measured the variation in a similar fashion which is represented with ``Same $V$". Here deviations are measured as $d(U, W) = \mid \mid P_{U} - P_{W} \mid \mid $ where $P_{U} =UU^{\T}$.}}
\label{validation_V}
\end{figure}

Another important modeling assumption in \eqref{HCovModel_decomp} that requires empirical justification is the use of common semi-orthogonal matrix $V$ across all the individuals as opposed to having individual specific semi-orthogonal matrices  in the independence model.  
In the following, we develop a simple diagnostic to this effect. 
\begin{figure}[h!]
\begin{subfigure}{0.45\textwidth}
\centering
\includegraphics[scale = 0.2]{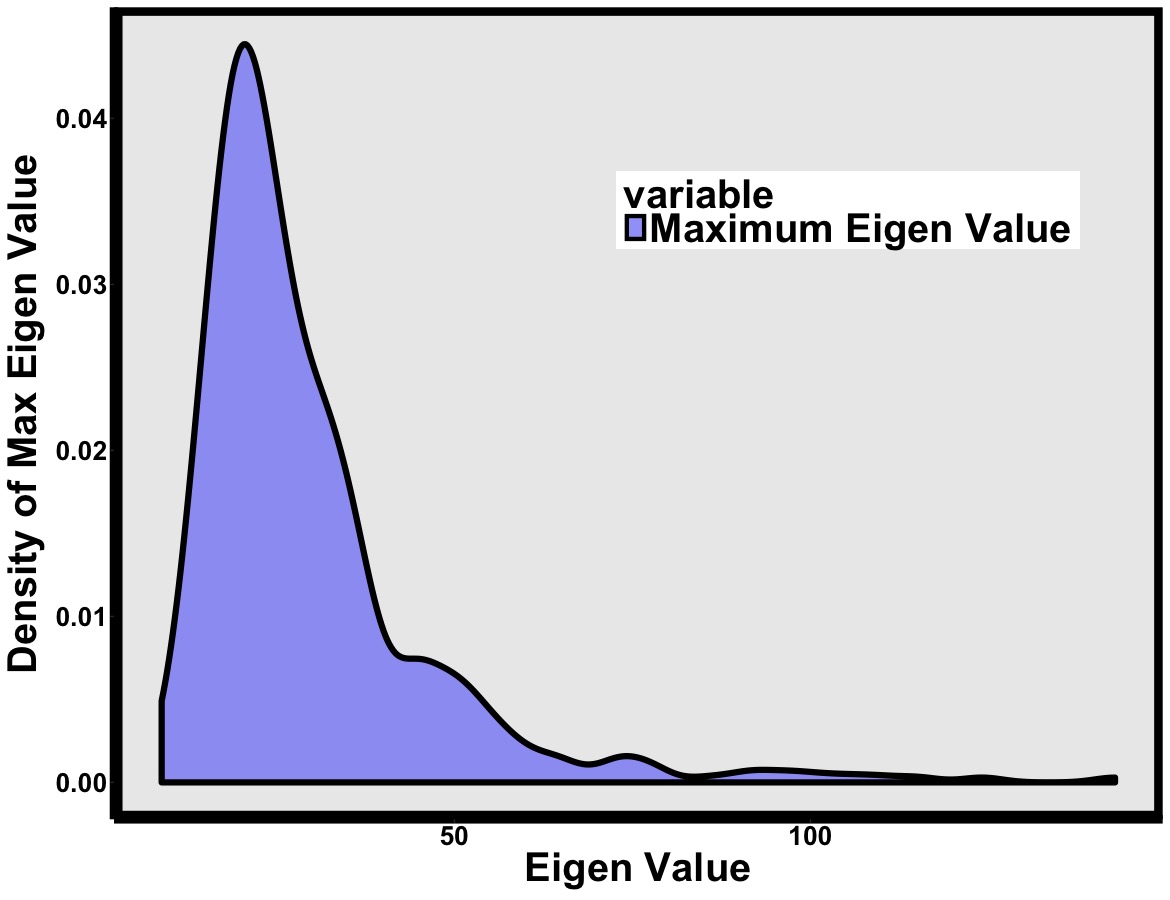}
\caption{} \label{validation_maxD}
\end{subfigure}\hfill
\begin{subfigure}{0.45\textwidth}
\centering
\includegraphics[scale = 0.2]{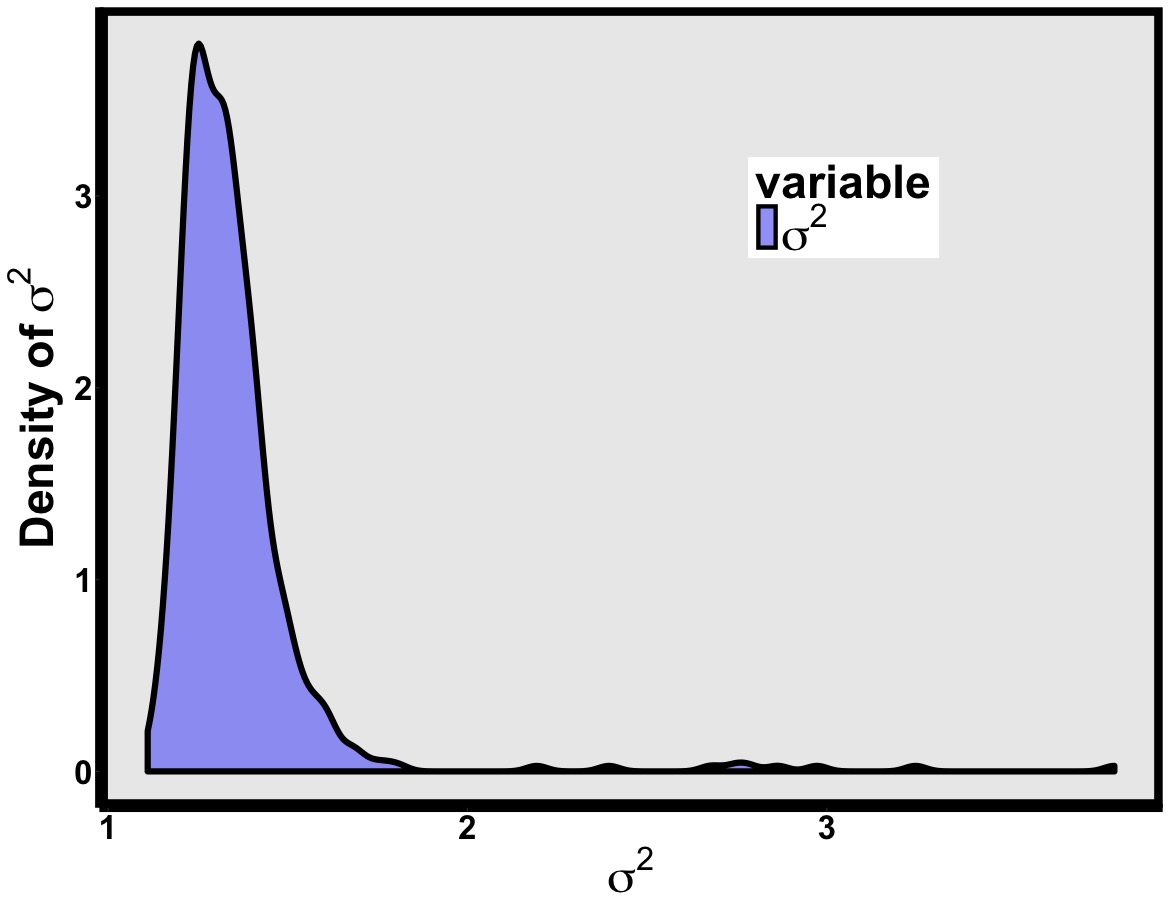}
\caption{} \label{validation_sigma_sq}
\end{subfigure}
\caption{ \em{Density plots of the estimated maximum eigenvalue (\ref{validation_maxD}) and $\sigma^2$ (\ref{validation_sigma_sq}) across individuals obtained from fitting the independence model separately for each individual in the HPC dataset. }}
\label{validation_maxD_sigmaSQ}
\end{figure}
First we fit the independence model separately for each individual and for each $i$ we calculate the Karcher mean  \citep{Marks_manifold} of posterior samples of $\{ V_{i} \}_{i=1}^{n}$ to obtain individual specific posterior estimate of the semi-orthogonal matrices, denoted $\{ \hat{V^{1}_{i}} \}_{i=1}^{n}$. For each $i$, variability of the estimate of the semi-orthogonal matrix is measured as $d( \hat{V}^{1}_{i}, \bar{V})$ where  $\bar{V} =$ Karcher mean of $\{ \hat{V}^{1}_{i} \}_{i=1}^{n}$ and $d(U, W) = \mid \mid P_{U} - P_{W} \mid \mid $ where $P_{U} =UU^{\T}$. In Figure \ref{validation_V}, ``Different V" shows the histogram of $ \{ d( \hat{V}^{1}_{i}, \bar{V}) \}_{i=1}^{n}$ describing the variability for the individual specific semi-orthogonal matrices. 
We generated data from \eqref{eqModel} with individual specific 
$\Omega_i$ set as $\bar{V}\hat{D}_i \bar{V}^{T} + \hat{\sigma}_i^2 I$ for all individuals $i=1, \ldots, n$ where $\hat{D}_i$ and $\hat{\sigma}_i$ are the individual specific posterior estimates from the independence model. We then refit the independence model to this new dataset and acquired individual specific posterior estimates of the semi-orthogonal matrices = $\{ \hat{V^{2}_{i}} \}_{i=1}^{n}$. Variability  of these semi-orthogonal matrices is measured as $d( \hat{V}^{2}_{i}, \bar{V})$ for each $i( = 1, \dots , n)$. ``Same V" in Figure \ref{validation_V} denotes the histogram of $ \{ d( \hat{V}^{2}_{i}, \bar{V}) \}_{i=1}^{n}$. Figure \ref{validation_V} clearly indicates a reduction in the variability of semi-orthogonal matrices if only a single semi-orthogonal matrix is considered across all individuals.

After fitting a complex Bayesian model, it is important to compare its predictive accuracy with other models, both simple and complex \citep{geisser1979predictive,hoeting1999bayesian,vehtari2012survey}. 
Cross-validation and information criteria are two approaches to estimate out-of-sample predictive accuracy using within-sample fits. The deviance information criterion (DIC) \citep{spiegelhalter2002bayesian} is a generalization of the Akaike information criterion (AIC) \citep{akaike1998information} for hierarchical settings. DIC has gained in popularity in recent years, in part through its implementation in the graphical modeling package BUGS, but it is known to suffer from issues from not being fully Bayesian.  The  
Watanabe-Akaike  information criterion (WAIC) \citep{watanabe2010asymptotic} can be viewed as an improvement on the deviance information criterion (DIC) for Bayesian models.   In this article, we use WAIC to compare the Independence model,  the hierarchical covariance model and the its extension to detect changepoints.  Computation time for
the WAIC is negligible compared to the cost of fitting the model and
obtaining posterior draws.   
\begin{table}[h!]
\begin{center}
\begin{tabular}{|c|c|} \hline
Model &WAIC \\ \hline
Independence Model &$37.7500$   \\ \hline
Hierarchical Covariance Model &$20.1703$   \\  \hline
Hierarchical Change-point Model &$19.6739$   \\  \hline
\end{tabular}
\end{center}
\caption{\em{ WAIC values for three models which are defined in \eqref{eqModel}, \eqref{HCovModel_decomp} and \eqref{HCpModel} respectively for HCP dataset under motor task with fitted rank value $5$. Reported WAIC values are in scale of $10^{4}$. }}
\label{WAIC_table}
 \end{table}
 The WAIC values suggest progressively better fits as we move from the independence model to the most complex change point model.

\section{Discussion}
To discover patterns within the connectivity matrix of human brain as subjects perform specific tasks, we start with a simple Wishart distribution with an approximate low rank structure on the mean for modeling the covariance valued data.  The methodology allows straightforward extension to a hierarchical model of multiple subjects where covariance valued time series is available for each subject.  Another important extension is to develop a method for detecting a single change point in the covariance time series.  Applying the methodology to the HCP data for the motor task reveals that the change point is associated with a particular regime switch of the experimental design. Also, the application to the resting state individuals in the ADNI study does not reveal any change point, which is in accordance with the expert opinions.

Another interesting application related to the HCP dataset  is where the subjects are performing psychometric tasks and the goal is to understand how the connectivity evolves  over time and whether a particular pattern in the time series motif is associated with the subjects ``intelligence'' or mental ability.  In this case, the goal is to understand how the connectivity changes with time and it is important to allow more complex time varying structure in the evolution of the covariance matrix.  Such applications also call for development of joint model of the mental ability scores and the connectivity matrices and is an interesting topic for future research. 

For simplicity, we focused on a single Wishart distribution as a model for the covariance value data. A more flexible alternative beyond the Wishart family is to consider a mixture of Wishart distributions, particularly to allow for departures that are not captured by a single scale parameter. However, this comes with an additional burden of identifying and interpreting the component specific mean parameters that are required to be properly regularized to get a meaningful inference.

\bibliography{hmcv}

%\begin{appendix}
\input{appendix_hmcv}
%\end{appendix}

\end{document}

%% file: appendix_hmcv.tex
\begin{appendix}
\addcontentsline{toc}{section}{Appendices}
\setcounter{table}{0}
\renewcommand{\thetable}{A\arabic{table}}
\counterwithin{figure}{section}
\part{Appendix} 
\parttoc 

In Appendix \ref{Dynamic_change_point_model_section}, we explore a dynamic extension of our hierarchical change point model \eqref{HCpModel}. Appendix \ref{Hcp_model_sensitivity} contains sensitivity analysis for the hierarchical change point model \eqref{HCpModel} and some additional simulations to study robustness. Appendix \ref{HCP_gambling} contains results for the gambling task for the HCP data. In Appendix \ref{variable_phi}, we validate our assumption of considering $\phi=p+1$. Appendix \ref{derivation_IND} \& \ref{derivation_HCM} contain details of posterior computation under the independence \& hierarchical models respectively.

\section{Dynamic Change Point Model} \label{Dynamic_change_point_model_section} 
 In our current change-point estimation framework, the observational units before and after the change-point are assumed to be independent and identically distributed. A dynamic extension of our model is certainty interesting to incorporate the dependence across the time-points.  To that end, we explore an auto-regressive Wishart process on the subject specific loadings. This is studied under the hierarchical change point model for one subject with a single change point: 
\begin{equation} \label{time_HCpModel}
\begin{split}
 & S_{it} \sim W(\phi_{1}, \phi_{1}^{-1} \Omega^{(t)}_{1_{i}}), \hspace{0.3cm}  \Omega^{(t)}_{1_{i}} = V_{1}D^{(t)}_{1_{i}}V_{1}^{\T} + \sigma_{1_{i}}^{2}I_{p}, \hspace{0.3cm} t = 1,\dots,c_{i} \hspace{0.1cm} ,\\
 & S_{it} \sim W(\phi_{2}, \phi_{2}^{-1} \Omega^{(t)}_{2_{i}}), \hspace{0.3cm} \Omega^{(t)}_{2_{i}} = V_{2}D^{(t)}_{2_{i}}V_{2}^{\T} + \sigma_{2_{i}}^{2}I_{p},  \hspace{0.3cm} t = c_{i}+1,\dots,T.
\end{split}
\end{equation}
We use the same orthogonal factor model type decomposition on $\Omega$ at each time point of an individual. As a parsimonious model for the time dependent covariance matrices, the orthogonal matrix $V$ is held constant with subject specific time varying process for the loadings $D_{i}^{(t)}$ and subject specific variance components. For computational convenience, we consider $\tilde{D}_{i}^{(t)} = D_{i}^{(t)} / \sigma^{2}$ which are assumed to evolve following an auto-regressive process: 
\begin{align}\label{AR1_hc}
\begin{rcases}
& \log \tilde{d}_{1_{ih}}^{(t)} = \rho_{1} \log \tilde{d}_{1_{ih}}^{(t-1)} + \epsilon_{1_{ih}}^{(t)},  \hspace{0.3cm} t = 2,\dots,c_{i} \\  
& \log \tilde{d}_{2_{ih}}^{(t)} = \rho_{2} \log \tilde{d}_{2_{ih}}^{(t-1)} + \epsilon_{2_{ih}}^{(t)},  \hspace{0.3cm} t = c_{i}+2,\dots,T
\end{rcases} 
\quad
\begin{array}{r@{\;}l}
h = 1, \ldots , r^{\ast}. 
\end{array}
\end{align}
The diagonal elements at the first time point are shrunk  using global and local shrinkage: 
\begin{align}\label{AR1_hc_1st_time}
\begin{split}
& \log \tilde{d}_{1_{ih}}^{(1)} =  \log \lambda_{1_{ih}}^{(1)} + \log \tau_{1_{i}}^{(1)} \\
& \log \tilde{d}_{2_{ih}}^{(c_{i}+1)} =  \log \lambda_{2_{ih}}^{(c_{i}+1)} + \log \tau_{2_{i}}^{(c_{i}+1)}.  
\end{split}
\end{align}
Here $\tau_{1_{i}}^{(1)}$, $\tau_{2_{i}}^{(c_{i}+1)}$ are the global shrinkage parameters while the $\lambda_{ih}^{(1)}$, $\lambda_{2_{ih}}^{(c_{i}+1)}$ parameters allow for coordinate specific deviations. We place independent half-Cauchy priors on the local parameters and half-Cauchy prior truncated to $(0, 1)$ on the global parameters. Also,  $\epsilon_{1_{ih}}^{(t)} \sim N(0,v_{1})$ and $\epsilon_{2_{ih}}^{(t)} \sim N(0,v_{2})$ with $v_{1}, v_{2} \overset{iid.}\sim \text{Gamma(1/2,2)}$. 

\begin{figure}[h!]
\centering
\includegraphics[scale = 0.2]{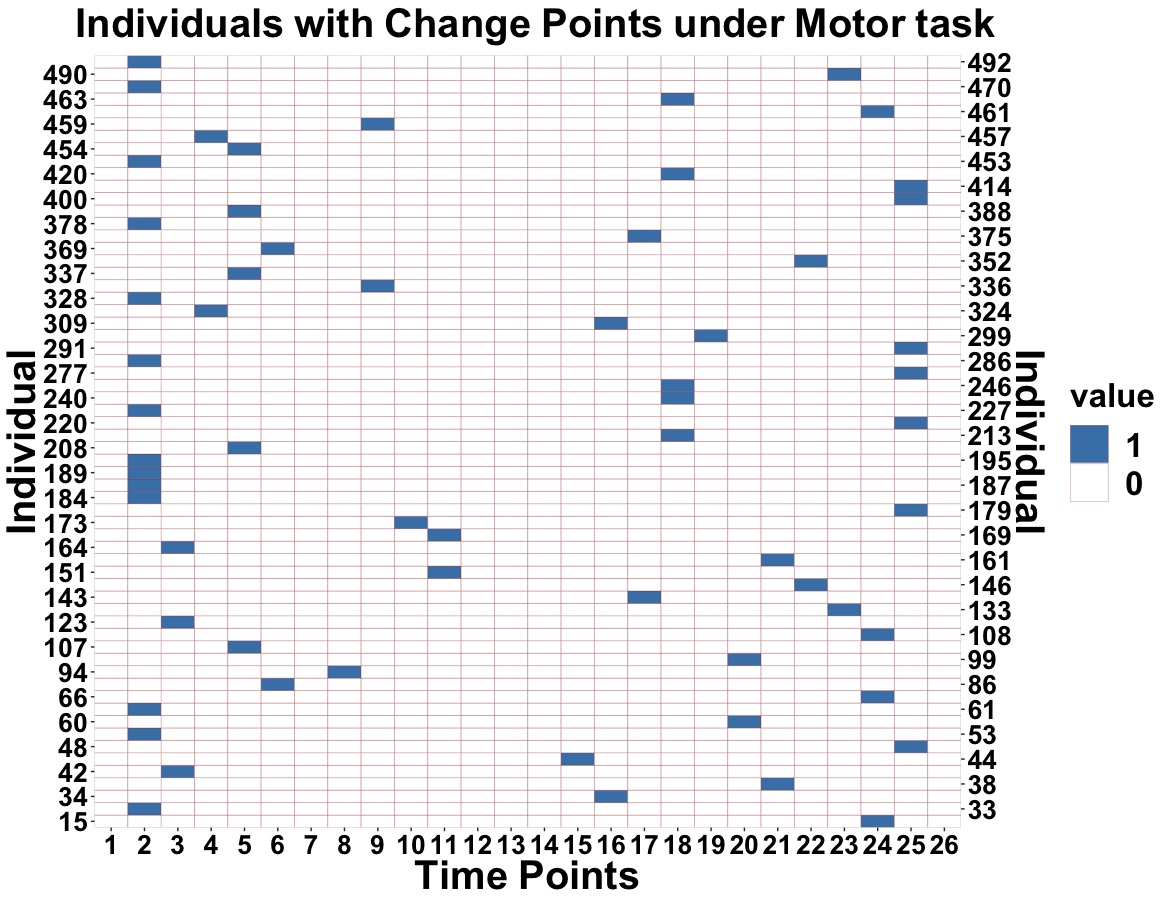}
\caption{\em{Detected change points for different individuals under Motor task from dynamic change point model \eqref{time_HCpModel}. On both sides of the y-axis of the plot we have labeled the individuals. Blue lines denote the individual specific change points.} }
\label{Motor_Dynamic_change_point}
\end{figure}

We detected change points of $62$ individuals from the  dynamic change point model which is provided in Figure \ref{Motor_Dynamic_change_point}.  There are indeed certain change points which are detected by both \eqref{HCpModel} and \eqref{time_HCpModel}. For example,  the change points for individual $66$, $133$ and $173$ are detected at $24$th, $23$rd and $10$th time point using both the models. 
Although the specific dynamic framework can be extended to a full-blown dynamic model, initial indications of the WAIC values suggest that it may not be a good fit for the current dataset. We leave this as a topic for future research.

\section{Sensitivity and robustness analysis of the hierarchical change point model}\label{Hcp_model_sensitivity}
We considered different values of the fitted rank $(r^{\ast} = 5,6,7,8)$ and ran the hierarchical change point model \eqref{HCpModel} on the HCP dataset under the Motor task to check the sensitivity of the results with respect to the fitted rank. The findings are provided in Table \ref{table:hcp_model_different_fitted_rank}. We obtained similar results for 26 individuals from different values of fitted ranks which are provided in the last column of Table \ref{table:hcp_model_different_fitted_rank}. Next we considered five different initializations for the model parameters in \eqref{HCpModel} and fixed $r^{\ast} = 5$. Detected change points with the corresponding individuals are shown in Table \ref{table:hcp_model_different_initial_setting} and the frequently appeared $34$ individuals for all those 5 settings are provided in last column of the table. The description of these initial values are provided next. For setting $1$, we set $\{ \sigma^{2}_{1_{i}}, \sigma^{2}_{2_{i}} \}_{i=1}^{n}$ to be $2.5$, $\{ \tau_{1_{i}}, \tau_{2_{i}} \}_{i=1}^{n} = 0.1$ and  $\{ \lambda_{1_{i}}, \lambda_{2_{i}} \}_{i=1}^{n}=2.5$. In case of setting $2$, we consider $\{ \sigma^{2}_{1_{i}}, \sigma^{2}_{2_{i}} \}_{i=1}^{n}$ to be $1.75$, $\{ \lambda_{1_{i}}, \lambda_{2_{i}} \}_{i=1}^{n}=2.25$ and values for all values of $\tau$ were same with setting $1$. Values of $\{ \sigma^{2}_{1_{i}}, \sigma^{2}_{2_{i}} \}_{i=1}^{n}$ was set to be $1.5$, $\{ \lambda_{1_{i}}, \lambda_{2_{i}} \}_{i=1}^{n}=3$ and same value for all $\tau$ parameters for setting $3$. In setting $4$, we have $\{ \sigma^{2}_{1_{i}}, \sigma^{2}_{2_{i}} \}_{i=1}^{n}$ to be $2.25$, $\{ \lambda_{1_{i}}, \lambda_{2_{i}} \}_{i=1}^{n}=3.5$ and $\{ \tau_{1_{i}}, \tau_{2_{i}} \}_{i=1}^{n} = 0.2$. In case of setting $5$, we consider $\{ \sigma^{2}_{1_{i}}, \sigma^{2}_{2_{i}} \}_{i=1}^{n}$ to be $2.5$, $\{ \lambda_{1_{i}}, \lambda_{2_{i}} \}_{i=1}^{n}=3.25$ and $\{ \tau_{1_{i}}, \tau_{2_{i}} \}_{i=1}^{n} = 0.25$. 

\begin{figure}[h!] 
\begin{subfigure}{0.45\textwidth}
\centering
\includegraphics[scale = 0.16]{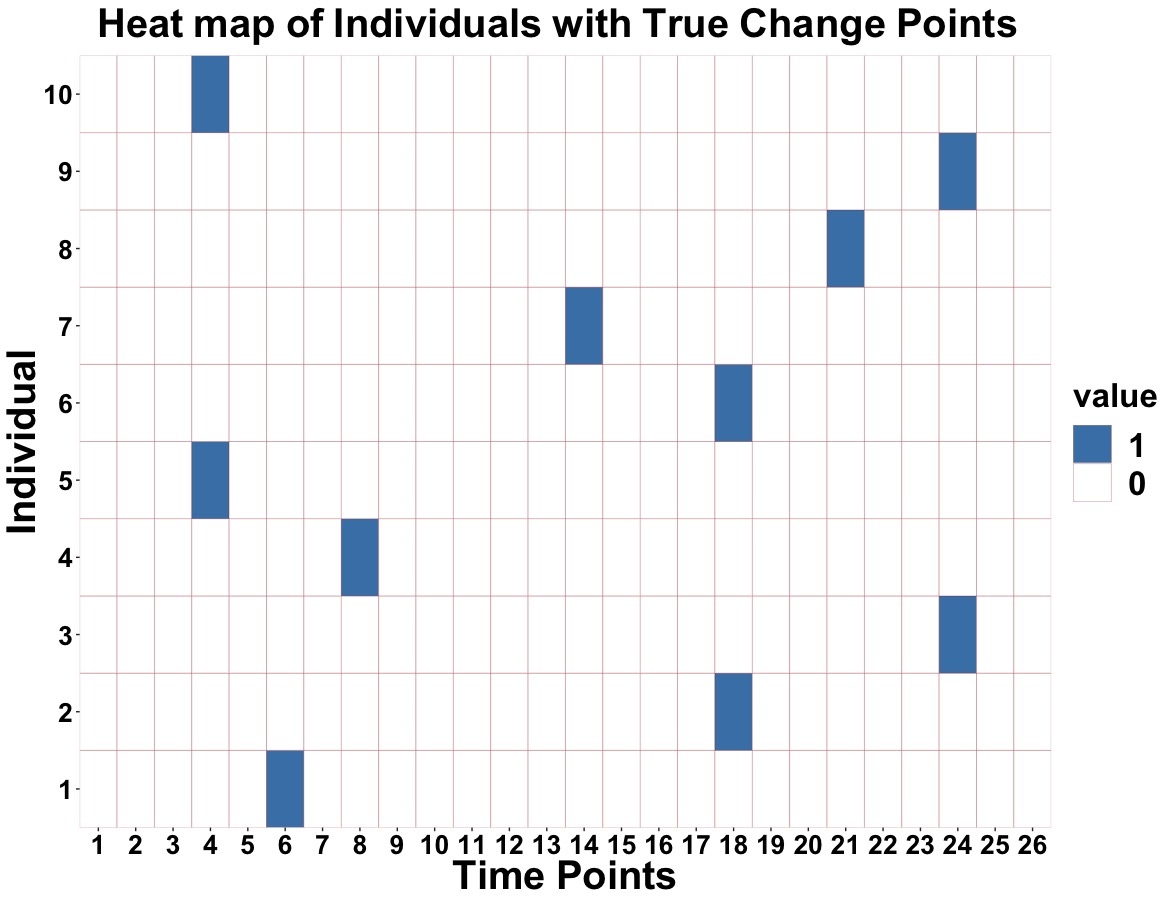}
\end{subfigure}\hfill
\begin{subfigure}{0.45\textwidth}
\centering
\includegraphics[scale = 0.16]{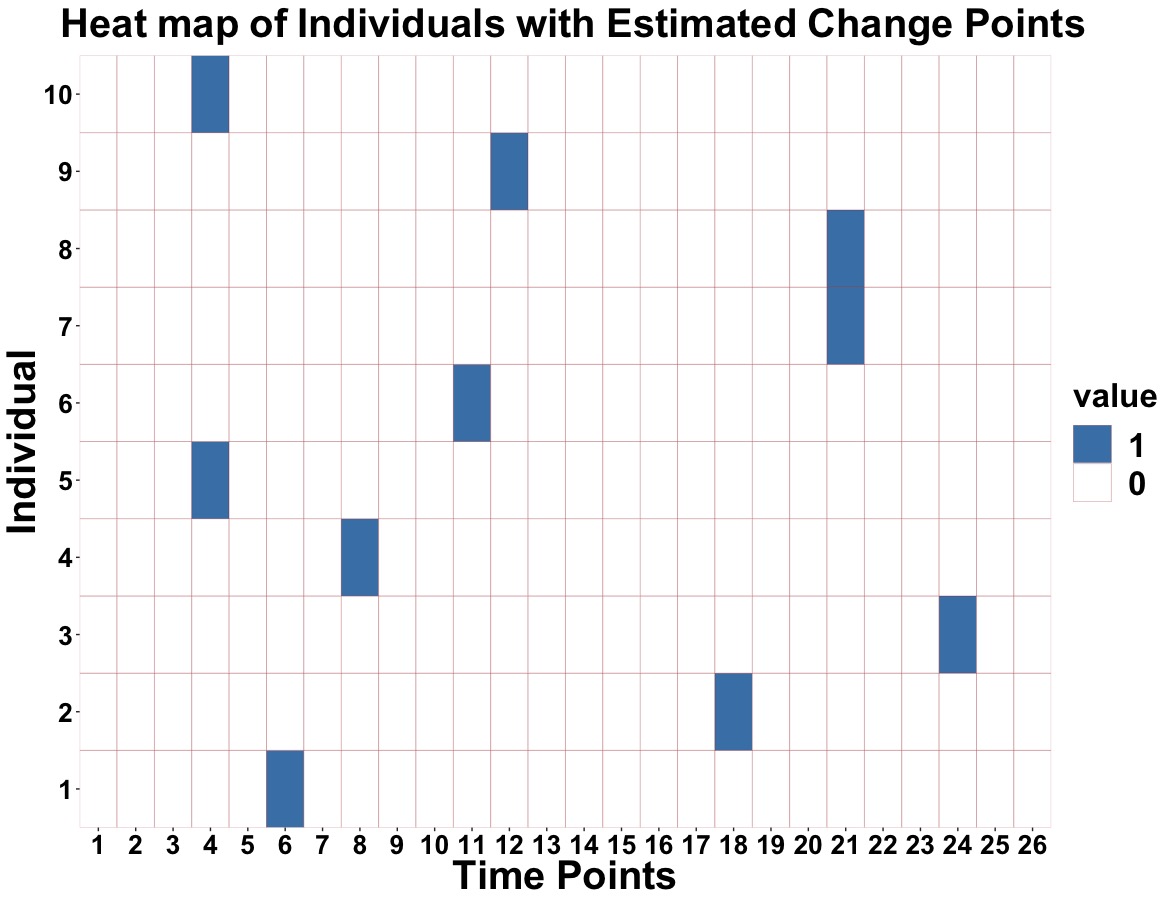}
\end{subfigure}
\caption{\em{Individual specific change point estimates obtained from hierarchical change point model (8) whereas the data is generated from time-varying hierarchical change point model \eqref{time_HCpModel}.}} 
\label{re_Hcp_robust}
\end{figure}

We conducted another experiment to test the robustness of the change point model when the data generating distribution indeed has correlation across time points. 
We simulated data from the time varying change point model \eqref{time_HCpModel} consisting of $10$ subjects with $26$ time points. All the subjects have individual specific change points which are generated uniformly in the interval $\{2, \dots , 25 \}$. For clarity, any parameter with subscript ``1" correspond to the pre-change-point regime (Group 1) and the ones with subscript ``2" corresponds to the post-change-point (Group 2). Individual specific variances $\{ \sigma^{2}_{01_{i}} \}_{i=1}^{n}$ and $\{ \sigma^{2}_{02_{i}} \}_{i=1}^{n}$ are generated from $\text{uniform}(0.25,0.50)$. Local parameters $\{ \lambda_{1_{i}}^{(1)}, \lambda_{2_{i}}^{(c_{i}+1)} \}_{i=1}^{n}$ and global parameters $\{ \tau_{1_{i}}^{(1)}, \tau_{2_{i}}^{(c_{i}+1)} \}_{i=1}^{n}$ are generated from half-Cauchy distribution. $\{ D^{(1)}_{01_{i}} \}_{i=1}^{n}$ and $\{ D^{(1)}_{02_{i}} \}_{i=1}^{n}$ are generated by following equation \eqref{AR1_hc_1st_time}. We use AR(1) process \eqref{AR1_hc} to generate rest of the columns of $D_{01}$ and $D_{02}$. Figure \ref{re_Hcp_robust} shows that our non-time varying model$(8)$ model is able to detect the individual specific change points for 60\% cases even though the data is generated from the time varying hierarchical change point model. 

\begin{table}[h!]
\begin{center}
\tiny{
\begin{tabular}{|cc|cc|cc|cc|cc|}
\hline
 \multicolumn{2}{|c|}{$r^{\ast}=5$}  &\multicolumn{2}{c|}{$r^{\ast}=6$} &\multicolumn{2}{c|}{$r^{\ast}=7$} &\multicolumn{2}{c|}{$r^{\ast}=8$} &\multicolumn{2}{c|}{Common Ind} \\  \hline
 Change Point &Ind &Change Points &Ind  &Change Points &Ind &Change Points &Ind &Change Points &Ind \\ \hline
$23$  &$11$    &$10$  &$13$    &$10$  &$13$     &$23$  &$11$      &$10$   &$13$  \\ \hline
$10$  &$13$    &$24$  &$15$    &$24$  &$15$     &$10$  &$13$     &$23$   &$15$  \\ \hline
$23$  &$15$    &$18$  &$56$    &$18$  &$56$     &$23$  &$15$     &$18$   &$56$  \\ \hline
$18$  &$56$    &$24$  &$66$    &$24$  &$66$     &$18$  &$56$     &$24$  &$66$  \\ \hline
$24$  &$66$    &$21$  &$69$    &$21$  &$69$     &$24$  &$66$     &$8$    &$99$  \\ \hline
$25$  &$90$    &$25$  &$90$    &$25$  &$90$     &$21$  &$69$     &$23$  &$108$  \\ \hline
$8$    &$99$    &$8$    &$99$    &$3$    &$93$     &$23$  &$82$     &$23$  &$118$  \\ \hline
$23$  &$108$  &$22$  &$108$  &$25$  &$101$   &$25$  &$90$     &$20$  &$129$  \\ \hline
$23$  &$118$  &$24$  &$118$  &$25$  &$108$   &$8$     &$99$     &$23$  &$133$  \\ \hline
$20$  &$129$  &$20$  &$129$  &$20$  &$129$   &$25$  &$101$   &$13$  &$147$   \\ \hline
$23$  &$133$  &$23$  &$133$  &$23$  &$133$   &$22$  &$108$   &$7$    &$165$   \\ \hline
$13$  &$147$  &$13$  &$147$  &$13$  &$147$   &$23$  &$118$   &$19$  &$177$   \\ \hline
$18$  &$165$  &$18$  &$165$  &$7$    &$165$   &$21$  &$122$   &$22$  &$201$   \\ \hline
$10$  &$173$  &$19$  &$179$  &$10$  &$173$   &$20$  &$129$   &$25$  &$220$   \\ \hline
$19$  &$179$  &$22$  &$201$  &$19$  &$179$   &$23$  &$133$   &$23$  &$240$   \\ \hline
$17$  &$201$  &$25$  &$211$  &$22$  &$201$   &$13$  &$147$   &$10$  &$257$   \\ \hline
$25$  &$220$  &$25$  &$220$  &$25$  &$220$   &$7$    &$165$   &$25$  &$277$   \\ \hline
$23$  &$240$  &$23$  &$240$  &$23$  &$240$   &$10$  &$173$   &$21$  &$329$   \\ \hline
$10$  &$257$  &$10$  &$257$  &$10$  &$257$   &$18$  &$177$   &$18$  &$333$   \\ \hline
$25$  &$270$  &$25$  &$277$  &$25$  &$270$   &$19$  &$179$   &$22$  &$400$   \\ \hline
$9$    &$274$  &$18$  &$333$  &$9$    &$274$   &$22$  &$201$   &$11$  &$405$   \\ \hline
$25$  &$277$  &$25$  &$335$  &$25$  &$277$   &$25$  &$220$   &$25$  &$414$   \\ \hline
$21$  &$329$  &$23$  &$380$  &$21$  &$329$   &$20$  &$224$   &$20$  &$420$   \\ \hline
$18$  &$333$  &$25$  &$383$  &$18$  &$333$   &$17$  &$225$   &$22$  &$467$   \\ \hline
$25$  &$335$  &$22$  &$400$  &$25$  &$383$   &$2$    &$240$   &$19$  &$482$   \\ \hline
$23$  &$380$  &$11$  &$405$  &$22$  &$400$   &$10$  &$257$   &$23$  &$490$  \\ \hline
$23$  &$381$  &$25$  &$414$  &$11$  &$405$   &$25$  &$264$   &          &           \\ \hline
$22$  &$400$  &$18$  &$420$  &$25$  &$414$   &$24$  &$267$   &          &           \\ \hline
$11$  &$405$  &$22$  &$467$  &$28$  &$420$   &$25$  &$270$   &          &           \\ \hline
$25$  &$414$  &$19$  &$482$  &$10$  &$427$   &$9$    &$274$   &          &           \\ \hline
$18$  &$420$  &$23$  &$490$  &$22$  &$467$   &$25$  &$277$   &          &            \\ \hline
$10$  &$427$  &          &            &$19$  &$482$   &$6$    &$291$   &          &            \\ \hline
$13$  &$439$  &          &            &$23$  &$490$   &$21$  &$329$   &          &            \\ \hline
$22$  &$467$  &          &            &          &             &$18$  &$333$   &          &            \\ \hline
$19$  &$482$  &          &            &          &             &$23$  &$380$   &          &            \\ \hline
$23$  &$490$  &          &            &          &             &$23$  &$381$   &          &            \\ \hline
          &            &          &            &          &             &$14$  &$399$   &          &            \\ \hline
          &            &          &            &          &             &$22$  &$400$   &          &            \\ \hline
          &            &          &            &          &             &$11$  &$405$   &          &            \\ \hline 
          &            &          &            &          &             &$25$  &$414$   &          &            \\ \hline
          &            &          &            &          &             &$18$  &$420$   &          &            \\ \hline
          &            &          &            &          &             &$18$  &$427$   &          &            \\ \hline
          &            &          &            &          &             &$22$  &$467$   &          &            \\ \hline 
          &            &          &            &          &             &$19$  &$482$   &          &            \\ \hline 
          &            &          &            &          &             &$23$  &$490$   &          &            \\ \hline                                                                   
\end{tabular}
}
\end{center}
\caption{{Detected change points and corresponding individuals from the hierarchical change point model\eqref{HCpModel} for different values of fitted rank.}}
\label{table:hcp_model_different_fitted_rank}
\end{table}

\begin{table}[h!]
\begin{center}
\tiny{
\begin{tabular}{|cc|cc|cc|cc|cc|cc|}
\hline
 \multicolumn{2}{|c|}{Setting 1}  &\multicolumn{2}{c|}{Setting 2} &\multicolumn{2}{c|}{Setting 3} &\multicolumn{2}{c|}{Setting 4}  &\multicolumn{2}{c|}{Setting 5}  &\multicolumn{2}{c|}{Common Ind}  \\  \hline
Change Point &Ind &Change Points &Ind  &Change Points &Ind &Change Points &Ind &Change Points &Ind &Change Points &Ind \\ \hline
$23$  &$11$     &$23$  &$11$      &$23$  &$11$      &$23$  &$11$      &$23$  &$11$      &$23$    &$11$   \\ \hline
$23$  &$15$     &$10$  &$13$      &$10$  &$13$     &$10$  &$13$      &$10$  &$13$      &$10$    &$13$   \\ \hline
$18$  &$56$     &$23$  &$15$      &$23$  &$15$     &$23$  &$15$      &$23$  &$15$      &$23$    &$15$         \\ \hline
$24$  &$66$     &$18$  &$56$      &$18$  &$56$     &$19$  &$23$      &$18$  &$56$      &$18$    &$56$       \\ \hline
$21$  &$69$     &$24$  &$66$      &$24$  &$66$     &$19$  &$32$      &$24$  &$66$      &$24$    &$66$      \\ \hline
$25$  &$90$     &$25$  &$90$      &$25$  &$90$     &$18$  &$56$      &$25$  &$90$      &$25$    &$90$       \\ \hline
$8$    &$99$     &$8$    &$99$      &$8$    &$99$     &$24$  &$66$      &$8$    &$99$      &$8$      &$99$      \\ \hline
$23$  &$108$   &$23$  &$108$    &$23$  &$108$   &$21$  &$69$      &$24$  &$108$    &$23$    &$108$       \\ \hline
$23$  &$118$   &$23$  &$118$    &$23$  &$118$    &$25$  &$90$      &$23$  &$118$    &$23$    &$118$       \\ \hline
$20$  &$129$   &$20$  &$129$    &$20$  &$129$   &$8$    &$99$      &$20$  &$129$    &$20$    &$129$      \\ \hline
$23$  &$133$   &$23$  &$133$    &$23$  &$133$   &$25$  &$101$    &$23$  &$133$    &$23$    &$133$       \\ \hline
$13$  &$147$   &$13$  &$147$    &$13$  &$147$   &$23$  &$108$    &$13$  &$147$    &$13$    &$147$         \\ \hline
$18$  &$165$   &$7$    &$165$    &$7$    &$165$   &$23$  &$118$    &$18$  &$165$    &$18$    &$165$       \\ \hline
$10$  &$173$   &$10$  &$173$    &$10$  &$173$   &$20$  &$129$    &$10$  &$173$    &$10$    &$173$          \\ \hline
$19$  &$179$   &$19$  &$179$    &$19$  &$179$   &$23$  &$133$    &$19$  &$179$    &$19$    &$179$          \\ \hline
$17$  &$201$   &$17$  &$201$    &$17$  &$201$   &$25$  &$139$    &$17$  &$201$    &$17$    &$201$           \\ \hline
$25$  &$220$   &$25$  &$220$    &$25$  &$220$   &$13$  &$147$    &$25$  &$220$    &$25$    &$220$           \\ \hline
$23$  &$240$   &$23$  &$240$    &$23$  &$240$   &$7$    &$165$    &$23$  &$240$    &$23$    &$240$           \\ \hline
$10$  &$257$   &$10$  &$257$    &$10$  &$257$   &$10$  &$173$    &$10$  &$257$    &$10$    &$257$           \\ \hline
$9$    &$274$   &$25$  &$270$    &$9$    &$274$   &$19$  &$179$    &$25$  &$270$    &$9$      &$274$            \\ \hline
$25$  &$277$   &$9$    &$274$    &$25$  &$277$   &$22$  &$201$    &$9$    &$274$    &$25$    &$277$           \\ \hline
$21$  &$329$   &$25$  &$277$    &$21$  &$329$   &$25$  &$220$    &$25$  &$277$    &$21$    &$329$           \\ \hline
$18$  &$333$   &$21$  &$329$    &$18$  &$333$   &$23$  &$240$    &$21$  &$329$    &$18$    &$333$           \\ \hline
$25$  &$335$   &$18$  &$333$    &$25$  &$335$   &$10$  &$257$    &$18$  &$333$    &$25$    &$335$           \\ \hline
$23$  &$380$   &$25$  &$335$    &$22$  &$400$   &$25$  &$270$    &$25$  &$335$    &$23$    &$380$           \\ \hline
$23$  &$381$   &$23$  &$380$    &$25$  &$414$   &$9$    &$274$    &$23$  &$380$    &$22$    &$400$           \\ \hline
$22$  &$400$   &$25$  &$381$    &$18$  &$420$   &$25$  &$277$    &$23$  &$381$    &$11$    &$405$           \\ \hline
$11$  &$405$   &$22$  &$400$    &$13$  &$439$   &$21$  &$329$    &$22$  &$400$    &$25$    &$414$           \\ \hline
$25$  &$414$   &$11$  &$405$    &$22$  &$467$   &$18$  &$333$    &$11$  &$405$    &$18$    &$420$           \\ \hline
$18$  &$420$   &$25$  &$414$    &$19$  &$482$   &$25$  &$335$    &$25$  &$414$    &$10$    &$427$           \\ \hline
$24$  &$427$   &$18$  &$420$    &$23$  &$490$   &$23$  &$380$    &$18$  &$420$    &$13$    &$439$         \\ \hline
$13$  &$439$   &$10$  &$427$    &          &             &$23$  &$381$    &$10$  &$427$    &$22$    &$467$         \\ \hline
$22$  &$467$   &$13$  &$439$    &          &             &$14$  &$399$    &$13$  &$439$    &$19$    &$482$         \\ \hline
$19$  &$482$   &$22$  &$467$    &          &             &$22$  &$400$    &$22$  &$467$    &$23$    &$490$          \\ \hline
$23$  &$490$   &$19$  &$482$    &          &             &$11$  &$405$    &$19$  &$482$    &          &                 \\ \hline
          &             &$23$  &$490$    &          &             &$25$  &$414$    &$23$  &$490$    &          &                \\ \hline
          &             &          &              &          &             &$18$  &$420$    &          &              &          &                \\ \hline
          &             &          &              &          &             &$10$  &$427$    &          &              &          &                \\ \hline     
          &             &          &              &          &             &$13$  &$439$    &          &              &          &               \\ \hline     
          &             &          &              &          &             &$22$  &$467$    &          &              &          &              \\ \hline 
          &             &          &              &          &             &$19$  &$482$    &          &              &          &              \\ \hline     
          &             &          &              &          &             &$23$  &$490$    &          &              &          &              \\ \hline                      
\end{tabular}
}
\end{center}
\caption{{Detected change points and corresponding individuals from the hierarchical change point model\eqref{HCpModel} for different initial setting.}}
\label{table:hcp_model_different_initial_setting}
\end{table}

\begin{figure}[h!]
\begin{subfigure}{0.45\textwidth}
\centering
\includegraphics[scale = 0.16]{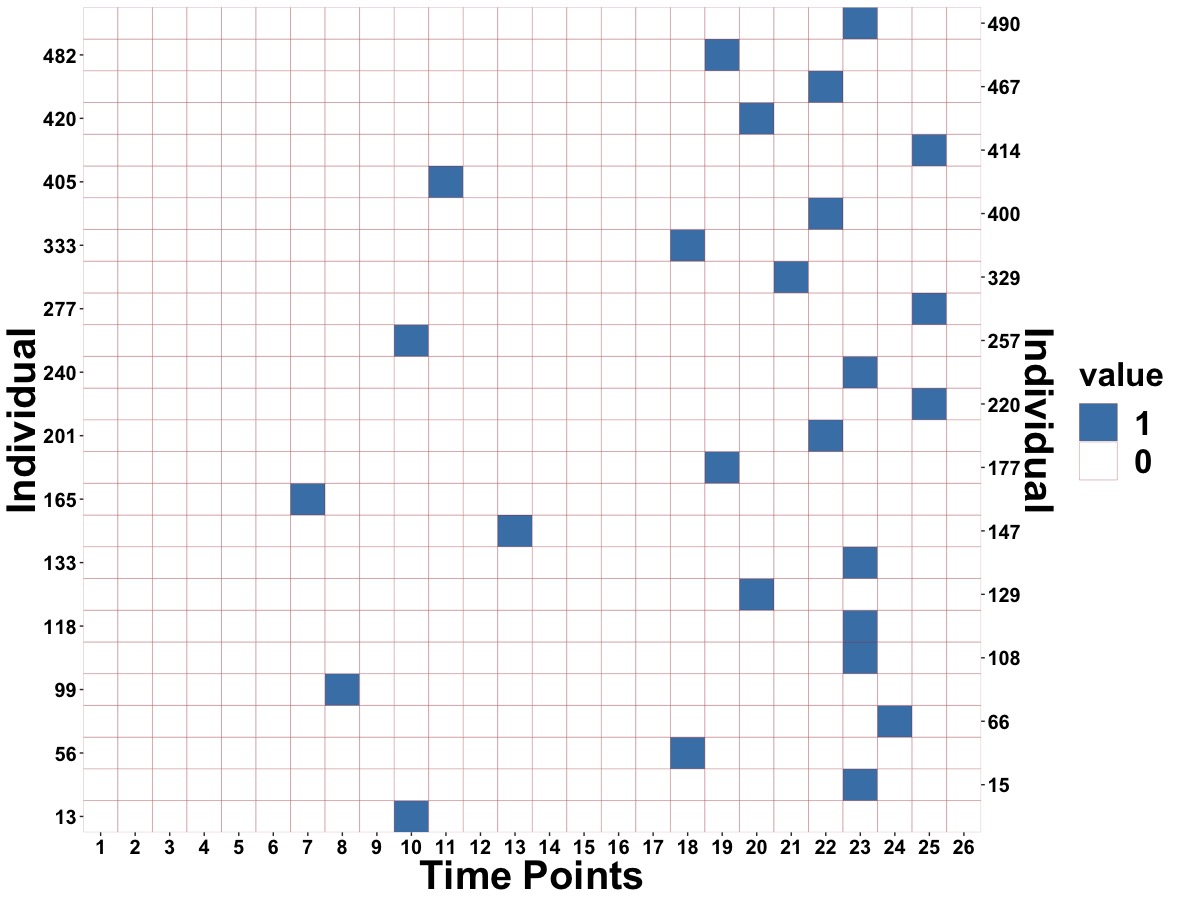}
\caption{} \label{Motor_Change_Point_diff_rank}
\end{subfigure}\hfill
\begin{subfigure}{0.45\textwidth}
\centering
\includegraphics[scale = 0.16]{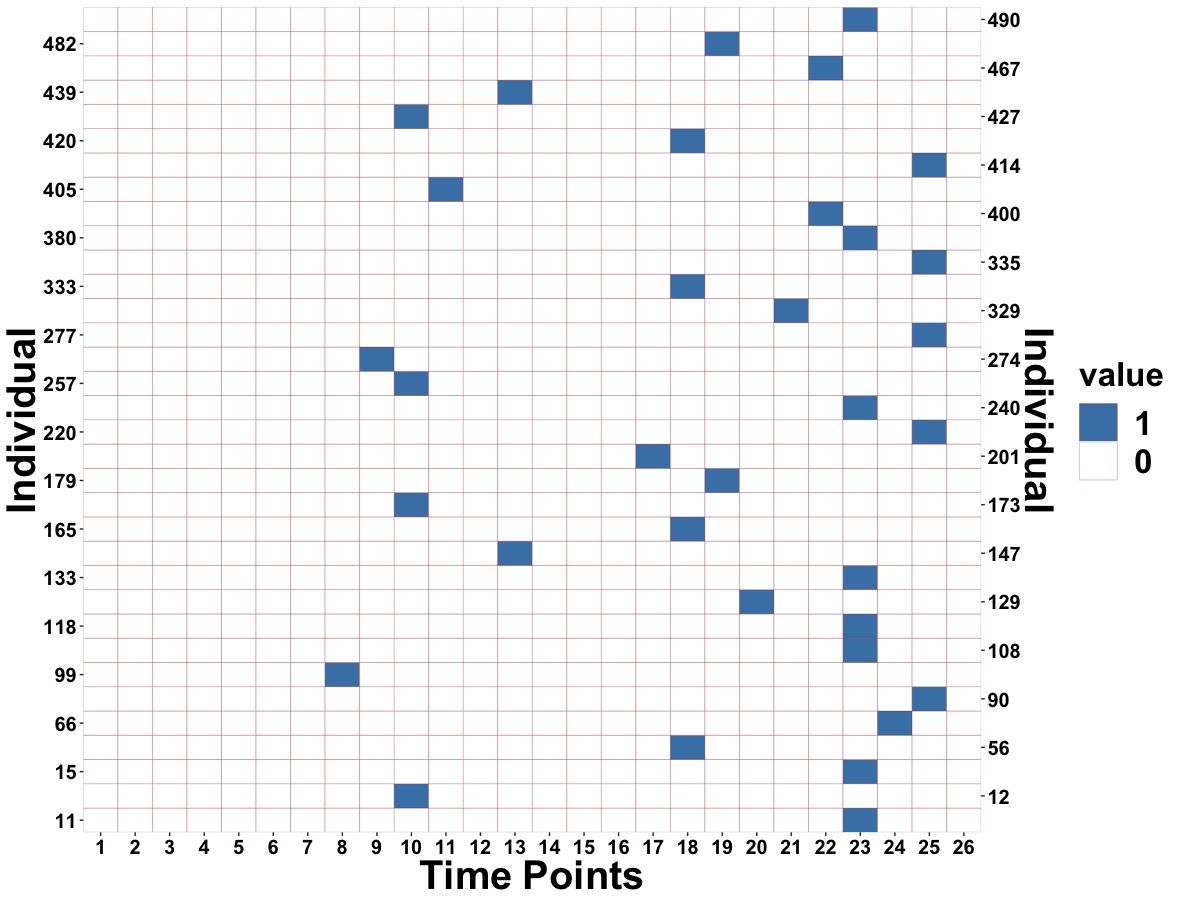}
\caption{} \label{hist_chp_Motor_diiferent_initial_set}
\end{subfigure}
\caption{\em{ Common change points and corresponding individuals detected from the hierarchical change point model\eqref{HCpModel} for different values of fitted rank (\ref{Motor_Change_Point_diff_rank}) and initial setting (\ref{hist_chp_Motor_diiferent_initial_set}).} }
\label{hcp_change_point_motor_diff_initial_set}
\end{figure}

\section{Case study for gambling task of HCP dataset}\label{HCP_gambling}
\begin{figure}[h!]
\begin{subfigure}{0.45\textwidth}
\centering
\includegraphics[scale = 0.16]{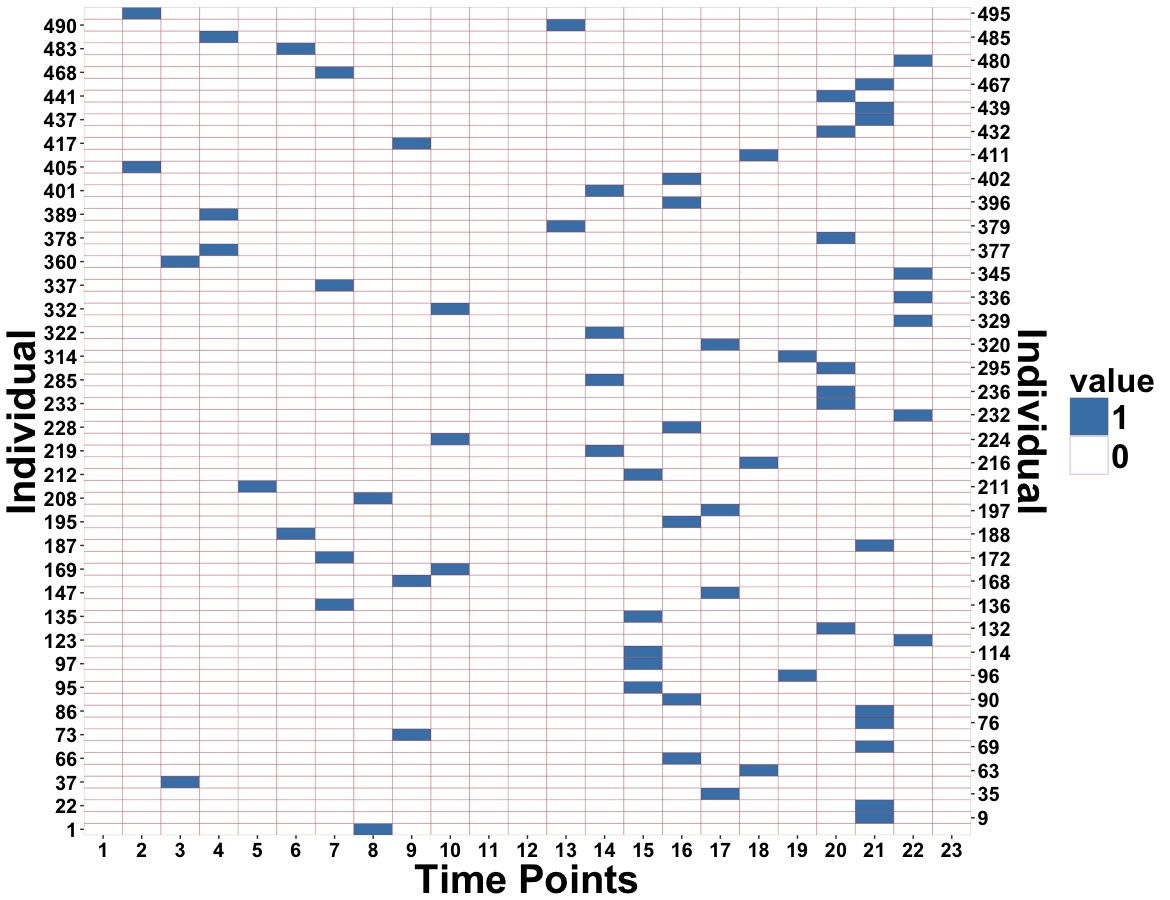}
\caption{} \label{Gambling_Change_Point}
\end{subfigure}\hfill
\begin{subfigure}{0.45\textwidth}
\centering
\includegraphics[scale = 0.16]{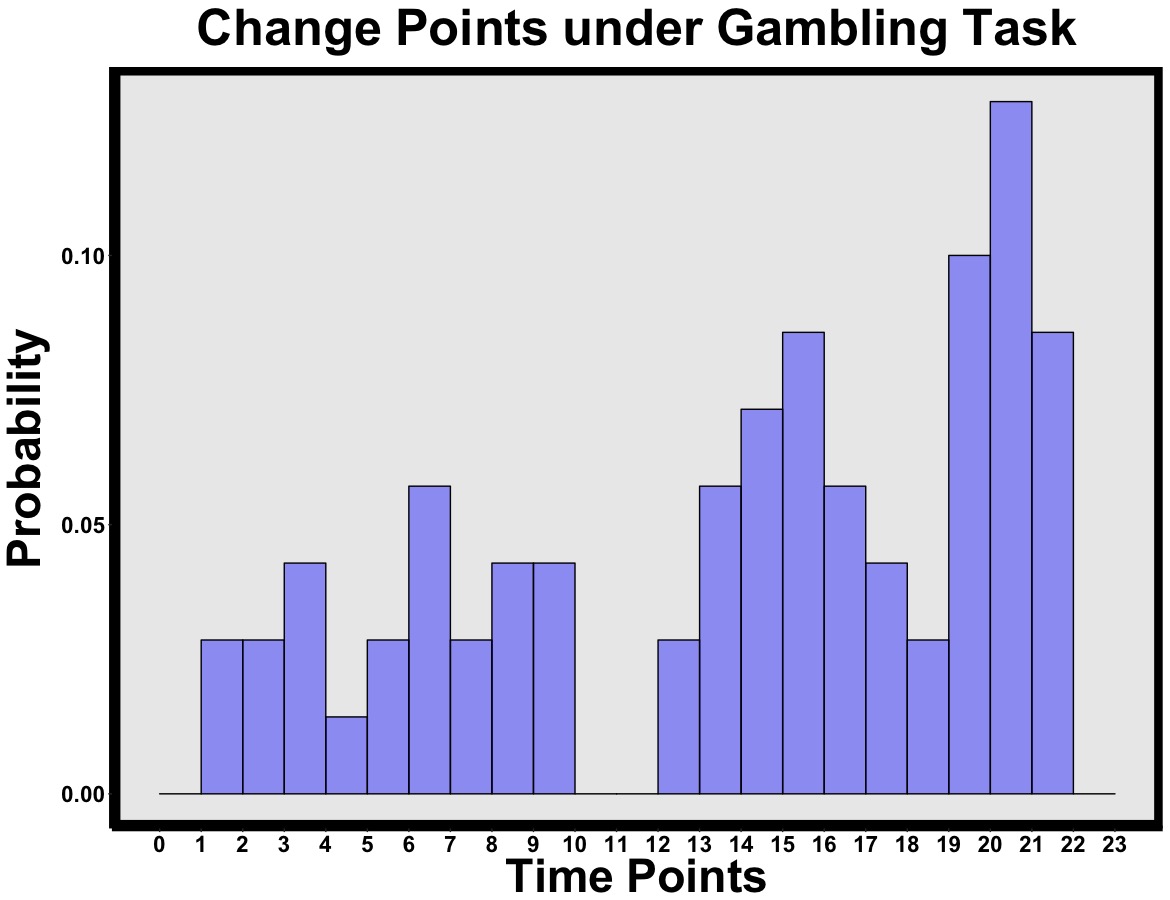}
\caption{} \label{hist_chp_gambling}
\end{subfigure}
\caption{\em{ Right panel shows heat map of binary matrix consisting of $1$ to $(i,j)^{\text{th}}$ position which corresponds to $i^{\text{th}}$ individual and $j^{\text{th}}$ $(j=1,\dots,23)$ time point which is a change point for the corresponding individual and $0$ otherwise. Individuals with one change point under gambling task are labeled on y-axis. In left panel we have histogram of change points under gambling task which shows most of the individuals have change point at $21$. }}
\label{Gambling_change_point}
\end{figure}

\begin{figure}[h!]
\centering
\includegraphics[scale = 0.18]{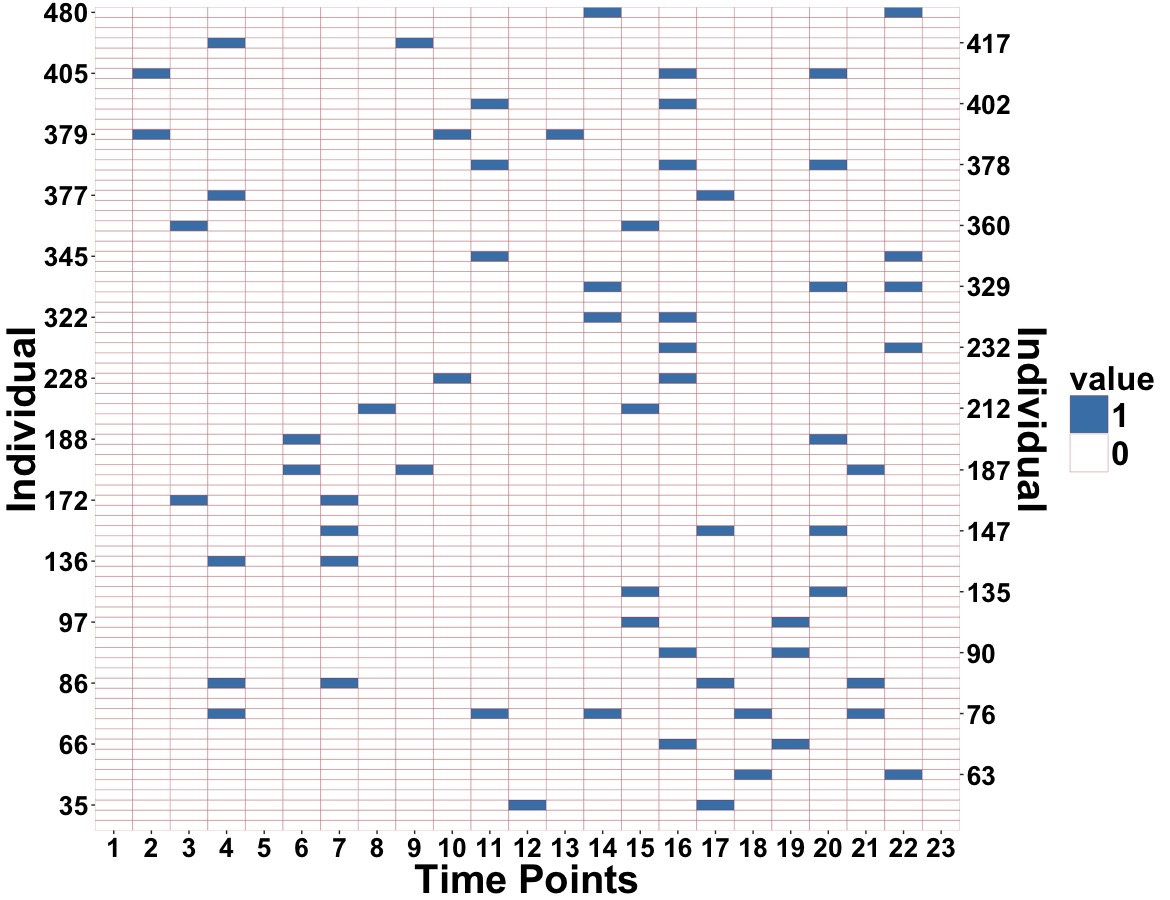}
\caption{\em{Multiple change points for different individuals under Gambling task. On y-axis of the plot we have labeled the individuals with multiple change points and on x-axis we have time points. Blue lines denote the individual specific change points.} }
\label{Gambling_Multiple_change_point}
\end{figure}

A detailed description of the gambling task corresponding to HCP is in \cite{Delgado2000}. We applied our hierarchical change point model on data which was scaled with subject specific minimum eigen value. We are able to detect $70$ individuals with change points under Gambling task. In Figure \ref{Gambling_change_point} we labeled the $70$ individuals with their corresponding change points. Figure \ref{Gambling_Change_Point} consists of 70 individuals with their most prominent change point detected through the hierarchical model. Figure \ref{hist_chp_gambling} shows the overall pattern of the most prominent change points across all the individuals.  The histogram in Figure \ref{hist_chp_gambling} shows that more than 10\% individuals have change points at $21$.  
Next we extended our study to detect multiple change points under gambling task. We applied the methodology discussed in section \ref{multiple_chp_HCP} on individuals under gambling task and detected multiple change points for different individuals. In Figure \ref{Gambling_Multiple_change_point}, we listed the individuals with at least two change points and their corresponding change points. 

\section{Validation of the value of $\phi$}\label{variable_phi}
In this section, we validate the choice of $\phi$ based on WAIC. Under the same simulation setting as in Section \ref{subsec:indep_sim}, the data is generated from the independence model \eqref{eqModel} by setting the true values of $\phi$ in  $\{11,15,20\}$. The fitted values of $\phi$ are varied over a range $\{11,\dots,20\}$ and corresponding WAIC values are reported in Table \eqref{table:vary_independence_model}. Irrespective of the true value of $\phi$, the lowest WAIC value is obtained when $\phi=11 \in \{11,\dots,20\}$. For the HCP data analysis, we compare the WAIC values from the hierarchical change point model \eqref{HCpModel} for several values of $\phi$ as shown in Table \eqref{table:Real_data_hcp_model_WAIC_vary_phi}. The lowest WAIC is attained at $\phi = 11$.

\begin{table}[h!]
\begin{center}
\begin{tabular}{|c|c||c|c||c|c||c|c||c|c||c|c|}
\hline
\multicolumn{4}{|c||}{True $\phi = 11$} &\multicolumn{4}{|c||}{True $\phi = 15$} &\multicolumn{4}{|c|}{True $\phi = 20$}\\
\hline
 $\phi$ &WAIC  &$\phi$ &WAIC  &$\phi$ &WAIC  &$\phi$ &WAIC   &$\phi$ &WAIC  &$\phi$ &WAIC \\ \hline
 11  &223980.5   &16   &982594.9       &11  &224046.5   &16   &891307.5      &11  &227826.6   &16    &832607.7  \\ \hline
 12  &350710.9   &17   &1161531.7     &12  &331893.7   &17   &1052557.1    &12  &326247.8   &17   &980285.6   \\ \hline
 13  &492837.7   &18   &1346223.5     &13  &454632.8   &18   &1219633.2    &13  &437647.2   &18   &1133538.3   \\ \hline
 14  &646175.0   &19   &1535622.7     &14  &590192.5   &19   &1391076.2    &14  &559376.4   &19   &1291091.1   \\ \hline
 15  &810530.4   &20   &1729035.9     &15  &736743.1   &20   &1566453.2    &15  &692127.0   &20   &1452505.0   \\ \hline
\end{tabular}
\end{center}
\caption{{WAIC values obtained from the independence model \eqref{eqModel} with true values of $\phi = \{11,15,20\}$ and fitted values are varied over a range $\{11,\dots,20 \}$. }}
\label{table:vary_independence_model}
\end{table}

\begin{table}[h!]
\begin{center}
\begin{tabular}{|c|c||c|c|}
\hline
\multicolumn{4}{|c|}{Real Data Analysis} \\
\hline
$\phi$ &WAIC  &$\phi$ &WAIC \\ \hline
 11  &196739   &16   &257329  \\ \hline
 12  &209801   &17   &271796   \\ \hline
 13  &218313   &18   &289761   \\ \hline
 14  &233022   &19   &297537   \\ \hline
 15  &242873   &20   &312590   \\ \hline
\end{tabular}
\end{center}
\caption{{WAIC values obtained by implementing hierarchical change point model on the HCP dataset with different values of $\phi \in \{ 11, \dots, 20 \}$. }}
\label{table:Real_data_hcp_model_WAIC_vary_phi}
\end{table}

\section{Posterior Computation under Independence Model}\label{derivation_IND}
We first record two useful identities. We have \\
 $ 
 |\Omega | = | \sigma^{2} I_{p} + VDV^{\T} | = (\sigma^{2})^{p} \bigg| I_{p} + V \frac{D}{\sigma^{2}} V^{\T} \bigg| = (\sigma^{2})^{p} | I_{p} + V \tilde{D} V^{\T} | $ \\
$ = (\sigma^{2})^{p} | I_{r^{\ast}} + (V^{\T}V) \tilde{D} | =  (\sigma^{2})^{p} \displaystyle \prod_{j=1}^{r^{\ast}} \bigg(1 + \frac{d_{j}}{\sigma^{2}} \bigg) = (\sigma^{2})^{p} \prod_{j=1}^{r^{\ast}} (1 + \tilde{d}_{j})
$ , \\
where $\tilde{D} = D/\sigma^{2}$. Moreover,

$ \Omega^{-1} = (\sigma^{2} I_{p} + VDV^{\T})^{-1} = \sigma^{-2}\bigg(I_{p} + V  \frac{D}{\sigma^{2}} V^{\T} \bigg)^{-1} = \sigma^{-2}(I_{p} + V  \tilde{D} V^{\T})^{-1} $ \\
$ = \sigma^{-2} \{ I_{p} - V (\tilde{D}^{-1} + V^{\T}V )^{-1} V^{\T} \} = \sigma^{-2} \{ I_{p} - V (\tilde{D}^{-1} + I_{r^{\ast}} )^{-1} V^{\T} \} = \sigma^{-2} \{ I_{p} - V E^{-1} V^{\T} \} $ ,  \\
where $E = \tilde{D}^{-1} + I_{r^{\ast}} $.

\begin{itemize}
\item \textbf{ Likelihood under Independence Model}: 
The joint likelihood of $(V,D,\sigma^{2})$ is 

$ L(V,D,\sigma^{2}) =  |\Omega|^{- \frac{N\phi}{2}} \prod_{j=1}^{N} \exp\bigg\{-\frac{1}{2} \tr(\Omega^{-1} S_{j})\bigg\} = |\Omega|^{- \frac{N\phi}{2}} \exp\bigg\{-\frac{1}{2} \tr(\Omega^{-1} \sum_{j=1}^{N} S_{j}) \bigg\}$ \\
$ = |\Omega|^{-  \frac{N\phi}{2}} \exp\bigg\{-\frac{1}{2} \tr(\Omega^{-1} S^{N}) \bigg\}$ \\
$ = (\sigma^{2})^{-\frac{N \phi p}{2}} \prod_{j=1}^{r^{\ast}}\bigg(1+\frac{d_{j}}{\sigma^{2}}\bigg)^{-\frac{N\phi}{2}} \exp\bigg[ -\frac{1}{2\sigma^{2}} \tr\{ ( I_{p} - V (\sigma^{2} D^{-1} + I_{r^{\ast}} )^{-1} V^{\T} ) S^{N} \} \bigg] $ , \\
where $\displaystyle S^{N} = \sum_{j=1}^{N} S_{j}$.

With the transformation $\tilde{D} = D/\sigma^{2}$, the joint likelihood of $(V,\tilde{D},\sigma^{2})$ is \\
$
L(V,\tilde{D},\sigma^{2}) = (\sigma^{2})^{-\frac{N \phi p}{2} + 1} \prod_{j=1}^{r^{\ast}}(1+\tilde{d_{j}})^{-\frac{N\phi}{2}} \exp\bigg[ -\frac{1}{2\sigma^{2}} \tr\{ ( I_{p} - V ( \tilde{D}^{-1} + I_{r^{\ast}} )^{-1} V^{\T} ) S^{N}\} \bigg]. 
$ \\

We now describe the full-conditional distributions that are used to implement a Metropolis within Gibbs sampler to sample from the joint posterior of $(V,\tilde{D},\sigma^{2})$. We use the notation $[\theta \mid -]$ to denote the full conditional distribution of a parameter. 

\vspace{0.5cm}
\item \textbf{ Full conditional of V}:

We have,

$ [V \mid -] \propto \exp\bigg[ \frac{1}{2\sigma^{2}} \tr\{  V E^{-1} V^{\T} S^{N}\}\bigg] = \exp\bigg[ \frac{1}{2\sigma^{2}} \tr\{  E^{-1} V^{\T} S^{N} V\}\bigg] $. \\
This is a Bingham$(S^{N},E^{-1}/2\sigma^{2})$ distribution, which can be sampled using the package Rstiefel in R.

\vspace{0.5cm}
\item \textbf{Full conditional of $\beta_{h}$}: 

We decomposed $\log \tilde{d}_h = \mu + \beta_h$ and considered $M = V^{T}S^{N}V$ and a transformation $w_{h} = (1+\beta_{h} \mu)^{-1}$ for $h = 1,\dots,r^{\ast}$. %Now the full conditional of $\beta_{h}$ is provided below. \\
The full-conditional of $w_{h}$ is give by \\
$ [w_{h} \mid -] \propto w_{h}^{\frac{N\phi}{2} -2} \exp\bigg( - \frac{\phi M_{hh}}{2\sigma^{2}} w_{h} \bigg)  \frac{1}{\mu^{2} + \bigg( \frac{1-w_{h}}{w_{h}} \bigg)^{2} } $, \\
To sample from the above, we consider a slice-sampling scheme. Specifically, augment a latent variable $u_{h}$ such that the joint distribution of $(w_{h}, u_{h})$ is \\
$ [w_{h},u_{h} \mid -] \propto w_{h}^{\frac{N\phi}{2} -2} \exp\bigg( - \frac{\phi M_{hh}}{2\sigma^{2}} w_{h} \bigg) I\bigg[0 < u_{h} < \frac{1}{\mu^{2} + \bigg( \frac{1-w_{h}}{w_{h}} \bigg)^{2} } \bigg] $. \\
It is clearly seen that the marginal distribution of $w_{h}$ is preserved under this joint distribution.  We then successively sample from the conditionals $[ u_{h} \mid w_{h}, - ]$ and $[ w_{h} \mid u_{h}, - ]$. We have \\
$[ u_{h} \mid w_{h}, - ] \sim$ Uniform$[ 0, \{ \mu^{2} + (( 1-w_{h})/w_{h} )^{2} \}^{-1} ]$ , \\
$[ w_{h} \mid u_{h}, - ] \sim$ Gamma(shape$=(N\phi/2) - 1$, rate $= \phi M_{hh} / 2 \sigma^{2}$)$I[ w_{h} >  \{1+ \sqrt{ (1/u_{h})-\mu^{2}} \}^{-1} ]$. \\

\item \textbf{Full conditional of $\mu$}:

The full conditional of $\mu$ is \\
$ [ \mu \mid - ] \propto \displaystyle \prod_{j=1}^{r^{\ast}} (1+\beta_{h} \exp(\mu))^{-\frac{N\phi}{2}} \exp \bigg[ \sum_{j=1}^{r^{\ast}} c_{h}( 1 + \frac{1}{\beta_{h} \exp(\mu) })^{-1} \bigg] \frac{\exp(\mu)}{1+ \exp(2\mu)} $. \\
where $c_{h} = \phi M_{hh}/(2\sigma^{2})$. We use a Metropolis--Hastings step to sample $\mu$ using a normal random-walk proposal as $\mu^{*} \sim N(\mu , s^{2})$ with standard deviation $s = 0.1$. We accept $\mu^{*}$ with probability $\min\{  \alpha(\mu,\mu^{*}) , 1 \}$ where
\begin{align*}
\alpha(\mu,\mu^{*}) = \frac{ \Pi(\mu^{*} \mid - ) }{ \Pi(\mu \mid - ) }.
\end{align*}

\item \textbf{Full conditional of $\sigma^{2}$:} 

$[ \sigma^{2} \mid - ] \propto (\sigma^{2})^{-\frac{N \phi p}{2} + 1} \exp\bigg[ -\frac{1}{2\sigma^{2}} \tr\{ Q S^{N}\} \bigg] . (\sigma^{2})^{\alpha_{\sigma} - 1} \exp\bigg(- \frac{\beta_{\sigma}}{\sigma^{2}} \bigg) $ \\
$ = (\sigma^{2})^{- \frac{N \phi p}{2} - \alpha_{\sigma}} \exp\bigg[ -\frac{1}{\sigma^{2}} \bigg\{  \beta_{\sigma} + \frac{tr(QS^{N})}{2} \bigg\}   \bigg] $ , \\
where $Q = ( V\tilde{D}V^{\T} + I_{p} )^{-1} V^{\T} $. This implies \\
$[ \sigma^{2} \mid - ] \sim$ Inverse-Gamma$(\alpha_{\sigma}-1 + Np\phi/2 , \beta_{\sigma} + \tr(QS^{N})/2)$.

\end{itemize}

\section{Posterior Computation and Algorithm under Hierarchical Model}\label{derivation_HCM}
The joint likelihood of $(V,D,\sigma^2)$ under the hierarchical covariance model \eqref{HCovModel_decomp} is 
\begin{align*}
\begin{split}
& L(V,D,\sigma^{2} ) \\
& = \bigg[ \prod_{i=1}^{n} \prod_{t=1}^{T}  |\Omega_{i}|^{- \frac{\phi}{2}} \exp\bigg\{-\frac{\phi}{2} \tr(\Omega_{i}^{-1} S_{it})\bigg\} \bigg] = \prod_{i=1}^{n} |\Omega_{i}|^{- \frac{ T \phi }{2}} \exp\bigg\{-\frac{\phi}{2} \tr(\Omega_{i}^{-1} \sum_{t=1}^{T} S_{it})\bigg\} \\
& = \bigg[ \prod_{i=1}^{n} \bigg\{ (\sigma_{i}^{2})^{-\frac{ pT \phi }{2}} \prod_{j=1}^{r^{\ast}}\bigg(1+\frac{d_{ij} }{\sigma_{i}^{2}}\bigg)^{-\frac{T\phi}{2}} \bigg\} \bigg]  \exp\bigg\{-\frac{\phi}{2} \tr \bigg( \sum_{i=1}^{n} \sigma_{i}^{-2} \{ I_{p} - V (\sigma_{i}^{2} D_{i}^{-1} + I_{r^{\ast}} )^{-1} V^{\T} \} \sum_{t=1}^{T} S_{it} \bigg)\bigg\}. 
\end{split}
\end{align*}
With the transformation $\tilde{D} = D/\sigma^{2}$, the joint likelihood of $(V,\tilde{D},\sigma^{2})$ is 
\begin{align*}
\begin{split}
& L(V,\tilde{D},\sigma^{2}) \\
& = \bigg[ \prod_{i=1}^{n} \bigg\{ (\sigma_{i}^{2})^{-\frac{ p T \phi }{2} + 1} \prod_{j=1}^{r^{\ast}}(1+\tilde{d}_{ij} )^{-\frac{T \phi}{2}} \bigg\} \bigg]  \exp\bigg\{-\frac{\phi}{2} \tr \bigg( \sum_{i=1}^{n} \sigma_{i}^{-2} \{ I_{p} - V ( \tilde{D}_{i}^{-1} + I_{r^{\ast}} )^{-1} V^{\T} \} \sum_{t=1}^{T} S_{it} \bigg)\bigg\}. \\ 
\end{split}
\end{align*}

Now we only describe detailed steps to derive the full-conditional of V. For the rest of the parameters, we cycled through the subject specific full-conditionals which are similarly derived following the posterior computation steps under the Independence model \eqref{eqModel}. We have
%\begin{itemize}
%\item \textbf{Full conditional of V}:
\begin{align*}
\begin{split}
& [ V \mid - ] \propto \exp\bigg[ \frac{\phi}{2} \sum_{i=1}^{n} \tr \bigg \{ V \frac{E_{i}^{-1} }{\sigma_{i}^{2}} V^{\T} \sum_{t=1}^{T} S_{it} \bigg \}\bigg] = \exp\bigg[ \frac{\phi}{2} \sum_{i=1}^{n} \tr \bigg \{ V \frac{E_{i}^{-1} }{\sigma_{i}^{2}} V^{\T} S_{i}^{\ast} \bigg \}\bigg]  \hspace{0.1cm},
\end{split}
\end{align*}
where $ \displaystyle  S_{i}^{\ast} = \sum_{t=1}^{T} S_{it}$ and  $E_{i} = ( \tilde{D}_{i}^{-1} + I_{r^{\ast}} )$.
The detailed steps of the identity used in \eqref{eqV_HCovM} are given as  
\begin{align*}
\begin{split}
& \bigg(V\frac{\phi E_{i}^{-1} }{2 \sigma_{i}^{2}} V^{\T} \bigg)S_{i}^{\ast} = \sum_{j=1}^{r^{\ast}} (v_{j} v_{j}^{\T}) \bigg(\frac{ \phi_{1} S_{i}^{\ast} }{ 2 e_{ij} \sigma_{i}^{2} } \bigg) , \\
& \sum_{i=1}^{n} \bigg( V\frac{\phi E_{i}^{-1} }{2 \sigma_{i}^{2}} V^{\T}\bigg) S_{i}^{\ast} = \sum_{i=1}^{n} \sum_{j=1}^{r^{\ast}} (v_{j} v_{j}^{\T}) \bigg(\frac{ \phi S_{i}^{\ast} }{ 2 e_{ij} \sigma_{i}^{2} } \bigg) \\
& = \sum_{j=1}^{r^{\ast}} (v_{j} v_{j}^{\T}) \sum_{i=1}^{n} \bigg(\frac{ \phi S_{i}^{\ast} }{ 2 e_{ij} \sigma_{i}^{2} } \bigg) = \sum_{j=1}^{r^{\ast}} (v_{j} v_{j}^{\T}) H_{j} = \sum_{j=1}^{r^{\ast}} (v_{j}^{\T} H_{j} v_{j}).\\
\end{split}
\end{align*}
The full-conditional of $V$ under the hierarchical covariance model \eqref{HCovModel_decomp} can thus be expressed as  
$$ 
[ V \mid - ] \propto \exp \bigg[ \sum_{j=1}^{r^{\ast}} v_{j}^{\T} H_{j} v_{j} \bigg] = \prod_{j=1}^{r^{\ast}} \exp( v_{j}^{\T} H_{j} v_{j} ) \hspace{0.2cm} \text{where} \hspace{0.2cm} H_{j} = \sum_{i=1}^{n} \bigg(\frac{ \phi S_{i}^{\ast} }{ 2 e_{ij} \sigma_{i}^{2} } \bigg).
$$
To sample from the above, we follow the steps in \S 3.3 of \cite{hoff2009simulation} and write $V = \{ V_{[,1]} , V_{[,-1]} \} = \{ Nz, V_{[,-1]} \}$ with $\|z\| = 1$. Here, $N$ is an $p \times (r^{\ast} - 1)$ orthonormal basis for the null space of $ V_{[,-1]} $ and $z$ is expressed as $z = N^{\T}V_{[,1]}$ because $N^{\T}N=I$. The conditional density of $z \mid V_{[,-1]}$ is derived as in \cite{hoff2009simulation} as
\begin{align*}
p(z \mid  V_{[,-1]} ) \propto \exp( z^{\T} N^{\T} H_{j} N z) = \exp( z^{\T} \tilde{H}_{j} z).
\end{align*}
We iterate through the steps 1--4 for each $j \in \{ 1 , \dots , r^{\ast} \}$ to obtain samples from the density \eqref{eqV_HCovM}:

1) $N$ = null space of $ V_{[,-j]} $ and $z_{j} = N^{T} V_{[,j]} $.

2) $\tilde{H}_{j} = N^{\T} H_{j} N$.

3) Sample $z_j$ from a vector Bingham($\tilde{H}_{j}$) density using the package \texttt{rsteifel} \citep{rstiefel_Hoff}. 

4) Set $v_{j} = N z_{j}$.

\vspace{1cm}
Next, we outline the full conditionals of the time points under hierarchical change point model \eqref{HCpModel}. Full conditional updates of the rest of the parameters are similar to full conditional under hierarchical covariance model \eqref{HCovModel_decomp}. We have 
\begin{align*}
\begin{split}
& P(c_{i} = k | -) = \frac{A_{k_{i}}}{\displaystyle \sum_{k=1}^{T} A_{k_{i}}}, \hspace{0.5cm}  k = 1,\dots,T , \\
& \text{where} \hspace{0.15cm} A_{k_{i}} = (\sigma_{1_{i}}^{2})^{-\frac{ p \phi_{1} k}{2} + 1} \prod_{j=1}^{r_{1}}(1+ \tilde{d}_{1_{ij}} )^{-\frac{ \phi_{1} k}{2}} \exp\bigg[ -\frac{\phi_{1}}{2\sigma_{1_{i}}^{2}} \tr\{ Q_{1_{i}} S_{1_{i}}^{'}\} \bigg] \times \\
& (\sigma_{2_{i}}^{2})^{-\frac{ p \phi_{2} (T-k)}{2} + 1} \prod_{j=1}^{r_{2}}(1+ \tilde{d}_{2_{ij}} )^{-\frac{\phi_{2} (T-k)}{2}} \exp\bigg[ -\frac{\phi_{2}}{2\sigma_{2_{i}}^{2}} \tr\{ Q_{2_{i}} S_{2_{i}}^{'}\} \bigg] ,
\end{split}
\end{align*}
with $\displaystyle S_{1_{i}}^{'} = \sum_{t=1}^{k} S_{it}$, $ \displaystyle S_{2_{i}}^{'} = \sum_{t=k+1}^{T} S_{it}$.

%\end{itemize}

\end{appendix}
	